\ifx\pdfoutput\undefined
\newcommand\texorpdfstring[2]{#1}
\else
\fi
\documentclass[prd,twocolumn,showpacs,showkeys,preprintnumbers,floatfix,nofootinbib,superscriptaddress,10pt,aps]{revtex4-1}
\usepackage{amsfonts} 
\usepackage{amssymb} 
\usepackage{amsmath} 
\usepackage{graphicx} 
\usepackage{subfigure} 
\usepackage{array} 
\usepackage{dcolumn} 
\usepackage{bm} 
\let\bold=\bm 
\usepackage{latexsym} 
\usepackage{longtable} 
\usepackage{hyperref} 
\usepackage{bm}
\usepackage{slashed}
\usepackage{bbold}
\usepackage{color}
\usepackage{multirow}
\usepackage{rotating}
\usepackage{comment}

\graphicspath{{./figsX/}}

\newcommand{\tskip}{\mathop{t_{\rm skip}}\nolimits}

\usepackage[margin=1in]{geometry}
\usepackage [english]{babel}
\usepackage [autostyle, english = american]{csquotes}
\MakeOuterQuote{"}
\usepackage{graphicx}
\usepackage{subfigure}
\usepackage{xcolor}
\usepackage[toc,page]{appendix}
\usepackage{setspace}
\usepackage{amsmath, amssymb}
\usepackage{slashed}
\newcommand{\be}{\begin{eqnarray}}
\newcommand{\ee}{\end{eqnarray}}

\newcommand{\ubpsp}{\ol{u}_N^{p}(s^\prime)}
\newcommand{\ups}{u_N^p(s)}

\newcommand{\gf}{\gamma^5}
\newcommand{\ra}{\rangle}
\newcommand{\la}{\langle}
\newcommand{\ripmom}{${\rm RI}^\prime{\rm -MOM}$ }
\newcommand{\MSbar}{\overline{\rm MS} }

\def\ol{\overline}

\def\nn{\nonumber}

\makeatletter
\newcommand{\overleftrightsmallarrow}{\mathpalette{\overarrowsmall@\leftrightarrowfill@}}
\newcommand{\overrightsmallarrow}{\mathpalette{\overarrowsmall@\rightarrowfill@}}
\newcommand{\overleftsmallarrow}{\mathpalette{\overarrowsmall@\leftarrowfill@}}
\newcommand{\overarrowsmall@}[3]{%
  \vbox{%
    \ialign{%
      ##\crcr
      #1{\smaller@style{#2}}\crcr
      \noalign{\nointerlineskip}%
      $\m@th\hfil#2#3\hfil$\crcr
    }%
  }%
}
\def\smaller@style#1{%
  \ifx#1\displaystyle\scriptstyle\else
    \ifx#1\textstyle\scriptstyle\else
      \scriptscriptstyle
    \fi
  \fi
}
\makeatother
\newcommand{\olra}[1]{\overleftrightsmallarrow{#1}}

\providecommand{\abs}[1]{\lvert#1\rvert}
\providecommand{\matrixe}[3]{\langle#1\lvert#2\rvert#3\rangle}

\definecolor{green}{rgb}{0.1, 0.8, 0.1}

\newcolumntype{.}[1]{D{.}{.}{#1}}

\allowdisplaybreaks

\begin{document}


\title{Moments of nucleon isovector structure functions in \texorpdfstring{$2+1+1$}{2+1+1}-flavor QCD}
%
%


\author{Santanu Mondal}
\email{santanu@lanl.gov}
\affiliation{Los Alamos National Laboratory, Theoretical Division T-2, Los Alamos, New Mexico 87545}

\author{Rajan Gupta}
\email{rajan@lanl.gov}
\affiliation{Los Alamos National Laboratory, Theoretical Division T-2, Los Alamos, New Mexico 87545}

\author{Sungwoo Park}
\email{sungwoo@lanl.gov}
\affiliation{Los Alamos National Laboratory, Theoretical Division T-2, Los Alamos, New Mexico 87545}

\author{Boram Yoon}
\email{boram@lanl.gov}
\affiliation{Los Alamos National Laboratory, Computer Computational and Statistical Sciences, CCS-7, Los Alamos, New Mexico 87545}

\author{Tanmoy Bhattacharya}
\email{tanmoy@lanl.gov}
\affiliation{Los Alamos National Laboratory, Theoretical Division T-2, Los Alamos, New Mexico 87545}


\author{Huey-Wen~Lin}
\email{hwlin@pa.msu.edu}
\affiliation{Department of Physics and Astronomy, Michigan State University, Michigan 48824, USA}
\affiliation{Department of Computational Mathematics,  Science and Engineering, Michigan State University, East Lansing, MI 48824}

\collaboration{Precision Neutron Decay Matrix Elements (PNDME) Collaboration}
\noaffiliation
\preprint{LA-UR-20-23801}
\preprint{MSUHEP-20-009}
\pacs{11.15.Ha, 
      12.38.Gc  
}
\keywords{nucleon momentum distribution, helicity, transversity, lattice QCD}
\date{\today}
\begin{abstract}
We present results on the isovector momentum fraction, $\langle x
\rangle_{u-d}$, helicity moment, $\langle x \rangle_{\Delta u-\Delta d}$, and the
transversity moment, $\langle x\rangle_{\delta u-\delta d}$, of the nucleon
obtained using nine ensembles of gauge configurations generated by the
MILC Collaboration using $2+1+1$-flavors of dynamical highly improved
staggered quarks (HISQ). The correlation functions are calculated
using the Wilson-Clover action and the renormalization of the three
operators is carried out nonperturbatively on the lattice in the
\ripmom\ scheme.  The data have been collected at lattice spacings $a
\approx 0.15,\ 0.12,\ 0.09,$ and 0.06~fm and $M_\pi \approx 310,\ 220$
and 135~MeV, which are used to obtain the physical values using a
simultaneous chiral-continuum-finite-volume fit. The final results, in
the $\MSbar$ scheme at 2~GeV, are $\langle x \rangle_{u-d} = 0.173(14)(07)$,
$\langle x \rangle_{\Delta u-\Delta d} = 0.213(15)(22)$ and $\langle x \rangle_{\delta
u-\delta d} = 0.208(19)(24)$, where the first error is the overall
analysis uncertainty and the second is an additional systematic
uncertainty due to possible residual excited-state
contributions. These results are consistent with other recent lattice
calculations and phenomenological global fit values.

\end{abstract}
\maketitle
%
%
%
%
\section{Introduction}
\label{sec:into}

\begin{table*}[tbhp]
\centering
\renewcommand{\arraystretch}{1.2}
\begin{tabular}{ |c|c|c|c|c|c|c|c|c|c| }
\hline
Ensemble&$a$&$M_\pi^{val}$& $L^3\times T$&$M_\pi^{\rm val}L$&$\tau/a$        &$aM_N$      &$N_{conf}$&$N_{HP}$ &$N_{LP}$\\
ID &[fm]&[MeV]&&&&&&&\\
\hline                                                                                                                   
\hline                                                                                                                   
$a15m310$&$0.1510(20)$&$320.6(4.3)$&$16^3\times48$&$3.93$&$\{5,6,7,8,9\}$    &$0.8287(24)$&$1917$    &$7,668$  &$122,688$   \\
\hline                                                                                                                
$a12m310$&$0.1207(11)$&$310.2(2.8)$&$24^3\times64$&$4.55$&$\{8,10,12,14\}$   &$0.6660(27)$&$1013$    &$8,104$  &$64,832$    \\
$a12m220$&$0.1184(09)$&$227.9(1.9)$&$32^3\times64$&$4.38$&$\{8,10,12,14\}$   &$0.6289(26)$&$1156$    &$4,624$  &$73,984$    \\
$a12m220L$&$0.1189(09)$&$227.6(1.7)$&$40^3\times64$&$5.49$&$\{8,10,12,14\}$  &$0.6125(21)$&$1000$    &$4,000$  &$128,000$   \\
                                                                                                                      
\hline                                                                                                                
$a09m310$&$0.0888(08)$&$313.0(2.8)$&$32^3\times96$&$4.51$&$\{10,12,14,16\}$  &$0.4951(13)$&$2263$    &$9,052$  &$144,832$   \\
$a09m220$&$0.0872(07)$&$225.9(1.8)$&$48^3\times96$&$4.79$&$\{10,12,14,16\}$  &$0.4496(18)$&$960$     &$7,668$  &$122,688$   \\
$a09m130$&$0.0871(06)$&$138.1(1.0)$&$64^3\times96$&$3.90$&$\{10,12,14,16\}$  &$0.4204(23)$&$1041$    &$8,328$  &$99,936$    \\
\hline                                                                                                                
$a06m310W$&$0.0582(04)$&$319.6(2.2)$&$48^3\times144$&$4.52$&$\{18,20,22,24\}$&$0.3304(23)$&$500$     &$-$      &$66,000$    \\
$a06m135$&$0.0570(01)$&$135.6(1.4)$&$96^3\times192$&$3.7$&$\{16,18,20,22\}$  &$0.2704(32)$&$751$     &$6,008$  &$48,064$    \\
\hline
\hline
\end{tabular}
\caption{Lattice parameters, nucleon mass $M_N$, number of
  configurations analyzed, and the total number of high precision (HP)
  and low precision (LP) measurements made. For the $a06m310W$
  ensemble, HP data were not collected, however, we note that the bias
  correction factor on all other eight ensembles was negligible. }
\label{tab:ensembles}
\end{table*}

The elucidation of the hadron structure in terms of quarks and gluons
is evolving from determining the charges and form factors of nucleons
to including more complex quantities such as parton distribution
functions (PDFs)~\cite{Brock:1993sz}, transverse momentum dependent
PDFs (TMDs)~\cite{Yoon:2017qzo}, and generalized parton distributions
(GPDs)~\cite{Diehl:2003ny} as experiments become more
precise~\cite{Accardi:2012qut,Boer:2011fh}. These distributions are
not measured directly in experiments, and phenomenological analyses
including different theoretical inputs are needed to extract them from
experimental data. Input from lattice QCD is beginning to play an
increasingly larger role in such analyses~\cite{Lin:2017snn}. In cases
where both lattice results and phenomenological analyses of
experimental data (global fits) exist, one can compare them to
validate the control over systematics in the lattice calculations, and, 
on the other hand, provide a check on the phenomenological process used
to extract these observables from experimental data. In other cases,
lattice results are predictions. The list of quantities for which good
agreement between lattice calculations and experimental results, and
their precision, has grown very significantly as discussed in the
recent Flavor Lattice Averaging Group (FLAG) 2019 report~\cite{Aoki:2019cca}.
While steady progress has been made in developing the framework for
calculating distribution functions using lattice
QCD~\cite{Cichy:2018mum,Karthik:2019}, even calculations of their
moments have had large statistical and/or systematic uncertainties
prior to 2018. This was the case even for the best studied quantity,
the isovector momentum fraction $\la x
\ra_{u-d}$~\cite{Lin:2017snn}. In this work, we show that the lattice
data for the momentum fraction, helicity and transversity moments are
now of quality comparable to that for nucleon charges (zeroth
moments). Together with much more precise data from the planned
electron-ion collider~\cite{Accardi:2012qut} and the Large Hadron
Collider, which will significantly improve the phenomenological global
fits, we anticipate steady progress toward a detailed description of
the hadron structure.

In this paper we present results on the three moments from high
statistics calculations done on nine ensembles generated using
2+1+1-flavors of highly improved staggered quarks
(HISQ)~\cite{Follana:2006rc} by the MILC
Collaboration~\cite{Bazavov:2012xda}.  The data at four values of
lattice spacings $a$, three values of the pion mass, $M_\pi$,
including two ensembles at the physical pion mass, and on a range of
large physical volumes, characterized by $M_\pi L$, allow us to carry
out a simultaneous fit in these three variables to address the
associated systematics uncertainties.  We also investigate the
dependence of the results on the spectra of possible excited states
included in the fits to remove excited-state contamination (ESC), and
assign a second error to account for the associated systematic
uncertainty. Our final results are $\la x \ra_{u-d} = 0.173(14)(07)$,
$\la x \ra_{\Delta u-\Delta d} = 0.213(15)(22)$ and $\la x \ra_{\delta
  u-\delta d} = 0.208(19)(24)$ in the $\MSbar$ scheme at 2~GeV. On comparing these with other lattice
and phenomenological global fit results in Sec.~\ref{sec:results}, we
find a consistent picture emerging.

The paper is organized as follows: In Sec.~\ref{sec:lattice}, we
briefly summarize the lattice parameters and methodology. The
definitions of moments and operators investigated are given in
Sec.~\ref{sec:moments}. The two- and three-point functions calculated,
and their connection to the moments, are specified in
Sec.~\ref{sec:correlators}, and the analysis of excited state
contributions to extract ground state matrix elements is presented in
Sec.~\ref{sec:ESC}.  Results for the moments after the
chiral-continuum-finite-volume (CCFV) extrapolation are given in
Sec.~\ref{sec:results}, and compared with other lattice calculations
and global fits.  We end with conclusions in
Sec.~\ref{sec:summary}. The data and fits used to remove excited-state
contamination are shown in Appendix~\ref{sec:ratios} and the results
for renormalization factors, $Z_{VD,AD,TD}$, for the three operators
in Appendix~\ref{sec:renormalization}.

\section{Lattice Methodology}
\label{sec:lattice}

The parameters of the nine HISQ ensembles are summarized in
Table~\ref{tab:ensembles}. They cover a range of lattice spacings
($0.057 \leq a \leq 0.15$ fm), pion masses ($135 \leq M_\pi \leq
310$)~MeV and lattice sizes ($3.7 \leq M_\pi L \leq 5.5$).  Most of
the details of the lattice methodology, the strategies for the
calculations and the analyses are already given in
Refs.~\cite{Bhattacharya:2015wna, Bhattacharya:2016zcn,
  Gupta:2018qil}.  We construct the correlation functions needed to
calculate the matrix elements using Wilson-clover fermions on these
HISQ ensembles. This mixed-action, clover-on-HISQ, formulation is
nonunitary and can suffer from the problem of exceptional
configurations at small, but {\it a priori} unknown, quark masses. We have
not found evidence for such exceptional configurations on any of the
nine ensembles analyzed in this work.

For the parameters used in the construction of the $2$- and $3$-point
functions with Wilson-clover fermion see Table II of
Ref.~\cite{Gupta:2018qil}. The Sheikholeslami-Wohlert coefficient
\cite{Sheikholeslami:1985ij} used in the clover action is fixed to its
tree-level value with tadpole improvement, $c_{sw} = 1/u_0$, where
$u_0$ is the fourth root of the plaquette expectation value calculated
on the hypercubic (HYP) smeared~\cite{Hasenfratz:2001hp} HISQ lattices.

The masses of light clover quarks were tuned so that the
clover-on-HISQ pion masses, $M_\pi^{\rm val}$, match the HISQ-on-HISQ
Goldstone ones, $M_\pi^{\rm sea}$. $M_\pi^{\rm val}$ values are given
in Table~\ref{tab:ensembles}. $M_\pi^{\rm sea}$ values are available
in Ref.~\cite{Gupta:2018qil}. All fits in $M_\pi^2$ to study the
chiral behavior are made using the clover-on-HISQ $M_\pi^{\rm val}$
since the correlation functions, and thus the chiral behavior of the
moments, have a greater sensitivity to it. Henceforth, for brevity, we
drop the superscript and denote the clover-on-HISQ pion mass as
$M_\pi$. The number of high precision (HP) and low precision (LP)
measurements made on each configuration in the truncated solver bias
corrected method~\cite{Bali:2009hu,Blum:2012uh} for cost-effective
increase in statistics are specified in Table~\ref{tab:ensembles}.

\section{Moments and Matrix elements}
\label{sec:moments}

In this work,  we calculate the first moments of spin independent (or
unpolarized), $q=q_\uparrow+q_\downarrow$, helicity (or polarized),
$\Delta q=q_\uparrow-q_\downarrow$, and transversity, 
$\delta q =q_\top+q_{\perp}$ distributions, defined as
\be
\langle x \rangle_q &=& \int_0^1~x~[q(x)+\ol{q}(x)]~dx \,, \\
\langle x \rangle_{\Delta q} &=& \int_0^1~x~[\Delta q(x)+\Delta \ol{q}(x)]~dx \,, \\
\langle x \rangle_{\delta q} &=& \int_0^1~x~[\delta q(x)+\delta \ol{q}(x)]~dx \,,
\ee
where $q_{\uparrow(\downarrow)}$ corresponds to quarks with helicity
aligned (antialigned) with that of a longitudinally polarized target,
and $q_{\top(\perp)}$ corresponds to quarks with spin aligned
(antialigned) with that of a transversely polarized target.

These moments, at leading twist, can be extracted from the hadron
matrix elements of one-derivative vector, axial-vector and tensor
operators at zero momentum transfer.  The unpolarized and polarized
moments $\langle x \rangle_q$ and $\langle x \rangle_{\Delta q}$ of
the nucleon are also obtained from phenomenological global fits while a
computation of the nucleon transversity $\langle x\rangle_{\delta q}$
using lattice QCD is still a prediction due to the lack of sufficient
experimental data~\cite{Lin:2017snn}.

We are interested in extracting the forward nucleon matrix elements
$\la N(p)|{\cal O }|N(p)\ra$, with the nucleon initial and final
3-momenta, $\vec p$, taken to be zero in this work. The complete set
of one-derivative vector, axial-vector, and tensor operators is the following:
\be
{\cal O}^{\mu \nu}_{V^a}&=&\ol{q} \gamma^{\{\mu}\olra{D}^{\nu\}} \tau^a q\nn \,, \\
{\cal O}^{\mu \nu}_{A^a}&=&\ol{q}\gamma^{\{\mu}  \olra{D}^{\nu\}} \gf \tau^a q\nn \,, \\
{\cal O}^{\mu \nu \rho}_{T^a}&=&\ol{q} \sigma^{[\mu\{\nu]} \olra{D}^{\rho\}} \tau^a q \,, 
\label{operators}
\ee 
where $q=\{u,d\}$ is the isodoublet of light quarks and  $\sigma^{\mu\nu} = (\gamma^\mu\gamma^\nu -
\gamma^\nu\gamma^\mu)/2$. The derivative 
$\olra{D}_{\nu}\equiv\frac{1}{2}(\overrightarrow{D}_\nu-\overleftarrow{D}_\nu)$ 
consists of four terms:
\be
\ol{\psi} (\Gamma \overrightarrow{D}_\nu-\Gamma \overleftarrow{D}_\nu) \psi(x) &\equiv& 
\frac{1}{2}\big[\ol{\psi}(x)\Gamma U_\nu(x)\psi(x+\nu) \nn\\
&-& \ol{\psi}(x)\Gamma U^{\dagger}_\nu(x-\nu)\psi(x-\nu) \nn\\
&+& \ol{\psi}(x-\nu)\Gamma U_\nu(x-\nu)\psi(x) \nn\\
&-& \ol{\psi}(x+\nu)\Gamma U^\dagger_\nu(x)\psi(x)\big] \,.
\label{eqcovD} 
\ee
Lorentz indices within $\{ ~\}$ in Eq.~\eqref{operators} are
symmetrized and within $[\, ]$ are antisymmetrized. It is also
implicit that, where relevant, the traceless part of the above
operators is taken.  Their renormalization is carried out
nonperturbatively in the regularization independent RI${}^\prime$-MOM
scheme as discussed in Appendix~\ref{sec:renormalization}. A more detailed
discussion of these twist-2 operators and their renormalization can be
found in Refs.~\cite{Gockeler:1995wg,Harris:2019bih}.

In this work, we consider only isovector quantities. These are
obtained from Eq.~\eqref{operators} by choosing $\tau^a = \tau^3$ for
the Pauli matrix.  The decomposition of the matrix elements of these
operators in terms of the generalized form factors at zero momentum
transfer is as follows:
\begin{flalign}
\la N(p,s')|{\cal O}^{\mu \nu }_{V^a} &| N(p,s) \ra =  \nonumber \\ &{}    \ubpsp A_{20}(0)\gamma^{\{ \mu} p^{\nu \}} \ups\\ 
\la N(p,s')|{\cal O}^{\mu \nu}_{A^a} &| N(p,s) \ra =   \nonumber \\ &{}  i \ubpsp \tilde{A}_{20}(0)\gamma^{\{ \mu} p^{\nu \}} \gf \ups  \\ 
\la N(p,s')|{\cal O}^{\mu \nu \rho}_{T^a} &| N(p,s) \ra = \nonumber \\ &{}  i \ubpsp A_{T20}(0)\sigma^{[ \mu\{ \nu]} p^{ \rho \}} \ups 
\end{flalign}
The relation between the momentum fraction, the helicity moment, and the transversity moment, 
and the generalized form factors is 
$\la x \ra_q = A_{20}(0)$, 
$\la x \ra_{\Delta q} = \tilde{A}_{20}(0)$ and 
$\la x \ra_{\delta q} = A_{T20}(0) $ respectively.

We end this discussion by mentioning that other approaches have been 
proposed to calculate the moments of PDFs from lattice QCD in recent
years~\cite{Davoudi:2012ya,Detmold:2005gg,Detmold:2018kwu}. 


\section{Correlation functions and Moments}
\label{sec:correlators}

We use the following interpolating operator ${\mathcal N}$ to
create/annihilate the nucleon state 
\be 
{\mathcal N} = \epsilon^{abc} \Big[  q_1^{aT} (x) C \gf \frac{(1\pm \gamma_4)}{2}q_2^b(x) \Big] q^c_1(x) \,, 
\label{nucop}
\ee 
where $\{a,b,c\}$ are color indices,
$q_1,q_2 \in \{u,d\}$ and $C=\gamma_0 \gamma_2$ is the charge conjugation
matrix.  The nonrelativistic projection $(1\pm \gamma_4)/2 $
is inserted to improve the signal, with the plus and minus signs
applied to the forward and backward propagation in Euclidean time,
respectively~\cite{Gockeler:1995wg}. At zero momentum, this operator
couples only to the spin $\frac{1}{2}$ state.  The zero momentum
2-point and 3-point nucleon correlation functions are defined as 
\begin{flalign}
\bold{C}^{2pt}_{\alpha \beta} (\tau ) &= \sum_{\bm x}\la 0 | {\mathcal N}_\alpha (\tau, {\bm x}) \ol{{\mathcal N}}_\beta(0,{\bm 0})| 0\ra\\ 
\bold{C}^{3pt}_{\mathcal{O},\alpha \beta}
(\tau,t ) &= \sum_{{\bm x}',{\bm x}} \la 0 | {\mathcal N}_\alpha (\tau, {\bm x}) {\cal O} (t, {\bm x}') \ol{{\mathcal N}}_\beta(0,{\bm 0})| 0\ra 
\end{flalign}
where $\alpha$ and $\beta$ are spin
indices. The source is placed at time slice 0, the sink is at $\tau$
and the one-derivative operators, defined in Sec.~\ref{sec:moments},
are inserted at time slice $t$.  Data have been accumulated for the
values of $\tau$ specified in Table~\ref{tab:ensembles}, and in each
case for all intermediate times $0 \leq t \leq \tau$.

To isolate the various operators, projected $2$- and $3$-point functions are constructed as 
\be
C^{2pt}&=& {\rm Tr} \big( {\cal P}_{2pt} \bold{C}^{2pt} \big)\\
C_\mathcal{O}^{3pt}&=& {\rm Tr} \big( {\cal P}_{3pt} \bold{C}^{3pt}_{\mathcal{O}} \big) \,.
\ee
The projector ${\cal P}_{2pt} = \frac{1}{2}\, (1 + \gamma_4)$ in the nucleon 
correlator gives the positive parity contribution for the nucleon propagating
in the forward direction.
For the connected $3$-point contributions ${\cal P}_{3pt}= \frac{1}{2}(1 + 
\gamma_4)(1+i \gf \gamma^3)$ is used. With these spin projections, 
the explicit operators used to calculate the forward matrix elements are: 
\be
\la x \ra_{u-d}&:& {\cal O}^{44}_{V^3} =  \ol{q} (\gamma^{4}\olra{D}^{4}  -\frac{1}{3}
{\bm \gamma} \cdot \olra{\bf D}) \tau^3 q 
\label{eq:finaloperatorV} \\
%
%
\la x \ra_{\Delta u-\Delta d}&:& {\cal O}^{34}_{A^3}=\ol{q} \gamma^{\{3}\olra{D}^{4\}} \gf \tau^3 q 
\label{eq:finaloperatorA} \\
\la x \ra_{\delta u-\delta d}&:& {\cal O}^{124}_{T^3}=\ol{q} \sigma^{[1\{2]}\olra{D}^{4\}} \tau^3 q \,.
\label{eq:finaloperatorT}
\ee
Our goal is to obtain the matrix elements, ($ME$), of 
these operators within the ground state of the nucleon.
These $ME$ are related to the moments as follows: 
\be
\la 0 | {\cal O}^{44}_{V^3}| 0 \ra &=&  -  M_N\, \la x \ra_{u-d} \,, 
\label{eq:me2momentV} \\
\la 0 | {\cal O}^{34}_{A^3}| 0 \ra &=&  - \frac{i  M_N}{2} \, \la x \ra_{\Delta u-\Delta d} \,, 
\label{eq:me2momentA} \\
\la 0 | {\cal O}^{124}_{T^3}| 0 \ra &=& - \frac{i M_N}{2} \, \la x \ra_{\delta u-\delta d} \,, 
\label{eq:me2momentT}
\ee 
where $M_N$ is the nucleon mass.  The three moments are dimensionless,
and their extraction on a given ensemble does not require knowing the
value of the lattice scale $a$. It enters only when performing the
chiral-continuum extrapolation to the physical point as discussed in
Sec.~\ref{sec:results}.

\begin{table*}[htbp]
\centering
\renewcommand{\arraystretch}{1.2}
\begin{tabular}{ |c|c|c|c|c|c|c|c|c|c| }
 \hline
\multicolumn{4}{|c|}{}&\multicolumn{3}{c|}{$\{4,3^*\}$} &\multicolumn{3}{c|}{$\{4,2^{\rm free}\}$} \\
\hline

Ensemble&$\tau/a$&$\tskip$&Observable&$ME$&$\langle x \rangle$&$\chi^2$/dof &$ME$&$\langle x \rangle$&$\chi^2$/dof\\
\hline
$a06m135$&$\{22,20,18\}$&$\{4,5\}$&$\la x\ra_{u-d}$&$-0.042(4)$&$0.155(14)$&$0.87$&$-0.045(5)$&$0.167(18)$&$0.99$\\
$a06m135$&$\{22,20,18\}$&$\{4,5\}$&$\la x\ra_{\Delta u-\Delta d}$&$-0.026(2)$&$0.191(12)$&$1.00$ &$-0.027(3)$&$0.198(22)$&$1.13$\\
$a06m135$&$\{22,20,18\}$&$\{4,5\}$&$\la x\ra_{\delta u-\delta d}$&$-0.025(2)$&$0.185(16)$&$1.32$&$-0.027(3)$&$0.202(23)$&$1.31$\\
\hline
\hline
$a06m310W$&$\{24,22,20\}*$&$\{6,6 \}$&$\la x\ra_{u-d}$&$-0.056(4)$&$0.170(13)$&$1.02$&$-0.063(3)$&$0.193(8)$&$1.10$\\
$a06m310W$&$\{24,22,20\}$&$\{6,6 \}$&$\la x\ra_{\Delta u-\Delta d}$&$-0.037(2)$&$0.223(15)$&$1.00$&$-0.038(1)$&$0.231(7)$&$1.33$\\
$a06m310W$&$\{24,22,20\}$&$\{6,6 \}$&$\la x\ra_{\delta u-\delta d}$&$-0.035(3)$&$0.213(18)$&$0.80$&$-0.037(1)$&$0.227(8)$&$0.83$\\
\hline
\hline
$a09m130$&$\{16,14,12\}$&$\{3,3 \}$&$\la x\ra_{u-d}$&$-0.074(3)$&$0.177(8)$&$0.93$&$-0.077(4)$&$0.184(9)$&$0.88$\\
$a09m130$&$\{16,14,12\}$&$\{3,3 \}$&$\la x\ra_{\Delta u-\Delta d}$&$-0.046(2)$&$0.218(7)$&$1.30$&$-0.048(1)$&$0.228(5)$&$1.33$\\
$a09m130$&$\{16,14,12\}$&$\{3,3 \}$&$\la x\ra_{\delta u-\delta d}$&$-0.045(2)$&$0.212(11)$&$1.30$&$-0.047(3)$&$0.225(14)$&$1.41$\\

\hline
\hline
$a09m220$&$\{16,14,12\}$&$\{3,3 \}$&$\la x\ra_{u-d}$&$-0.082(3)$&$0.184(5)$&$0.89$&$-0.086(2)$&$0.191(4)$&$0.78$\\

$a09m220$&$\{16,14,12\}$&$\{3,3 \}$&$\la x\ra_{\Delta u-\Delta d}$&$-0.051(1)$&$0.227(4)$&$0.92$&$-0.053(1)$&$0.235(3)$&$0.60$\\

$a09m220$&$\{16,14,12\}$&$\{3,3 \}$&$\la x\ra_{\delta u-\delta d}$&$-0.053(1)$&$0.234(6)$&$1.29$&$-0.055(1)$&$0.243(4)$&$1.26$\\

\hline
\hline
$a09m310$&$\{16,14,12\}$&$\{3,3 \}$&$\la x\ra_{u-d}$&$-0.097(2)$&$0.196(4)$&$1.25$&$-0.094(2)$&$0.190(5)$&$1.16$\\

$a09m310$&$\{16,14,12\}$&$\{3,3 \}$&$\la x\ra_{\Delta u-\Delta d}$&$-0.058(1)$&$0.233(3)$&$1.24$&$-0.059(1)$&$0.238(3)$&$1.25$\\

$a09m310$&$\{16,14,12\}$&$\{3,3 \}$&$\la x\ra_{\delta u-\delta d}$&$-0.059(1)$&$0.239(4)$&$0.78$&$-0.060(1)$&$0.241(4)$&$0.79$\\

\hline
\hline
$a12m220$&$\{14,12,10\}$&$\{3,3 \}$&$\la x\ra_{u-d}$&$-0.125(5)$&$0.199(8)$&$1.32$&$-0.130(5)$&$0.207(8)$&$1.24$\\

$a12m220$&$\{14,12,10\}$&$\{3,3 \}$&$\la x\ra_{\Delta u-\Delta d}$&$-0.074(3)$&$0.234(9)$&$0.92$&$-0.077(2)$&$0.245(6)$&$0.87$\\

$a12m220$&$\{14,12,10\}$&$\{3,3 \}$&$\la x\ra_{\delta u-\delta d}$&$-0.077(4)$&$0.246(11)$&$1.24$&$-0.080(6)$&$0.254(17)$&$1.20$\\

\hline
\hline

$a12m220L$&$\{14,12,10\}$&$\{3,2\}$&$\la x\ra_{u-d}$&$-0.117(6)$&$0.191(9)$&$1.44$&$-0.120(4)$&$0.196(7)$&$1.35$\\

$a12m220L$&$\{14,12,10\}$&$\{3,3\}$&$\la x\ra_{\Delta u-\Delta d}$&$-0.073(2)$&$0.240(7)$&$1.33$&$-0.074(4)$&$0.241(14)$&$1.43$\\

$a12m220L$&$\{14,12,10\}$&$\{3,3\}$&$\la x\ra_{\delta u-\delta d}$&$-0.073(3)$&$0.237(10)$&$1.25$&$-0.075(4)$&$0.244(14)$&$1.28$\\

\hline
\hline
$a12m310$&$\{14,12,10\}$&$\{3,3\}$&$\la x\ra_{u-d}$&$-0.130(8)$&$0.195(11)$&$1.66$&$-0.137(5)$&$0.206(8)$&$1.54$\\

$a12m310$&$\{14,12,10\}$&$\{3,3\}$&$\la x\ra_{\Delta u-\Delta d}$&$-0.079(5)$&$0.238(16)$&$0.76$&$-0.083(4)$&$0.250(13)$&$0.77$\\

$a12m310$&$\{14,12,10\}$&$\{3,3\}$&$\la x\ra_{\delta u-\delta d}$&$-0.084(6)$&$0.251(16)$&$0.69$&$-0.087(3)$&$0.261(9)$&$0.66$\\

\hline
\hline
$a15m310$&$\{9,8,7\}\dagger$&$\{2,3\}$&$\la x\ra_{u-d}$&$-0.177(5)$&$0.214(6)$&$1.94$&$-0.191(3)$&$0.231(3)$&$1.90$\\

$a15m310$&$\{9,8,7\}$&$\{2,2\}$&$\la x\ra_{\Delta u-\Delta d}$&$-0.110(3)$&$0.266(7)$&$0.76$&$-0.111(3)$&$0.267(7)$&$0.69$\\

$a15m310$&$\{9,8\}$&$\{2,2\}$&$\la x\ra_{\delta u-\delta d}$&$-0.122(5)$&$0.293(12)$&$0.66$&$-0.119(4)$&$0.286(9)$&$0.98$\\

\hline
\hline
\end{tabular}
\caption{Our best estimates of the unrenormalized moments from 
  the two fit strategies, $\{4,3^\ast \}$ and $\{4,2^{\rm free} \}$, used to analyze 
  the two- and three-point functions.  The
  second column gives the values of $\tau$ used in the fits and the third
  column lists $\tskip=\{i,j\}$, the number of time slices from the
  source and sink omitted for each $\tau$ for the two fit types to the
  three-point functions. For each fit-type we give the result for the
  ground state matrix element, $ME$, the moment $\langle x \rangle$ 
  obtained from it using Eqs.~\protect\eqref{eq:me2momentV}--\protect\eqref{eq:me2momentT}, and the 
  $\chi^2$/dof of the fit to the three-point function. In two cases, the values of 
  $\tau/a$ included are different: the $*$ in the second
  column denotes $\tau/a =\{22, 20, 18\}$ and $\dagger$ denotes
  $\tau/a =\{9,8\}$ were used for the $\{4,2^{\rm free} \}$ fits.  }
\label{tab:best-fits}
\end{table*}


\begin{table*}[htbp]
\centering
\setlength{\tabcolsep}{2pt}
\renewcommand{\arraystretch}{1.1}
\begin{tabular}{|c|c|c|c|c|c|c|c|c|c| }
\hline

\multicolumn{10}{|c|}{$\langle x \rangle_{u-d}$} \\
\hline
Ensemble & Fit-type&$a \Delta M_1$&$a \Delta M_2$&$\la 0|{\cal O}|0 \ra$
&$\frac{\la 1|{\cal O}| 1\ra}{\la 0|{\cal O}|0 \ra}$&$\frac{\la 1|{\cal O}|0 \ra}{\la 0|{\cal O}|0 \ra}$ 
&$\frac{\la 2|{\cal O}| 0\ra}{\la 0|{\cal O}|0 \ra}$ &$\frac{\la 2|{\cal O}|1 \ra}{\la 0|{\cal O}|0 \ra}$
&$\chi^2$/dof\\
\hline
\hline

$a09m310$ & $\{4,2\}$&$0.434(58)$&                  &$0.0982(26)$&$4.90(3.34)$&$0.73(7)$& & &$1.31$\\

$a09m310$ & $\{4^{N\pi},2\}$&$0.343(44)$&           &$0.0928(35)$&$1.45(1.81)$&$0.91(14)$& & &$1.12$\\

$a09m310$ & $\{4,3^*\}$&$0.434(58)$&$0.697(132)$&$0.0971(21)$&$4.50(3.66)$&$0.83(7)$&$-0.27(31)$&$-4.5(14)$&$1.25$\\

$a09m310$ & $\{4^{N\pi},3^*\}$&$0.343(44)$&$0.555(69)$&$0.0933(25)$&$1.1(2.0)$&$0.89(10)$&$-0.01(17)$&$2.3(4.2)$&$1.20$\\

$a09m310$ & $\{4,2^{\rm free}\}$&$0.358(33)$& &$0.0941(24)$&$0.78(1.44)$&$0.76(8)$& & &$1.16$\\
\hline
\hline
$a06m135$ & $\{4,2\}$       &$0.197(37)$ & &$0.0402(56)$&$2.6(1.5)$&$1.12(0.31)$& & &$0.95$\\

$a06m135$ & $\{4^{N\pi},2\}$&$0.0846(84)$& & -- & -- & -- & & & -- \\

$a06m135$ & $\{4,3^*\}$&$0.197(37)$&$0.287(49)$&$0.0418(40)$&$3.2(1.9)$&$0.89(24)$&$0.44(32)$&$-2(5)$&$0.87$\\

$a06m135$ & $\{4^{N\pi},3^*\}$&$0.0846(84)$&$0.201(23)$&$0.038(15)$&$3.4(2.8)$&$0.11(1)$&$1.2(4)$&$-0.3(2.2)$&$0.90$\\

$a06m135$ & $\{4,2^{\rm free}\}$         &$0.241(49)$& &$0.0452(47)$&$6(6)$&$0.99(16)$& & &$0.99$\\
\hline
\end{tabular}
\caption{Comparison of fits using five strategies, $\{4,2\}$,
  $\{4^{N\pi},2\}$, $\{4,3^*\}$, $\{4^{N\pi},3^*\}$ and $\{4,2^{\rm
    free}\}$, for the momentum fraction $\langle x \rangle_{ u-  d}$ on
  two ensembles $a09m310$ (highest statistics and $M_\pi \sim
  310$~MeV) and $a06m135$ (physical $M_\pi \sim 135$~MeV).  In the
  $\{4,2^{\rm free}\}$ fit, the excited state mass gap, $\Delta M_1$,
  is left as a free parameter that is determined from the fit to the
  three-point function.  The values of $\tau/a$ and $\tskip$ used are the same
  as listed in Table~\protect\ref{tab:best-fits}. We could not find a $\{4^{N\pi},2\}$
  fit to the  $a06m135$ data that gave reasonable values. 
}
\label{tab:Npi-fits-momfrac}
\end{table*}


\begin{table*}[htbp]
\centering
\setlength{\tabcolsep}{3pt}
\renewcommand{\arraystretch}{1.1}
\begin{tabular}{|c|c|c|c|c|c|c|c|c|c| }
\hline

\multicolumn{10}{|c|}{$\langle x \rangle_{\Delta u-\Delta d}$} \\
\hline
Ensemble & Fit-type&$a \Delta M_1$&$a \Delta M_2$&$\la 0|{\cal O}|0 \ra$
&$\frac{\la 1|{\cal O}| 1\ra}{\la 0|{\cal O}|0 \ra}$&$\frac{\la 1|{\cal O}|0 \ra}{\la 0|{\cal O}|0 \ra}$ 
&$\frac{\la 2|{\cal O}| 0\ra}{\la 0|{\cal O}|0 \ra}$ &$\frac{\la 2|{\cal O}|1 \ra}{\la 0|{\cal O}|0 \ra}$
&$\chi^2$/dof\\
\hline
\hline

$a09m310$ & $\{4,2\}$       &$0.434(58)$         &&$0.115(26)$&$2.6(2.6)$&$0.72(5)$& & &$1.15$\\

$a09m310$ & $\{4^{N\pi},2\}$&$0.343(44)$&        &$0.110(33)$&$0.33(1.5)$&$0.85(11)$& & &$1.43$\\

$a09m310$ & $\{4,3^*\}$&$0.434(58)$&$0.697(132)$&$0.115(19)$&$3.46(2.6)$&$0.63(7)$&$0.50(20)$&$-3(12)$&$1.24$\\

$a09m310$ & $\{4^{N\pi},3^*\}$&$0.343(44)$&$0.555(69)$&$0.113(24)$&$1.0(2.3)$&$0.54(15)$&$0.49(46)$&$7(10)$&$1.16$\\

$a09m310$ & $\{4,2^{\rm free}\}$&$0.539(40)$&       &$0.118(15)$&$14(10)$&$0.83(9)$& & &$1.25$\\

\hline
\hline

$a06m135$ & $\{4,2\}$       &$0.197(37)$ &           &$0.0468(61)$&$1.07(1.09)$&$1.07(29)$& & &$1.29$\\

$a06m135$ & $\{4^{N\pi},2\}$&$0.0846(84)$&           &$0.004(14)$ & $ -23(110)$&$28(115)$& & &$0.93$\\

$a06m135$ & $\{4,3^*\}$&$0.197(37)$&$0.287(49)$&$0.0517(36)$&$4.01(1.84)$&$0.45(21)$&$1.28(26)$&$-7(6)$&$1.00$\\

$a06m135$ & $\{4^{N\pi},3^*\}$&$0.0846(84)$&$0.201(23)$&$0.075(21)$&$6(3)$&$-1.3(7)$&$1.6(3)$&$-3.4(1.5)$&$1.06$\\
$a06m135$ & $\{4,2^{\rm free}\}$         &$0.260(67)$& &$0.0535(60)$&$5(7)$&$0.98(14)$& & &$1.18$\\

\hline
\hline
\end{tabular}
\caption{Comparison of fits using five strategies, $\{4,2\}$,
  $\{4^{N\pi},2\}$, $\{4,3^*\}$, $\{4^{N\pi},3^*\}$ and $\{4,2^{\rm
    free}\}$, for the helicity moment $\langle x \rangle_{\Delta
    u-\Delta d}$.  The rest is the same as in
  Table~\protect\ref{tab:Npi-fits-momfrac}.  }
\label{tab:Npi-fits-helfrac}
\end{table*}


\begin{table*}[htbp]
\centering
\setlength{\tabcolsep}{3pt}
\renewcommand{\arraystretch}{1.1}
\begin{tabular}{|c|c|c|c|c|c|c|c|c|c| }
\hline

\multicolumn{10}{|c|}{$\langle x \rangle_{\delta u-\delta d}$} \\
\hline
Ensemble & Fit-type&$a \Delta M_1$&$a \Delta M_2$&$\la 0|{\cal O}|0 \ra$
&$\frac{\la 1|{\cal O}| 1\ra}{\la 0|{\cal O}|0 \ra}$&$\frac{\la 1|{\cal O}|0 \ra}{\la 0|{\cal O}|0 \ra}$ 
&$\frac{\la 2|{\cal O}| 0\ra}{\la 0|{\cal O}|0 \ra}$ &$\frac{\la 2|{\cal O}|1 \ra}{\la 0|{\cal O}|0 \ra}$
&$\chi^2$/dof\\
\hline
\hline

$a09m310$ & $\{4,2\}$       &$0.434(58)$ &&$0.117(36)$&$2.4(3.1)$&$0.92(10)$& & &$0.84$\\

$a09m310$ & $\{4^{N\pi},2\}$&$0.343(44)$& &$0.109(49)$&$-0.83(1.9)$&$1.17(19)$& & &$1.45$\\

$a09m310$ & $\{4,3^*\}$&$0.434(58)$&$0.697(132)$&$0.118(24)$&$1.3(3.0)$&$0.84(8)$&$0.04(33)$&$18(15)$&$0.78$\\

$a09m310$ & $\{4^{N\pi},3^*\}$&$0.343(44)$&$0.555(69)$&$0.115(27)$&$-0.8(1.8)$&$0.82(10)$&$0.28(19)$&$10(6)$&$0.77$\\

$a09m310$ & $\{4,2^{\rm free}\}$&$0.486(37)$& &$0.120(19)$&$8(6)$&$0.93(10)$& & &$0.79$\\

\hline
\hline

$a06m135$ & $\{4,2\}$       &$0.197(37)$ & &$0.0385(97)$&$0.69(1.75)$&$2.00(81)$& & &$1.70$\\

$a06m135$ & $\{4^{N\pi},2\}$&$0.0846(84)$& & -- & -- & -- & & & -- \\

$a06m135$ & $\{4,3^*\}$&$0.197(37)$&$0.287(49)$&$0.0500(44)$&$3.6(2.2)$&$0.61(35)$&$1.30(43)$&$-1(6)$&$1.32$\\

$a06m135$ &$\{4^{N\pi},3^*\}$&$0.0846(84)$&$0.201(23)$&$0.082(30)$&$6(3)$&$-1.3(8)$&$1.5(3)$&$-1.8(1.7)$&$1.34$\\
$a06m135$ & $\{4,2^{\rm free}\}$         &$0.306(81)$& &$0.0545(62)$&$17(26)$&$1.29(14)$& & &$1.31$\\

\hline
\hline
\end{tabular}
\caption{Comparison of fits using five strategies, $\{4,2\}$,
  $\{4^{N\pi},2\}$, $\{4,3^*\}$, $\{4^{N\pi},3^*\}$ and $\{4,2^{\rm
    free}\}$, for the transversity moment $\langle x \rangle_{\delta u-\delta d}$. The rest is the
  same as in Table~\protect\ref{tab:Npi-fits-momfrac}.
}
\label{tab:Npi-fits-transfrac}
\end{table*}


\begin{figure*}[htbp]
\begin{subfigure}
\centering
\includegraphics[angle=0,width=0.48\textwidth]{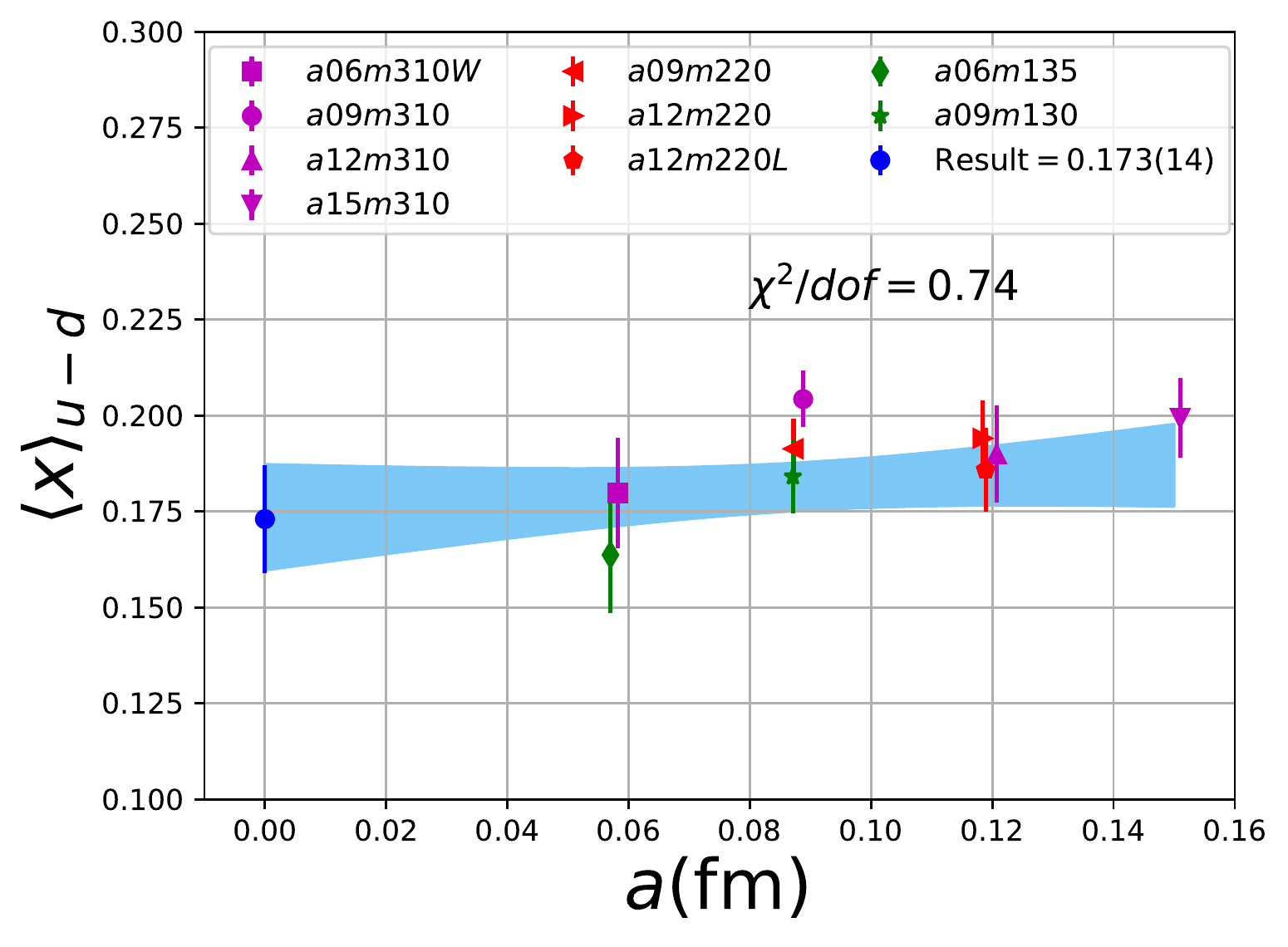}
\end{subfigure}
\begin{subfigure}
\centering
\includegraphics[angle=0,width=0.48\textwidth]{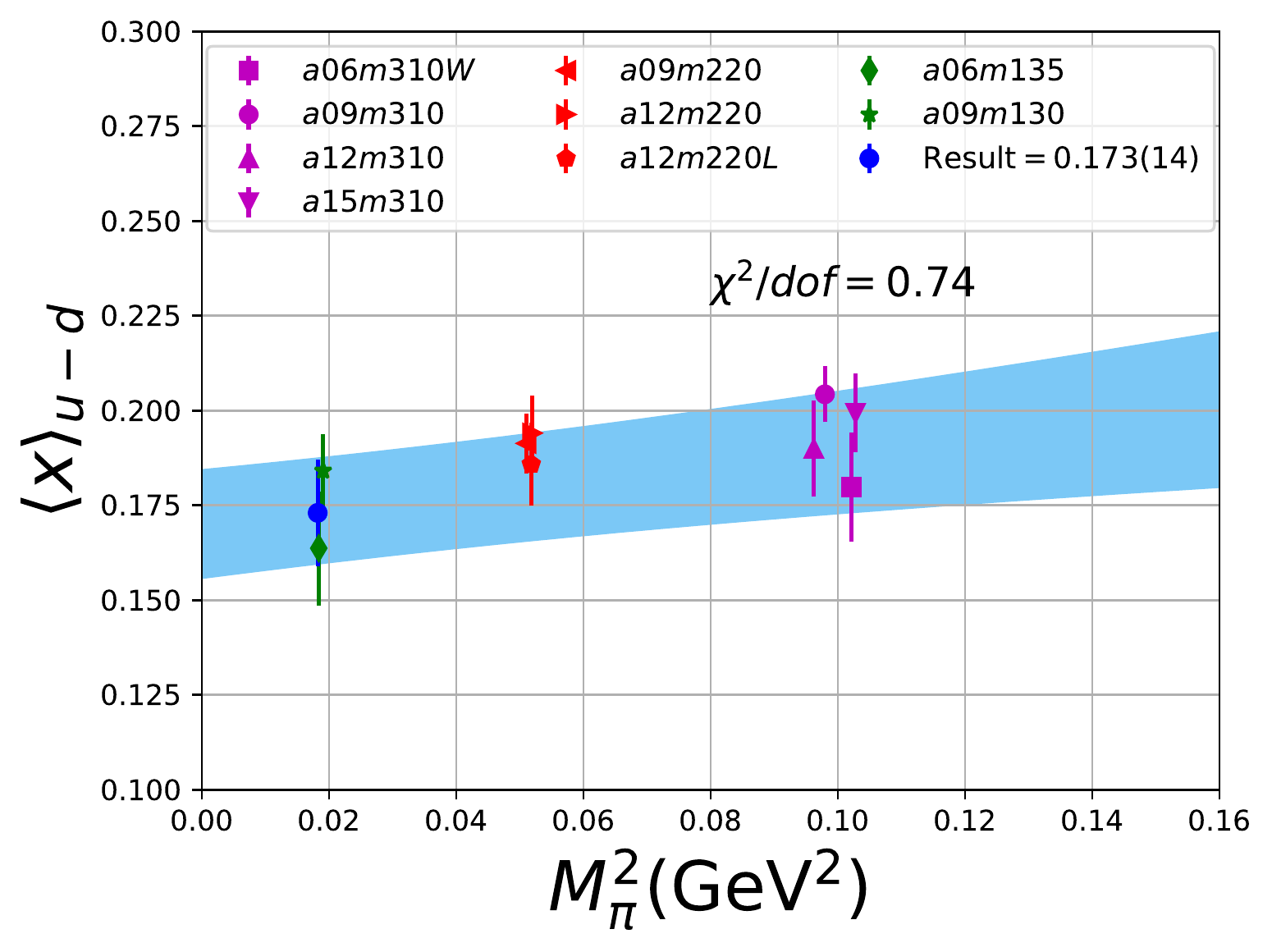}
\end{subfigure}
\vspace{-0.08in}
\caption{Data for $\la x \ra_{u-d}$, renormalized in
  the $\MSbar$ scheme at $\mu=2$ GeV, for all nine ensembles. 
  The blue band in the left panel shows the CC
  fit result evaluated at $M_\pi = 135$~MeV and 
  plotted versus $a$, while in the right panel it shows the result versus
  $M_\pi^2$ evaluated at $a = 0$.  }
\label{fig:momfrac-vs-a}
\end{figure*}
\begin{figure*}[htbp]
\begin{subfigure}
\centering
\includegraphics[angle=0,width=0.48\textwidth]{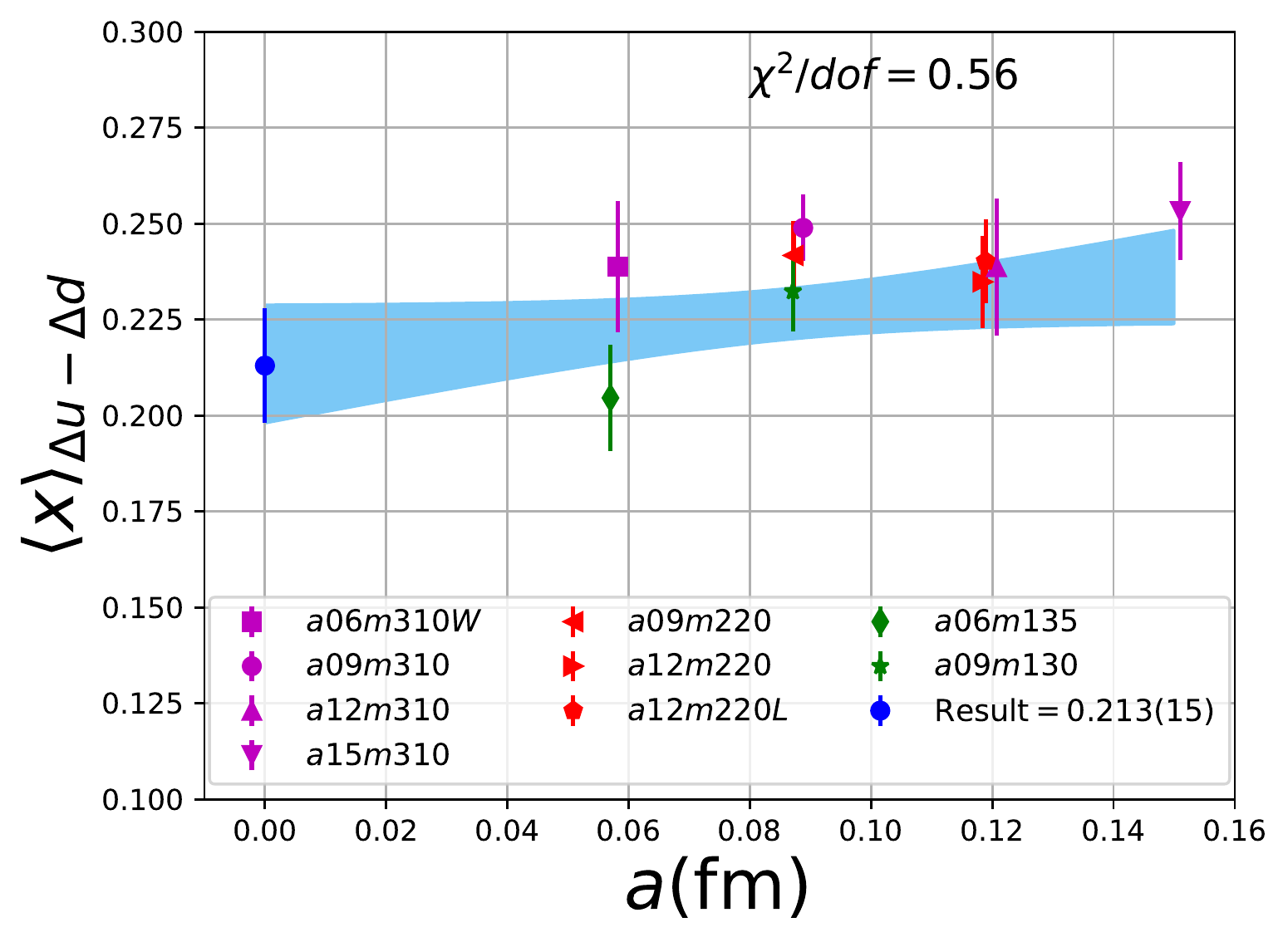}
\end{subfigure}
\begin{subfigure}
\centering
\includegraphics[angle=0,width=0.48\textwidth]{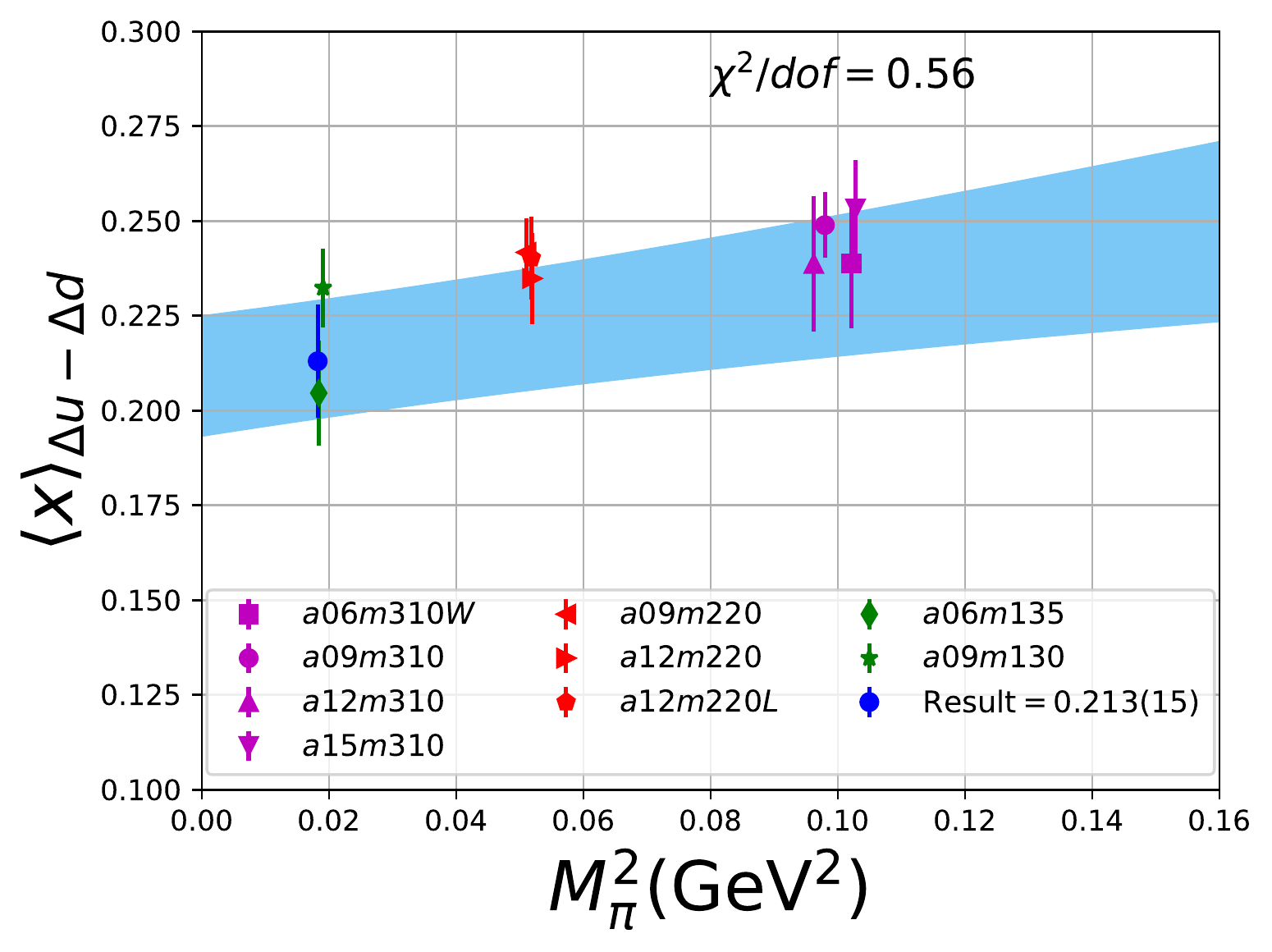}
\end{subfigure}
\vspace{-0.08in}
\caption{ Data for $\la x \ra_{\Delta u- \Delta d}$, renormalized in
  the $\MSbar$ scheme at $\mu=2$ GeV, for all nine ensembles
  plotted as a function of $a$ (left panel) and $M_\pi^2$ (right
  panel). The rest is the same as in Fig.~\protect\ref{fig:momfrac-vs-a}.
}
\label{fig:helfrac-vs-a}
\end{figure*}
\begin{figure*}[htbp]
\begin{subfigure}
\centering
\includegraphics[angle=0,width=0.48\textwidth]{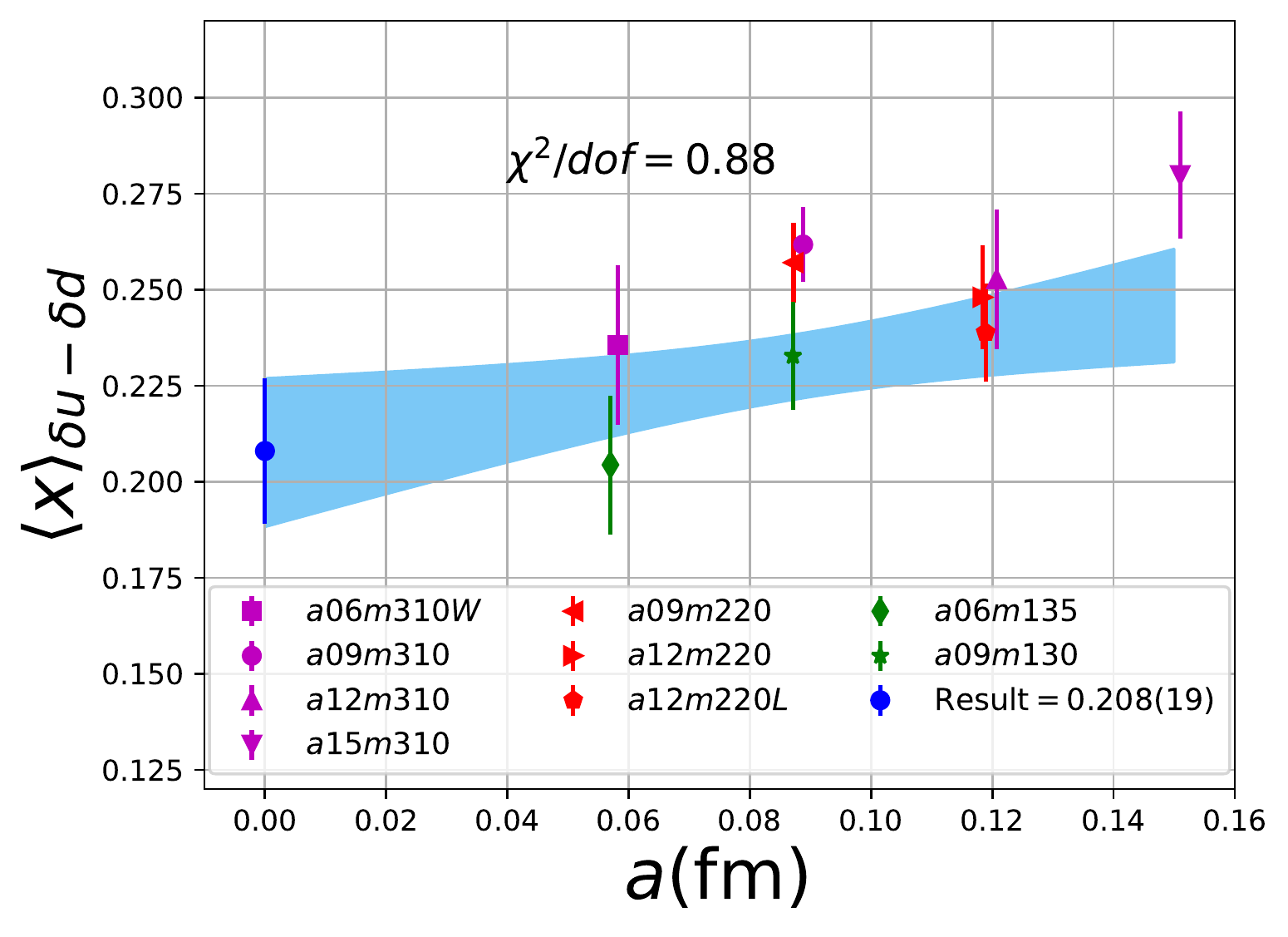}
\end{subfigure}
\begin{subfigure}
\centering
\includegraphics[angle=0,width=0.48\textwidth]{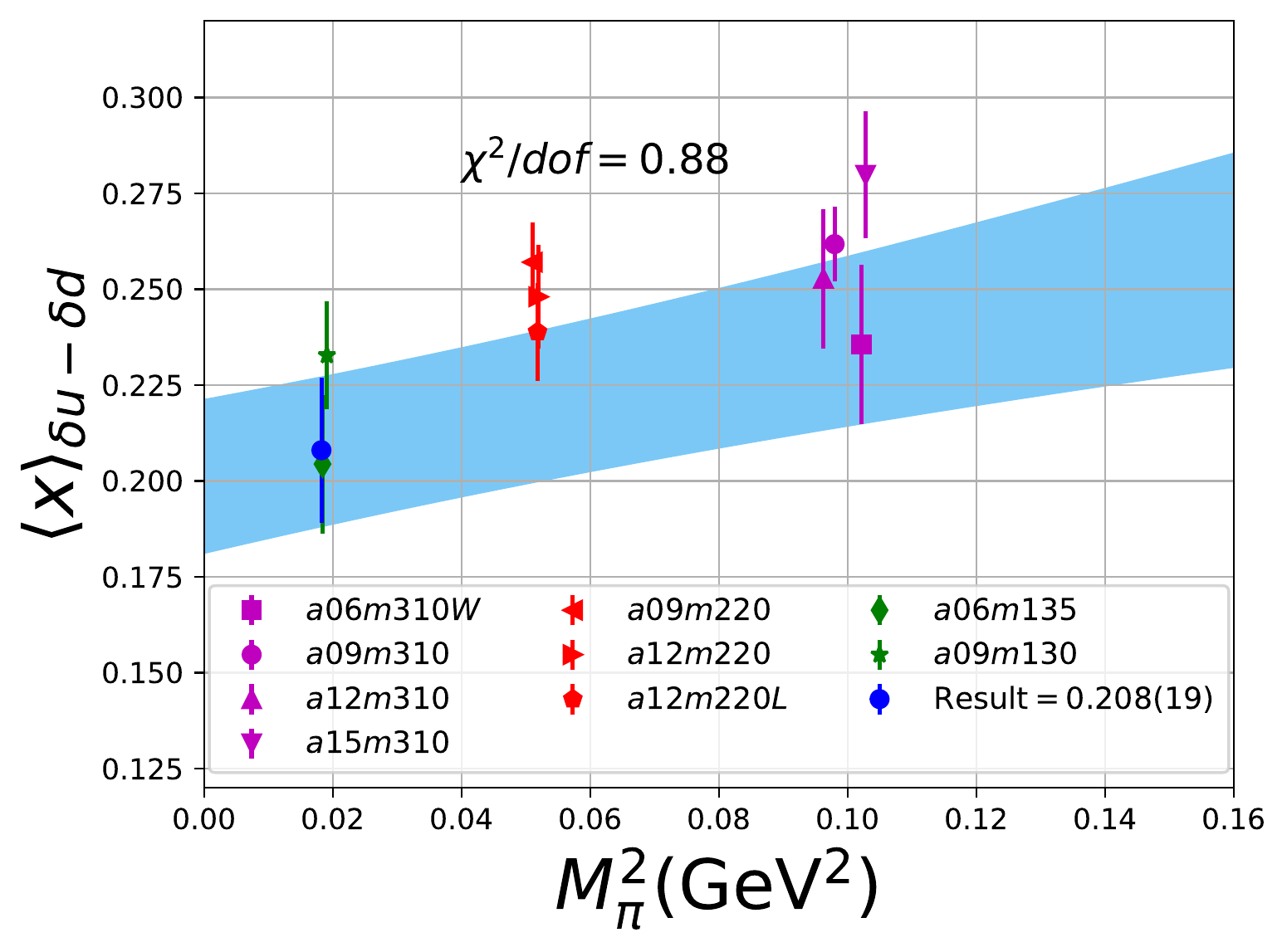}
\end{subfigure}
\vspace{-0.08in}
\caption{ Data for $\la x \ra_{\delta u- \delta d}$,
  renormalized in the $\MSbar$ scheme at $\mu=2$ GeV, for all
  nine ensembles plotted as a function of $a$ (left panel) and
  $M_\pi^2$ (right panel). The rest is the same as in
  Fig.~\protect\ref{fig:momfrac-vs-a}.  }
\label{fig:transversity-vs-a}
\end{figure*}


\section{Controlling excited state contamination}
\label{sec:ESC}

To calculate the matrix elements of the operators defined in
Sec.~\ref{sec:moments} between ground-state nucleons, contributions of
all possible excited states need to be removed.  The lattice nucleon
interpolating operator ${\mathcal N}$ given in Eq.~\eqref{nucop},
however, couples to the nucleon, all its excitations and multiparticle
states with the same quantum numbers. Previous lattice calculations
have shown that the ESC can be large~\cite{Bhattacharya:2013ehc,Bali:2014gha, Bali:2012av}.  In our earlier
works \cite{Bhattacharya:2015wna, Bhattacharya:2016zcn, Gupta:2018qil,
  Yoon:2016dij}, we have shown that this can be controlled to within a
few percent. We use the same strategy here.  In particular, we use HYP
smearing of the gauge links before calculating Wilson-clover quark
propagators with optimized Gaussian smeared sources using the
multigrid algorithm~\cite{Babich:2010qb,Clark:2009wm}.  Correlation
functions constructed from these smeared source propagators have
smaller excited state contamination~\cite{Yoon:2016dij}. To
extract the ground state matrix elements from these, we fit the three-point data
at several $\tau$-values (listed in Table \ref{tab:ensembles})
simultaneously using the spectral decomposition given in
Eq.~\eqref{eq:3pt}.

Fits to the zero-momentum two-point functions, ${\rm C_{2pt}}$, were carried
out keeping up to four states in the spectral decomposition:
\begin{equation}
C_{2pt}(\tau) = \sum_{i=0}^3  |{\cal A}_i|^2e^{-M_i \tau} \,.
\label{eq:2pt}
\end{equation}
Fits are made over a range $\{\tau_{min}-\tau_{max}\}$ to extract
$M_i$ and ${\cal A}_i$, the masses and the amplitudes for the
creation/annihilation of these states by the interpolating operator
${\mathcal N}$.  In fits with more than two states, estimates of the
amplitudes ${\cal A}_i$ and masses $M_i$ for $i \geq 2$ were sensitive to the
choice of the starting time slice $\tau_{min}$. We used the largest
time interval allowed by statistics, i.e., by the stability of the
covariance matrix.  We perform two types of 4-state fits.  In the fit
denoted $\{4\}$, we use the empirical Bayesian technique described in
Ref.~\cite{Yoon:2016jzj} to stabilize the three excited-state
parameters. In the second fit, denoted $\{4^{N\pi}\}$, we use as prior
for $M_1$ either the noninteracting energy of $N({-\bm 1}) \pi({\bm
  1})$ or the $N({\bm 0})\pi({\bm 0}) \pi({\bm 0})$ state, which are both lower
than the $M_1$ obtained from the $\{4\}$ fit, and roughly equal for the
nine ensembles. The lower energy $N({-\bm 1}) \pi({\bm 1})$ state has
been shown to contribute in the axial channel~\cite{Jang:2019vkm},
whereas for the vector channel the $N({\bm 0}) \pi({\bm 0}) \pi({\bm
  0})$ state is expected to be the relevant one. We find that these
two fits to the two-point function cannot be distinguished on the
basis of the $\chi^2/$dof, in fact, the full range of $M_1$ between the
two estimates from $\{4\}$ and $\{4^{N\pi}\}$ are viable first-excited-state
masses on the basis of $\chi^2/$dof alone. The same is true of the values for $M_2$. 
We therefore, investigate the
dependence of the results for moments on the exited-state spectra by
doing the full analysis with multiple strategies as discussed below.
The ground-state nucleon mass obtained from the various fits is
denoted by the common symbol $M_N \equiv M_0$ and the mass gaps by
$\Delta M_i \equiv M_i - M_{i-1}$. 

The analysis of the zero-momentum three-point functions,
$C_\mathcal{O}^{3\text{pt}}$, is performed retaining up to three states $|i\rangle$ 
in the spectral decomposition: 
\begin{equation}
  C_\mathcal{O}^{3\text{pt}}(\tau;t) = 
   \sum_{i,j=0}^2 \abs{\mathcal{A}_i} \abs{\mathcal{A}_j}\matrixe{i}{\mathcal{O}}{j} e^{-M_i t - M_j(\tau-t)}\,.
   \label{eq:3pt}
\end{equation}
%
%
The operators, $\mathcal{O}$, are defined in
Eqs.~\eqref{eq:finaloperatorV},~\eqref{eq:finaloperatorA} and~\eqref{eq:finaloperatorT}.  By fixing the momentum at the sink to
zero and inserting the operator at zero momentum transfer we get the
forward matrix element. The practical challenge discussed above is
determining the relevant $M_1$ and $M_2$ to use, and, failing that, to
investigate the sensitivity of $\matrixe{0}{\mathcal{O}}{0}$ to
possible values of $M_1$ and $M_2$ and including that variation as a
systematic uncertainty.

\begin{table*}[htbp]
\setlength{\tabcolsep}{6pt}
\renewcommand{\arraystretch}{1.5}
\centering
\begin{tabular}{ |c|c|c|c|c|c|c| }
\hline

Fit-type&Observable&$c_1$&$c_2$&$c_3$&$c_4$&$\chi^2$/dof\\
\hline
CC&$\langle x \rangle_{u-d}$               &$0.170(14)$&$0.09(14)$&$0.19(11)$&&$0.74$\\
CC&$\langle x \rangle_{\Delta u-\Delta d}$ &$0.209(16)$&$0.15(16)$&$0.24(13)$&&$0.56$\\
CC&$\langle x \rangle_{\delta u-\delta d}$ &$0.201(20)$&$0.26(20)$&$0.35(16)$&&$0.88$\\
\hline
\hline
CCFV &$\langle x \rangle_{u-d}$              &$0.167(16)$&$0.12(16)$&$0.24(17)$&$-9(23)$ &$0.85$\\
CCFV &$\langle x \rangle_{\Delta u-\Delta d}$&$0.206(16)$&$0.18(17)$&$0.32(19)$&$-15(25)$&$0.59$\\
CCFV &$\langle x \rangle_{\delta u-\delta d}$&$0.202(21)$&$0.25(20)$&$0.34(24)$&$ 3(31)$ &$1.06$\\
\hline
\end{tabular}
\caption{Results for the fit parameters in the CCFV ansatz given in
  Eq.~\eqref{eq:CCFV} and used for the chiral, continuum and finite
  volume (CCFV) extrapolation of the $\{4,3^\ast \}$ data. The CC and
  CCFV fit-types correspond to fits with $c_4=0$ or $c_4 \neq 0$.}
\label{tab:CCFV}
\end{table*}

For all the strategies used to determine $M_1$ and $M_2$, we extract the
desired ground state matrix element $\la 0| {\cal O}| 0 \ra $ by
fitting the three-point correlators
$C_\mathcal{O}^{3\text{pt}}(t;\tau)$ for a subset of values of $t$ and
$\tau$ simultaneously. This subset is chosen to reduce ESC---we select the
largest values of $\tau$ and discard $\tskip$ number of points next to the source
and sink for each $\tau$. These values of $\tau$ and of $\tskip$ are given in
Table~\ref{tab:best-fits}.

The data for the ratio
$C_\mathcal{O}^{3\text{pt}}(\tau;t)/C^{2\text{pt}}(\tau)$ are shown in
Figs.~\ref{fig:Ratio1} and~\ref{fig:Ratio2} in the
Appendix~\ref{sec:ratios} for all nine ensembles. The signal in the
three-point correlators decreases somewhat from momentum fraction to
helicity moment to transversity moment.  Nevertheless, we are able to
make $3^\ast$state (3-state with $\matrixe{2}{\mathcal{O}}{2} = 0$)
fits in all cases.  The spectral decomposition predicts that the data
for all three quantities is symmetric about $t=\tau/2$, however, on
some of the ensembles, and for some of the larger values of $\tau$,
the data show some asymmetry, which is indicative of the size of
statistical fluctuations that are present.

The fits to $C^{2\text{pt}}(\tau)$ and
$C_\mathcal{O}^{3\text{pt}}(\tau;t)$ are carried out within a
single-elimination jackknife process, which is used to get both the
central values and the errors.

We have investigated five fit types, $\{4,2\}$, $\{4^{N\pi},2\}$,
$\{4,3^*\}$, $\{4^{N\pi},3^*\}$ and $\{4,2^{\rm free}\}$, based on the
spectral decomposition to understand and control ESC. The labels
$\{m,n\}$ denote an $m$-state fit to the two-point function and an
$n$-state fit to the three-point function.  In the $2^{\rm {free}}$-fit 
to the three-point function, $M_1$ is left as a free parameter,
while a $3^\ast$-fit is a 3-state fit with $\langle 2 | O | 2\rangle =
0$.  The results from the five strategies for the momentum fraction, $\la x \ra_{u-d}$, in
Table~\ref{tab:Npi-fits-momfrac}, for the helicity moment, $\la x
\ra_{\Delta u-\Delta d}$, in Table~\ref{tab:Npi-fits-helfrac}, and for
the transversity moment, $\la x \ra_{\delta u-\delta d}$, in
Table~\ref{tab:Npi-fits-transfrac} illustrate the observed behavior
for the $a09m310$ ensemble, which has the highest statistics, and the
physical mass ensemble $a06m135$ at the smallest value of $a$.

For all three observables, the five results in
Tables~\ref{tab:Npi-fits-momfrac},~\ref{tab:Npi-fits-helfrac},
and~\ref{tab:Npi-fits-transfrac} for the ground state matrix element,
$\la 0 | O | 0 \ra$, are consistent within $ 2\sigma$ on the $a09m310$
ensemble. On the $a06m135$ ensemble, the difference in $\Delta M_1
\equiv M_1 - M_0$ between $\{4\}$ and $\{4^{N\pi}\}$ analyses becomes
roughly a factor of 2, and $\Delta M_1$ from the $\{2^{\rm free}\}$
fit is larger than even the $\{4\}$ value, i.e., the $\{2^{\rm
free}\}$ fit does not prefer the small $\Delta M_1$ given by
$\{4^{N\pi}\}$. On the other hand, the $\Delta M_1$ from a two-state
fit is expected to be larger since it is an effective combination of
the mass gaps of the full tower of excited states. Due to a small $\Delta M_1$, fits
with the spectrum from $\{4^{N\pi}\}$ fail on $a06m135$, whereas, on both ensembles,
the $\{4,3^*\}$ and $\{4,2^{\rm free}\}$ fits give results consistent
within $ 2\sigma$.  The estimates from these two fit-types are given
in Table~\ref{tab:best-fits}.  To summarize, our overall strategy is
to keep as many excited states as possible without
overparameterization of the fits. We, therefore, choose, for the
central values, the $\{4,3^*\}$ results, and to take into account the
spread due to the fit-type, we add a second, systematic, uncertainty
to the final results in Table~\ref{tab:finalresults}. This is taken to
be the difference between the results obtained by doing the full
analysis with the $\{4,3^*\}$ and $\{4,2^{\rm free}\}$ strategies.

The renormalization of the matrix elements is
carried out using estimates of $Z_{VD},~Z_{AD}$, and $ Z_{TD}$
calculated on the lattice in the \ripmom scheme and then converted to
the $\MSbar$ scheme at 2~GeV as described in 
Appendix~\ref{sec:renormalization}. The final values of
$Z_{VD},~Z_{AD}$, and $ Z_{TD}$ used in the analysis are given in
Table \ref{tab:Z-fac}.

\begin{table}[htbp]
\centering
\setlength{\tabcolsep}{4pt}
\renewcommand{\arraystretch}{1.5}
\begin{tabular}{ |c|c|c|c| }
\hline
Observable & $\{4,3^\ast \}$  &  $\{4,2^{\rm free}\}$  & Best estimate \\
\hline
$\langle x \rangle_{u-d}^{{\rm \ol{MS}}}$               & 0.173(14) & 0.180(14)  & 0.173(14)(07) \\
$\langle x \rangle_{\Delta u-\Delta d}^{{\rm \ol{MS}}}$ & 0.213(15) & 0.235(15)  & 0.213(15)(22) \\
$\langle x \rangle_{\delta u-\delta d}^{{\rm \ol{MS}}}$ & 0.208(19) & 0.236(18)  & 0.208(19)(24) \\
\hline
\hline
\end{tabular}
\caption{Results for the three moments from the two strategies
  $\{4,3^\ast \}$ and $\{4,2^{\rm free}\}$. For our best estimates, we take the
  $\{4,3^\ast \}$ values and assign a second, systematic, error that is the
  difference between the two results. The results are in the ${{\rm \ol{MS}}}$ scheme at 
scale 2~GeV. }
\label{tab:finalresults}
\end{table}

\section{Chiral, continuum and infinite volume extrapolation}
\label{sec:results}

To obtain the final, physical results at $M_\pi=135$~MeV, $M_\pi L \to
\infty$ and $a=0$, we make a simultaneous CCFV fit keeping only the
leading correction term in each variable:
\begin{flalign}
\la x \ra (M_\pi; a;L) &= c_1 + c_2 a +c_3 M_\pi^2  
+ c_4 \frac{M_\pi^2~ e^{-M_\pi L}}{\sqrt{M_\pi L}} \,. 
\label{eq:CCFV}
\end{flalign}
Note that, since the operators are not $O(a)$ improved and we used the
Clover-on-HISQ formulation, we take the discretization errors to start
with a term linear in $a$. The fits to the $\{4,3^\ast\}$ data from
the nine ensembles are shown in
Figs.~\ref{fig:momfrac-vs-a},~\ref{fig:helfrac-vs-a}
and~\ref{fig:transversity-vs-a}.  The fit parameters are summarized in
Table~\ref{tab:CCFV}.  

\begin{figure*}[htbp]
\begin{subfigure}
\centering
\includegraphics[angle=0,width=0.32\textwidth]{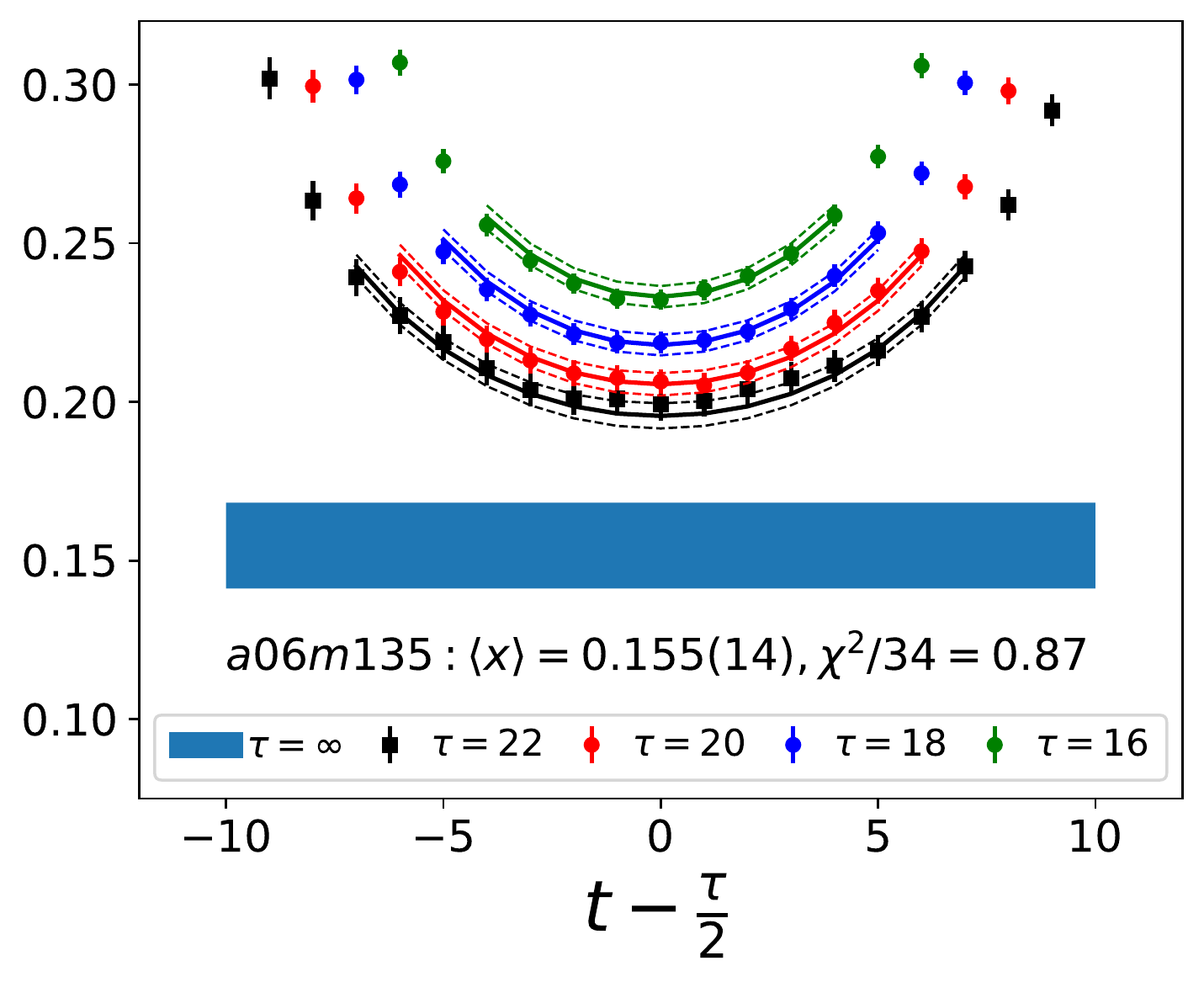}
\includegraphics[angle=0,width=0.32\textwidth]{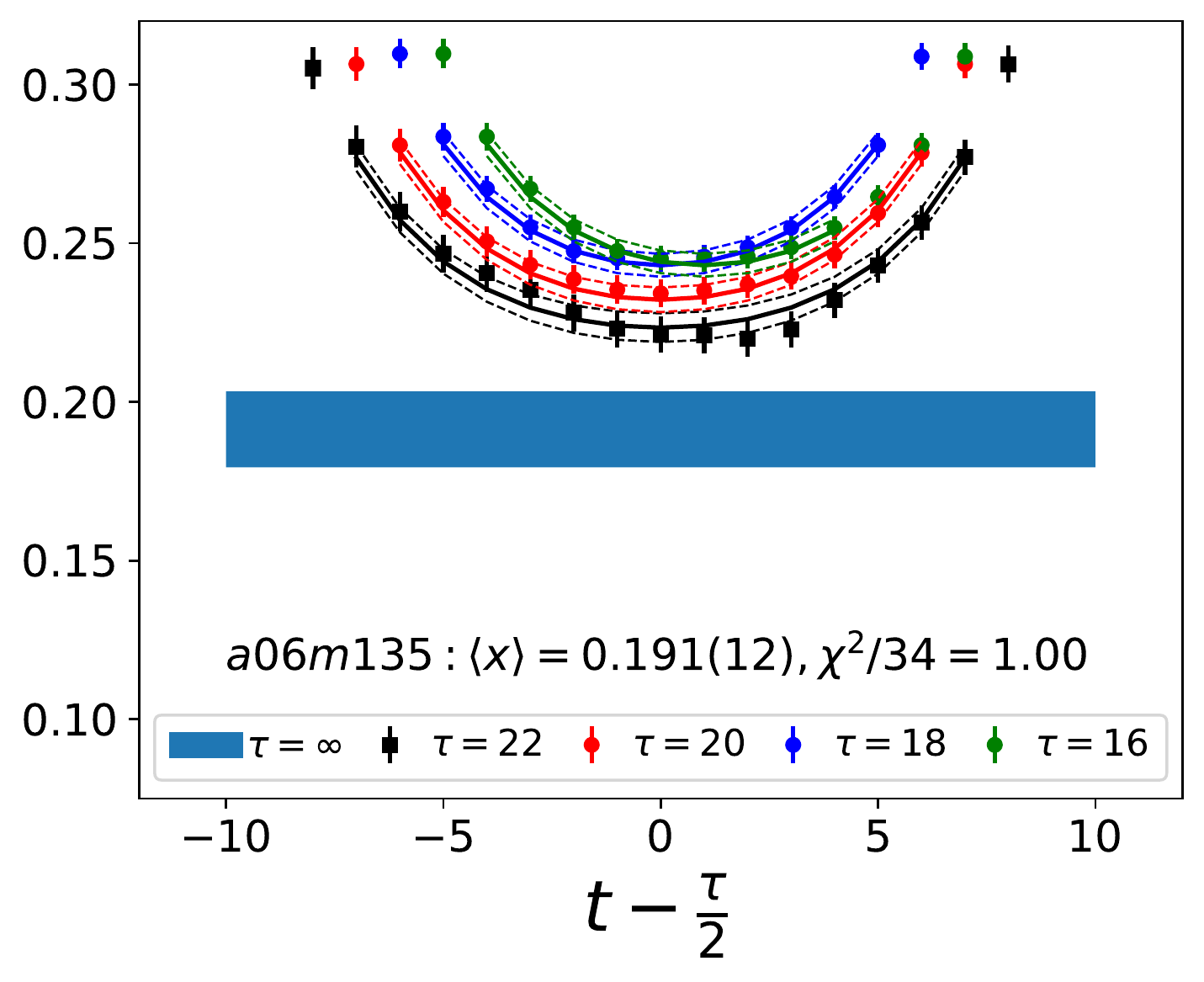}
\includegraphics[angle=0,width=0.32\textwidth]{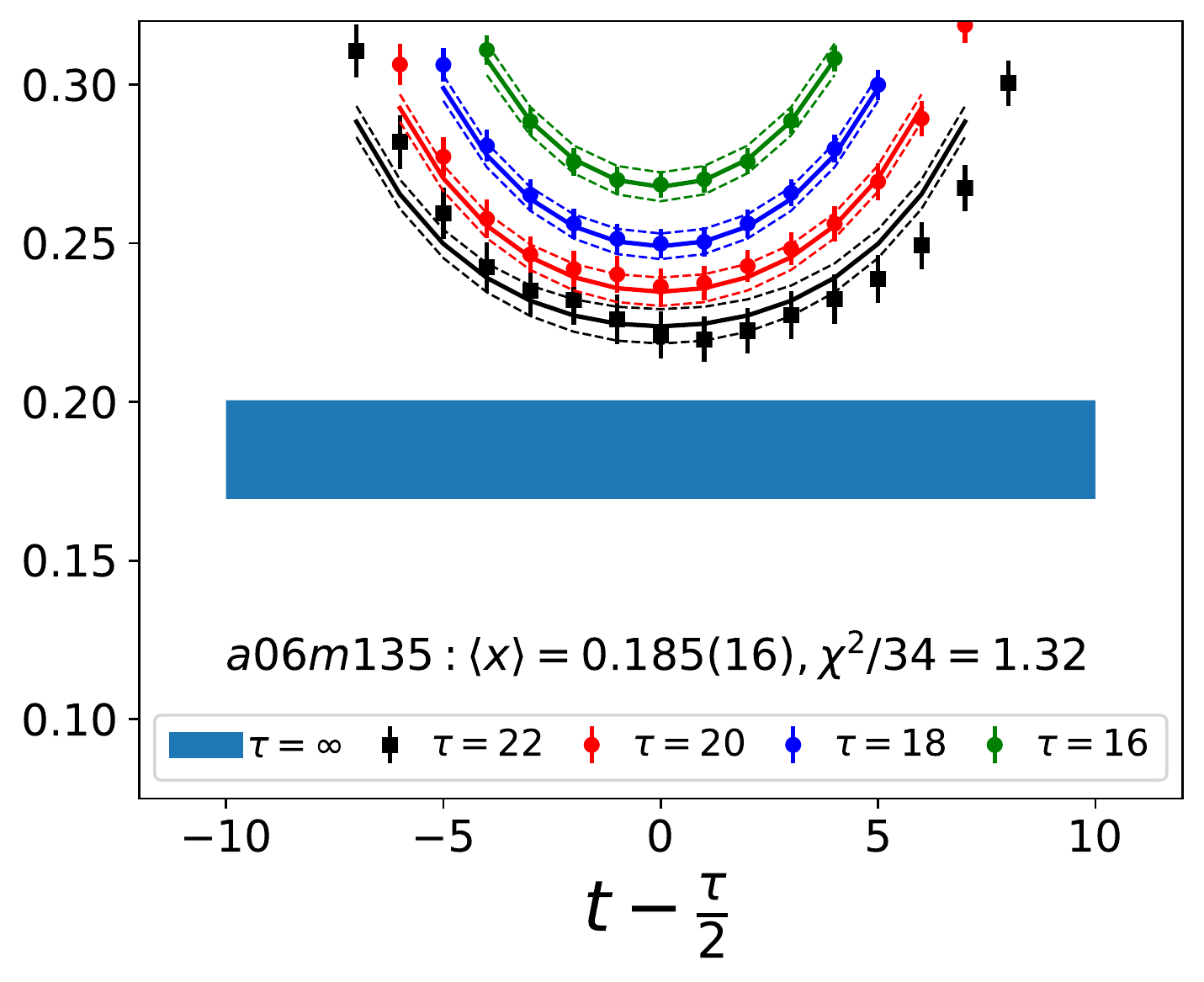}
\end{subfigure}
\begin{subfigure}
\centering
\includegraphics[angle=0,width=0.32\textwidth]{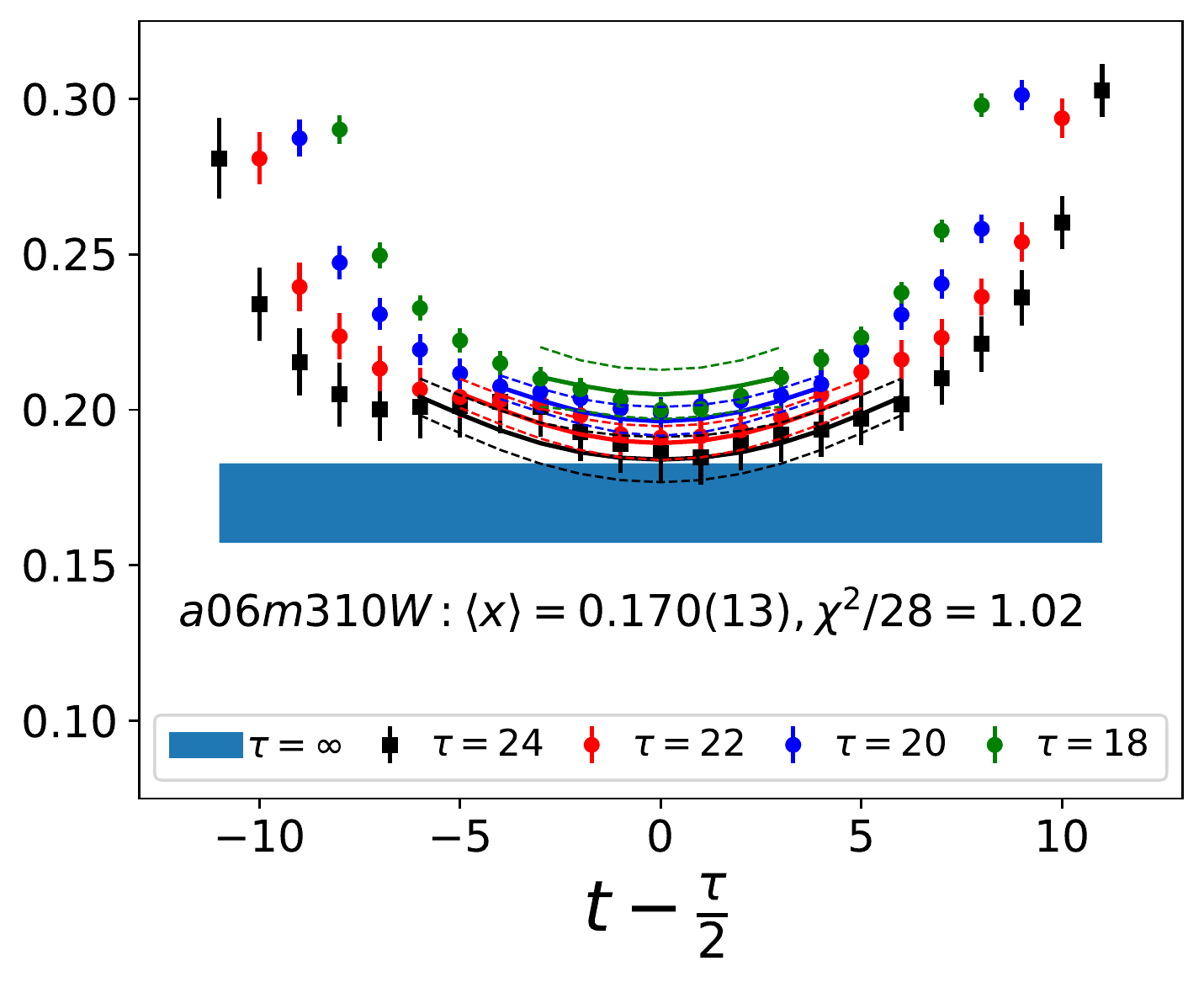}
\includegraphics[angle=0,width=0.32\textwidth]{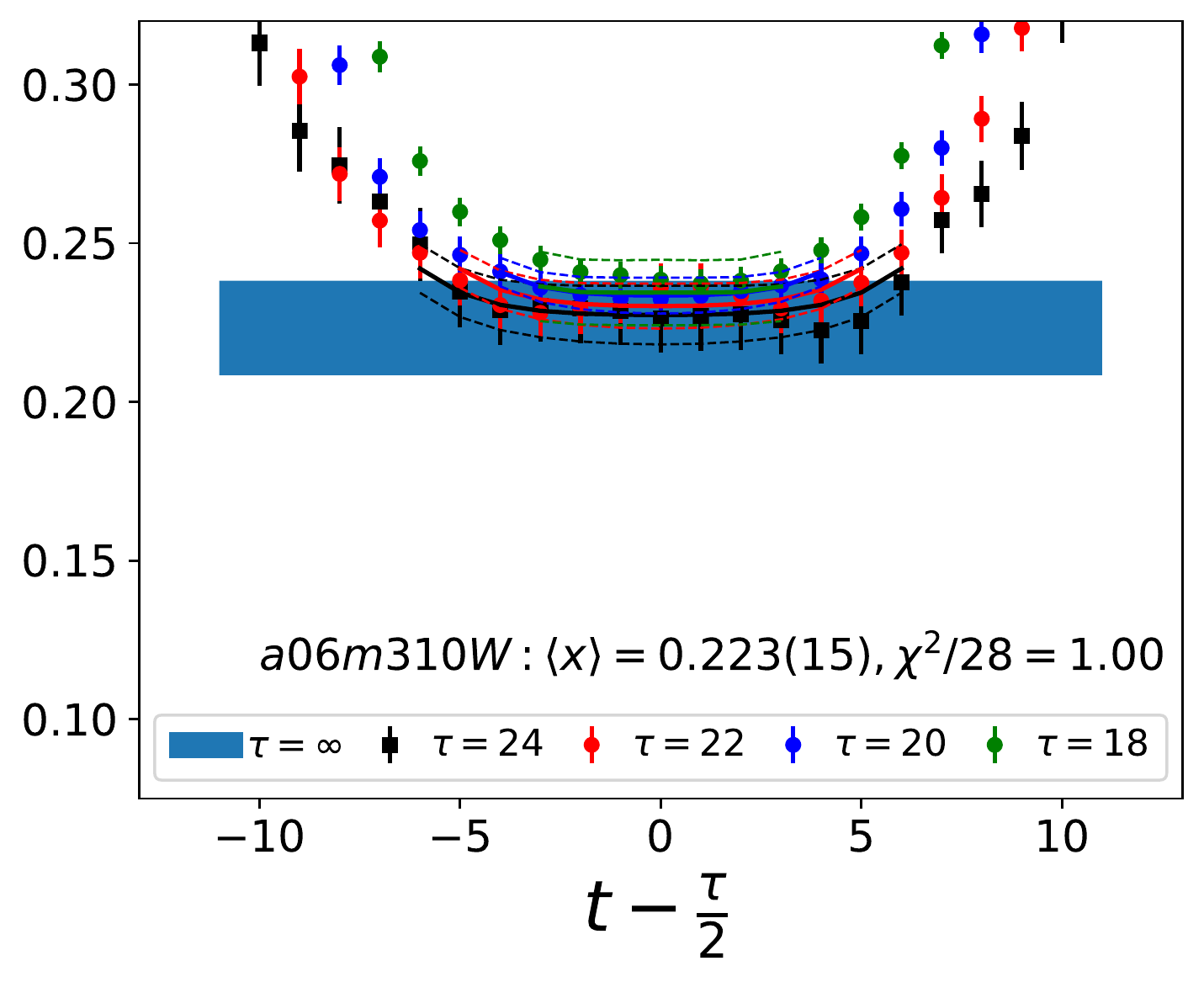}
\includegraphics[angle=0,width=0.32\textwidth]{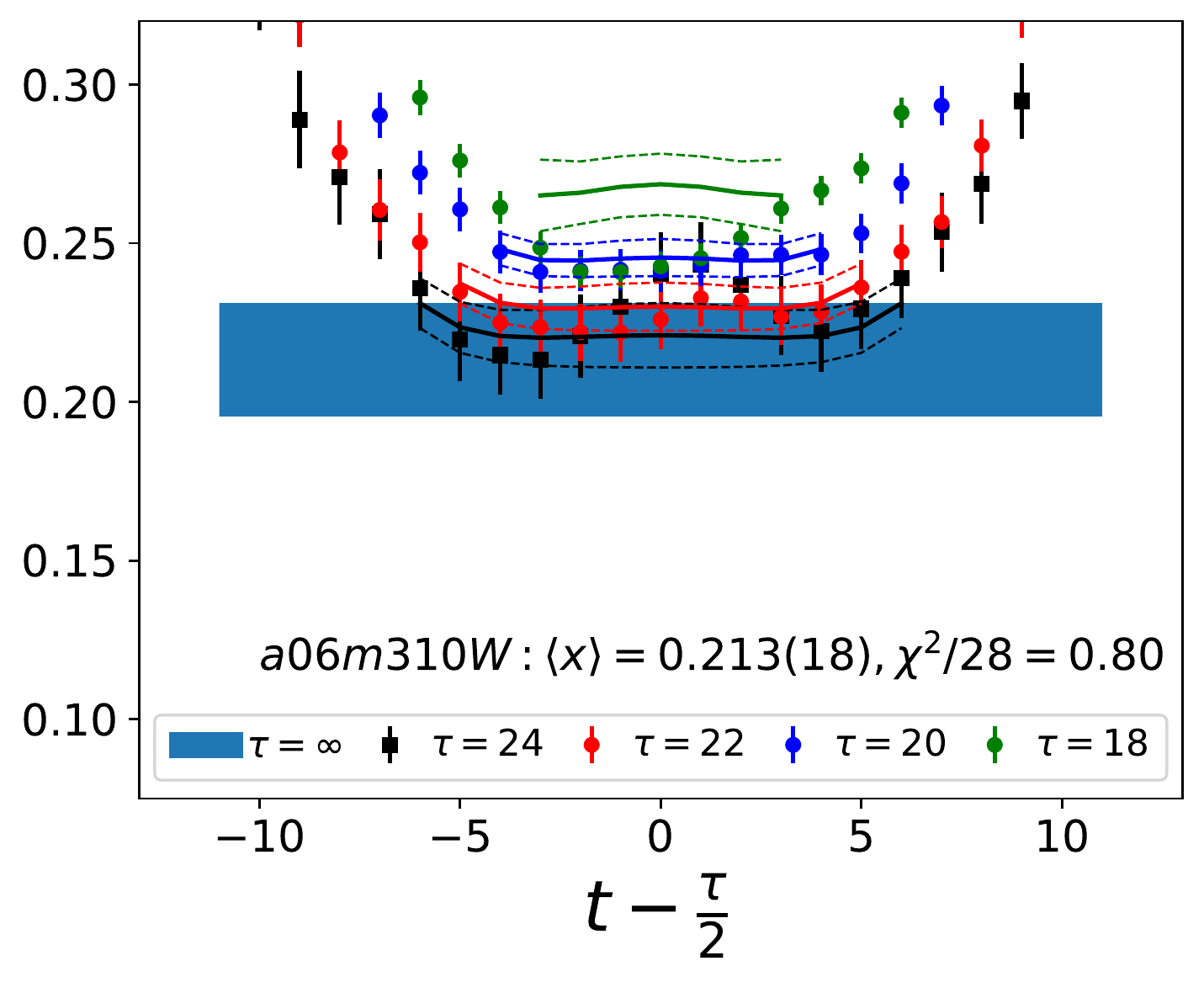}
\end{subfigure}
\begin{subfigure}
\centering
\includegraphics[angle=0,width=0.32\textwidth]{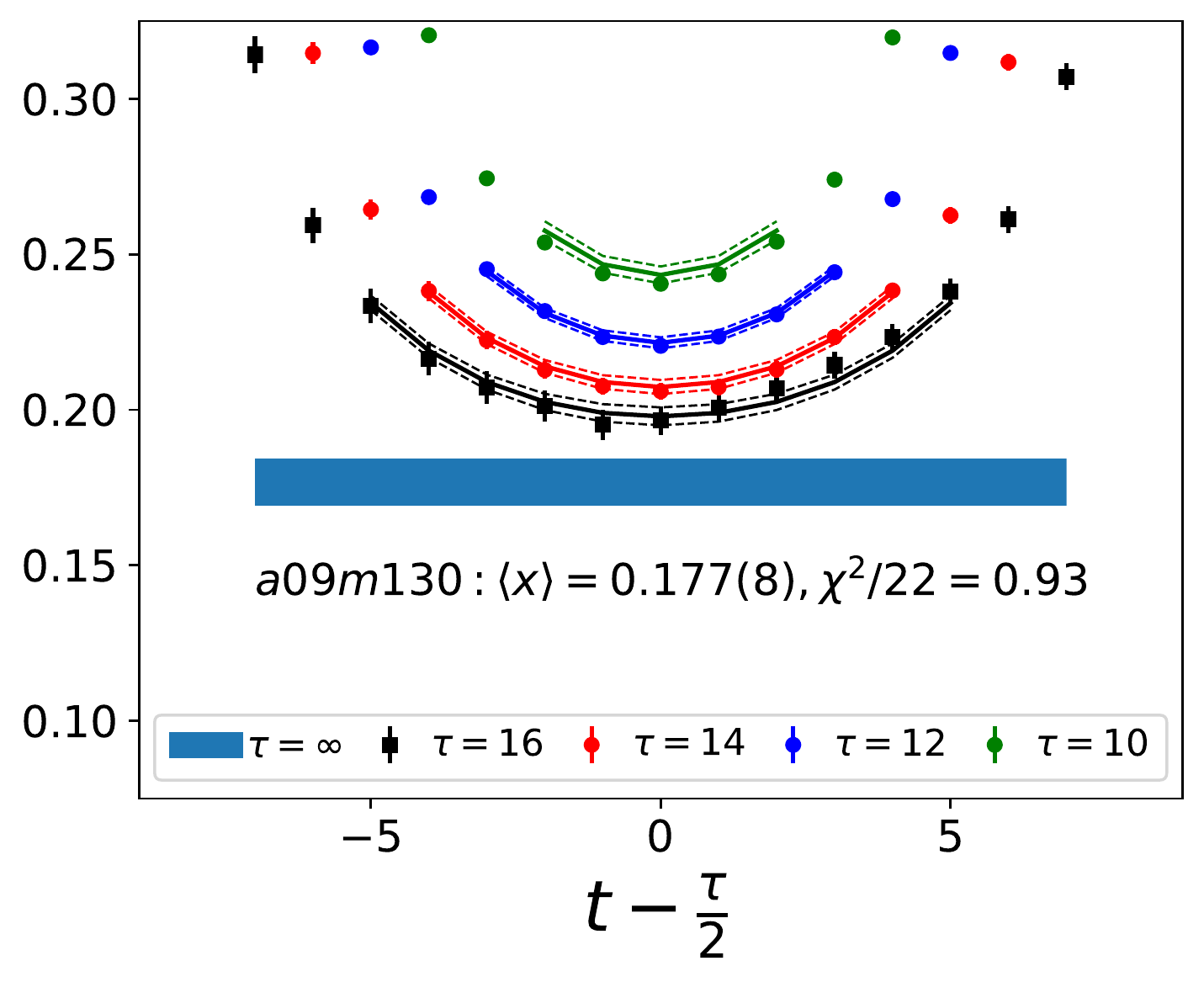}
\includegraphics[angle=0,width=0.32\textwidth]{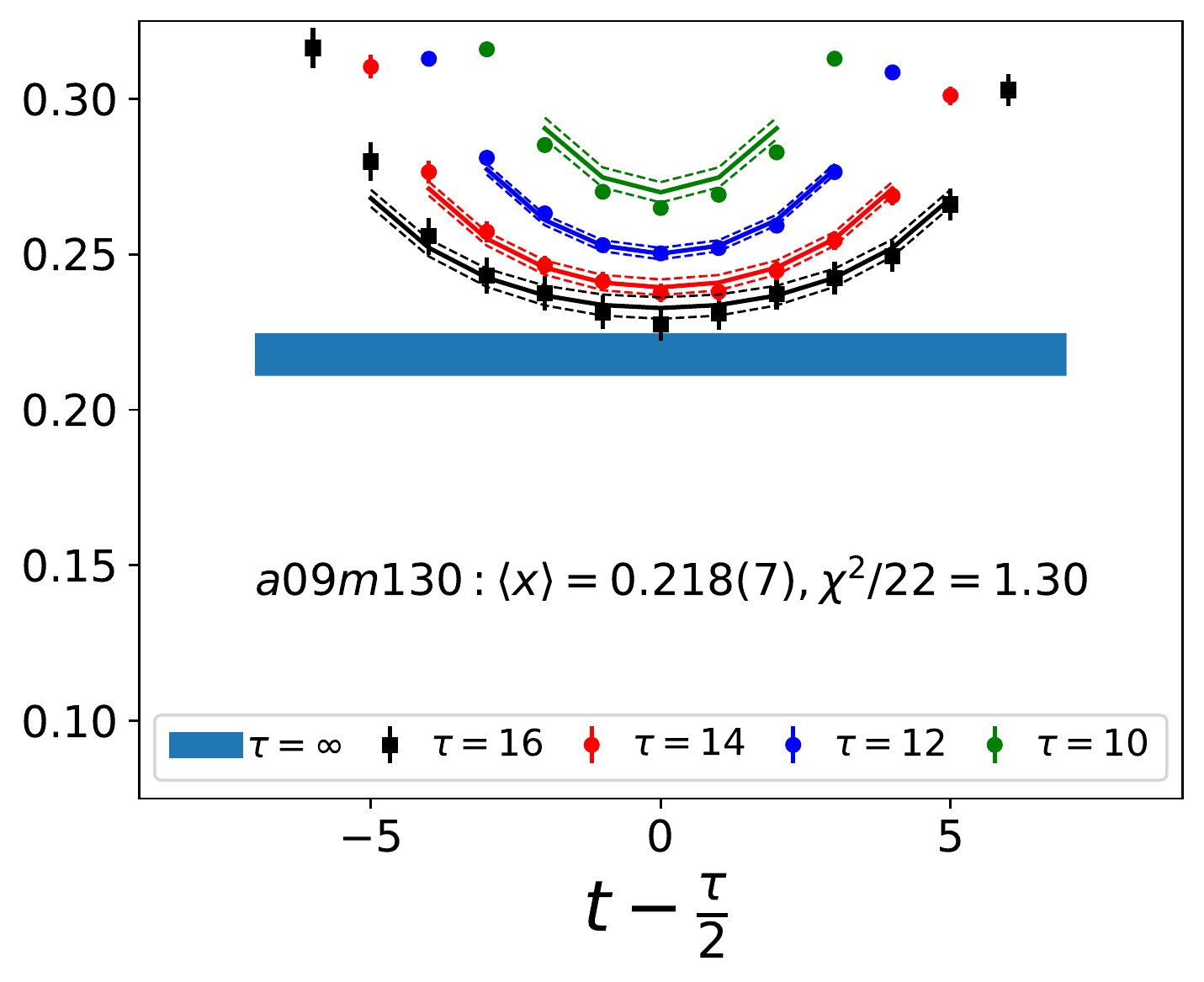}
\includegraphics[angle=0,width=0.32\textwidth]{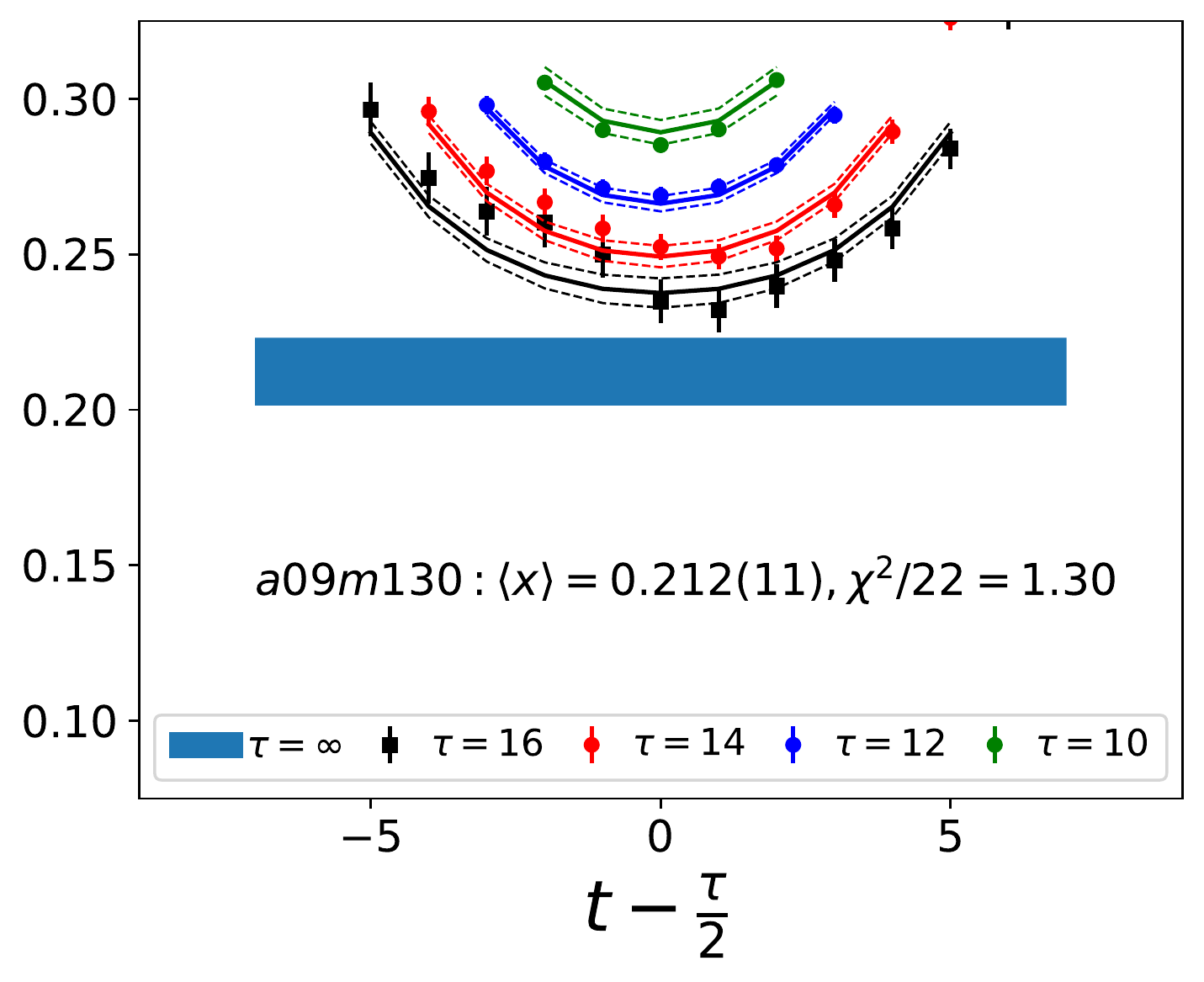}
\end{subfigure}
\begin{subfigure}
\centering
\includegraphics[angle=0,width=0.32\textwidth]{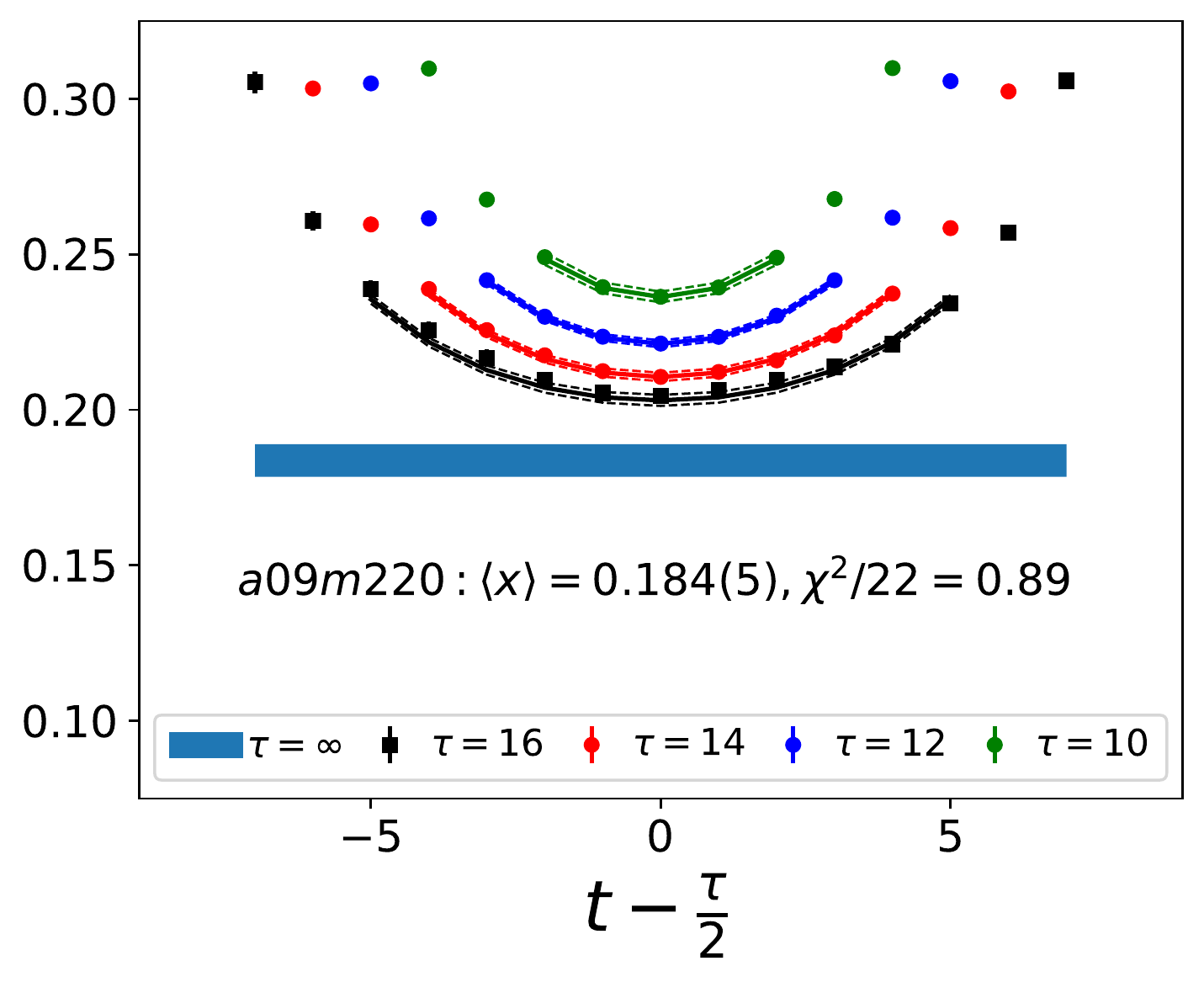}
\includegraphics[angle=0,width=0.32\textwidth]{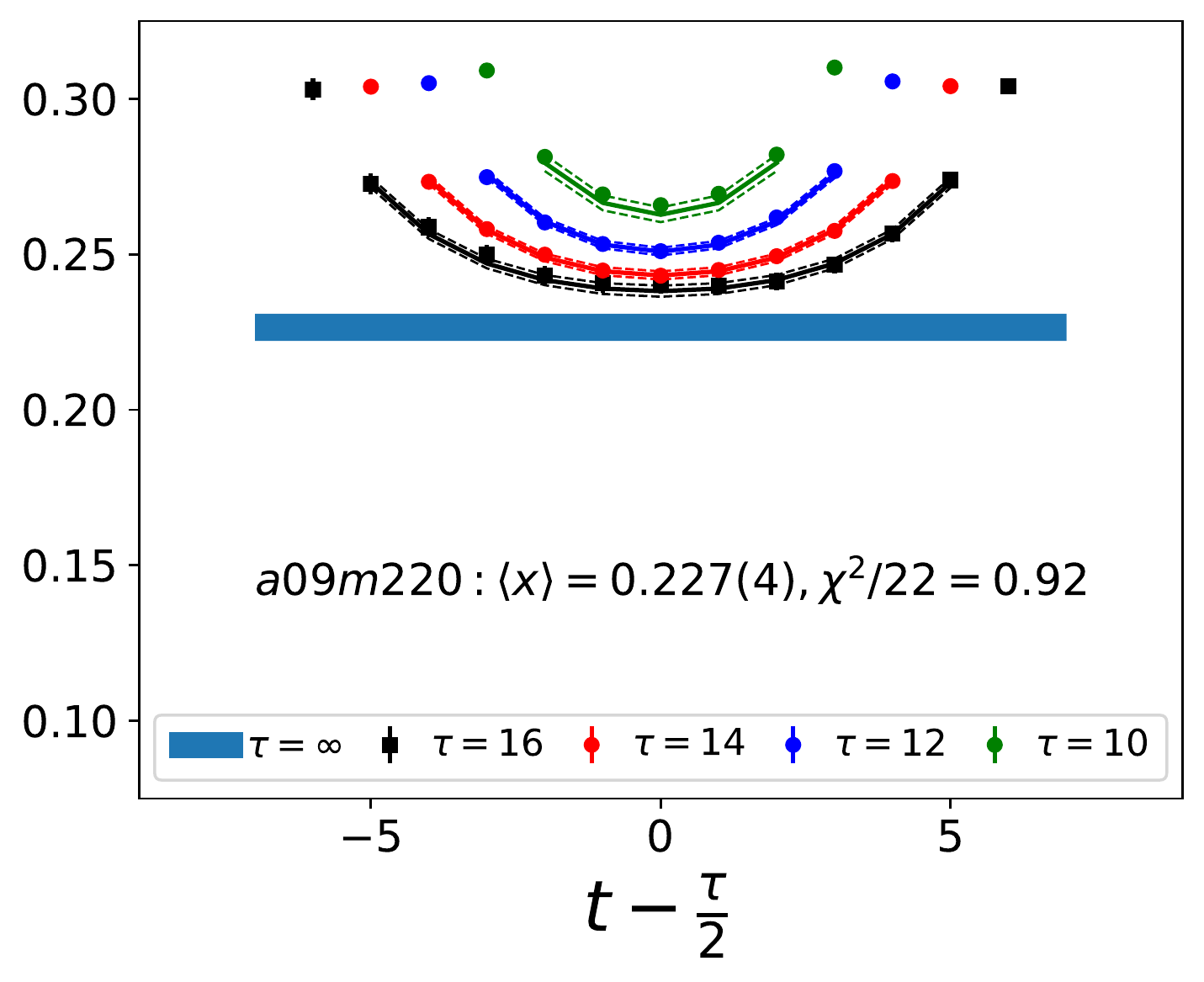}
\includegraphics[angle=0,width=0.32\textwidth]{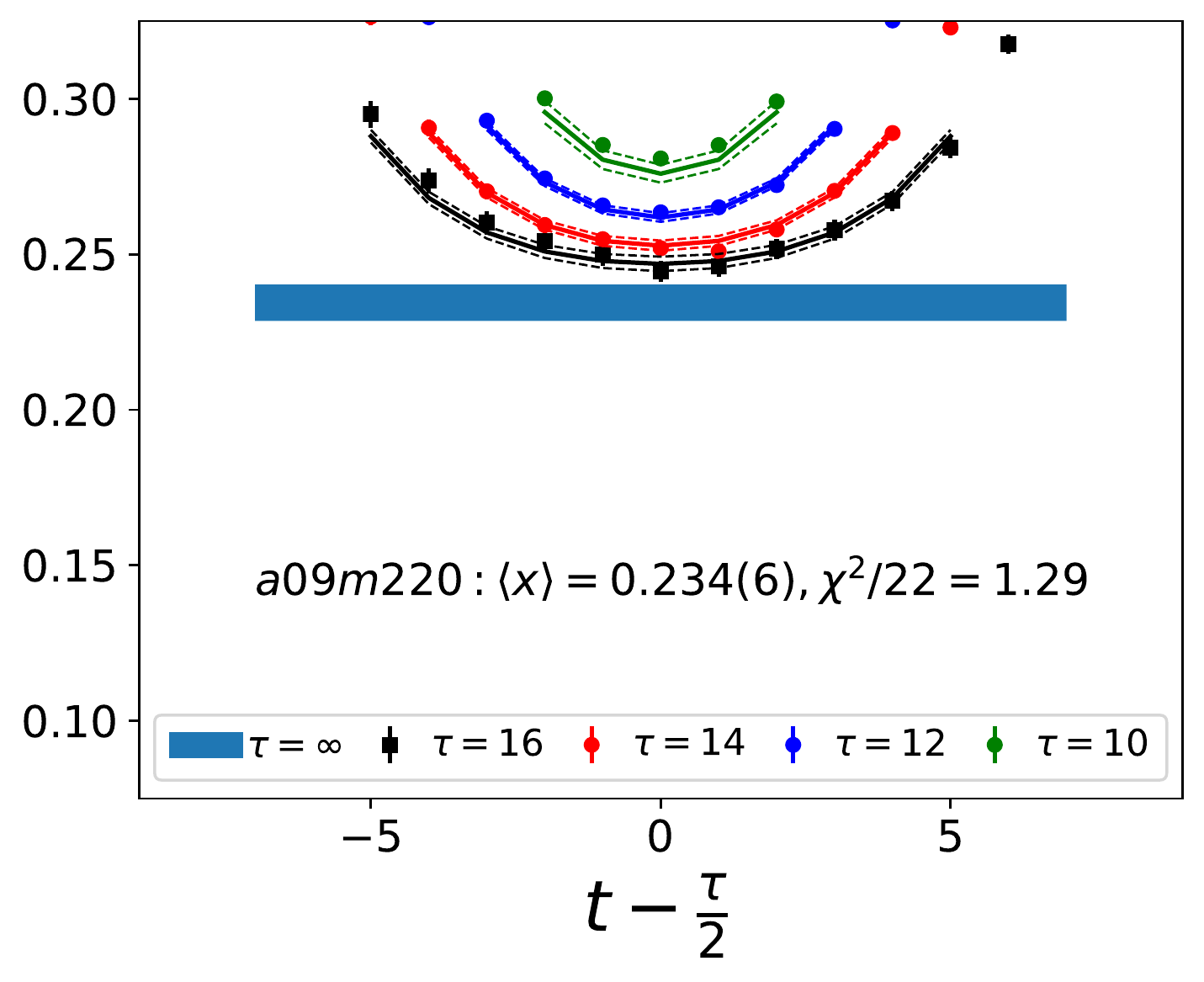}
\end{subfigure}
\caption{Data and fits for $a06m135$ (top row), $a06m310W$ (second
  row), $a09m130$ (third row) and $a09m220$ (last row).  In each row,
  the three panels show the ratio
  $C_\mathcal{O}^{3\text{pt}}(\tau;t)/C^{2\text{pt}}(\tau)$ scaled
  according to
  Eq.~\protect\eqref{eq:me2momentV}--\protect\eqref{eq:me2momentT} to
  give $\langle x \rangle_{u-d}$ (left), $\langle x \rangle_{\Delta
  u-\Delta d}$ (middle), and $\langle x \rangle_{\delta u-\delta d}$
  (right).  For each $\tau$, the line in the same color as the data
  points is the result of the $\{4,3^\ast\}$ fit (see
  Sec.~\protect\ref{sec:ESC}) used to obtain the ground state matrix
  element. The ensemble ID, the final result $\langle x \rangle$ (also
  shown by the blue band and summarized in
  Table~\protect\ref{tab:best-fits}), the values of $\tau$, and
  $\chi^2/$dof of the fit are also given in the legends. The interval
  of the y axis is selected to be the same for all the panels to
  facilitate comparison. }
\label{fig:Ratio1}
\end{figure*}

\begin{figure*}[ht]
\begin{subfigure}
\centering
\includegraphics[angle=0,width=0.3\textwidth]{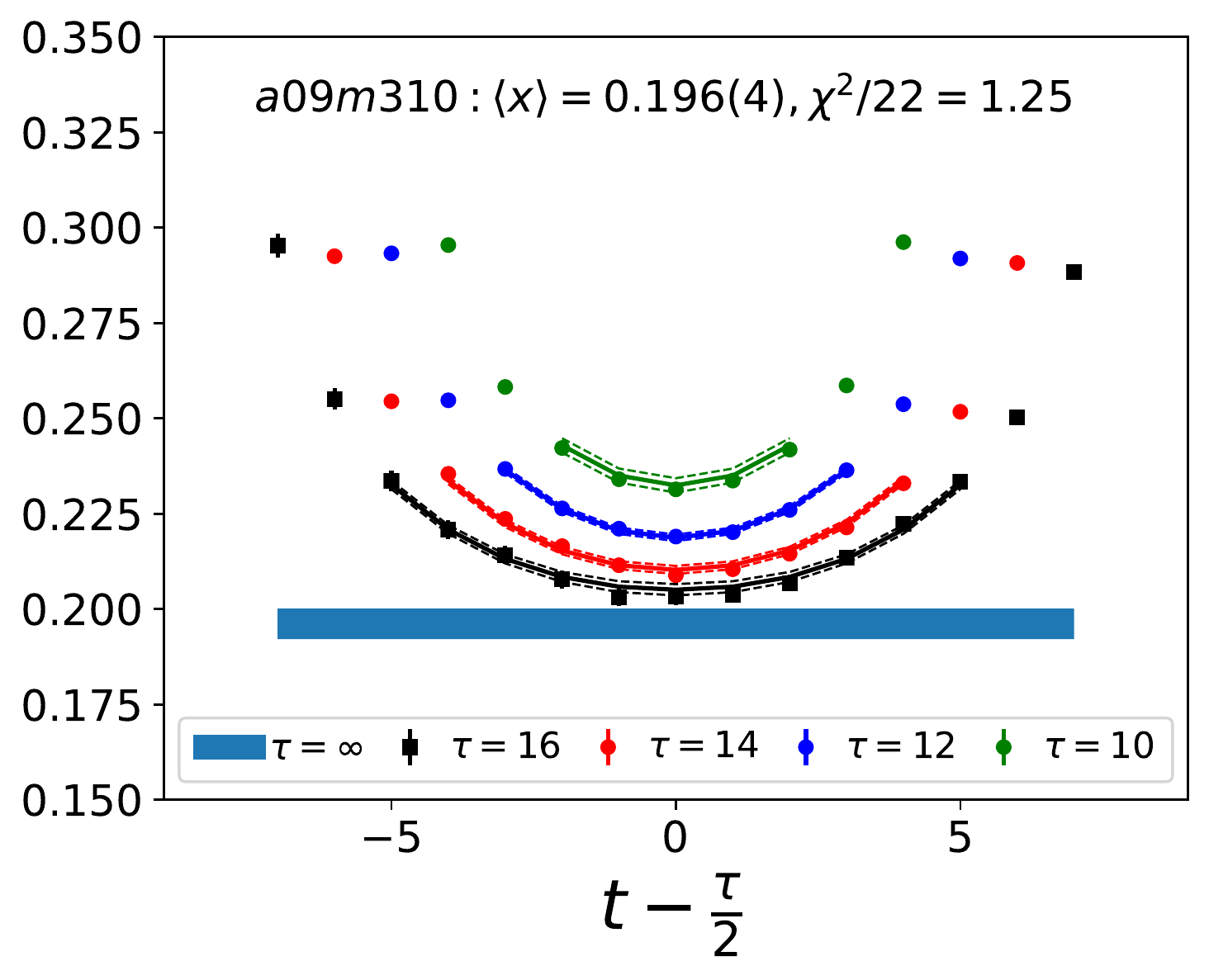}
\includegraphics[angle=0,width=0.3\textwidth]{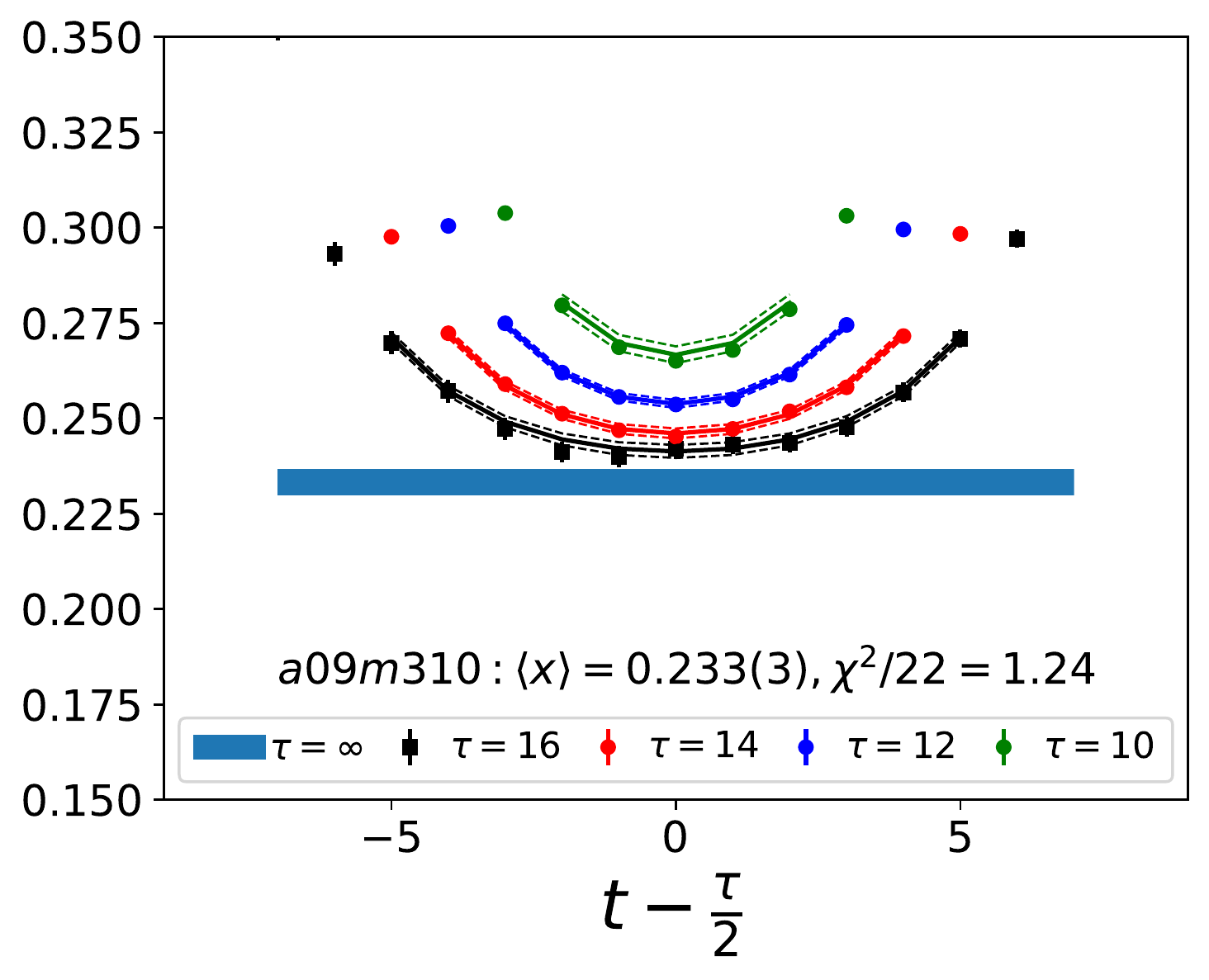}
\includegraphics[angle=0,width=0.3\textwidth]{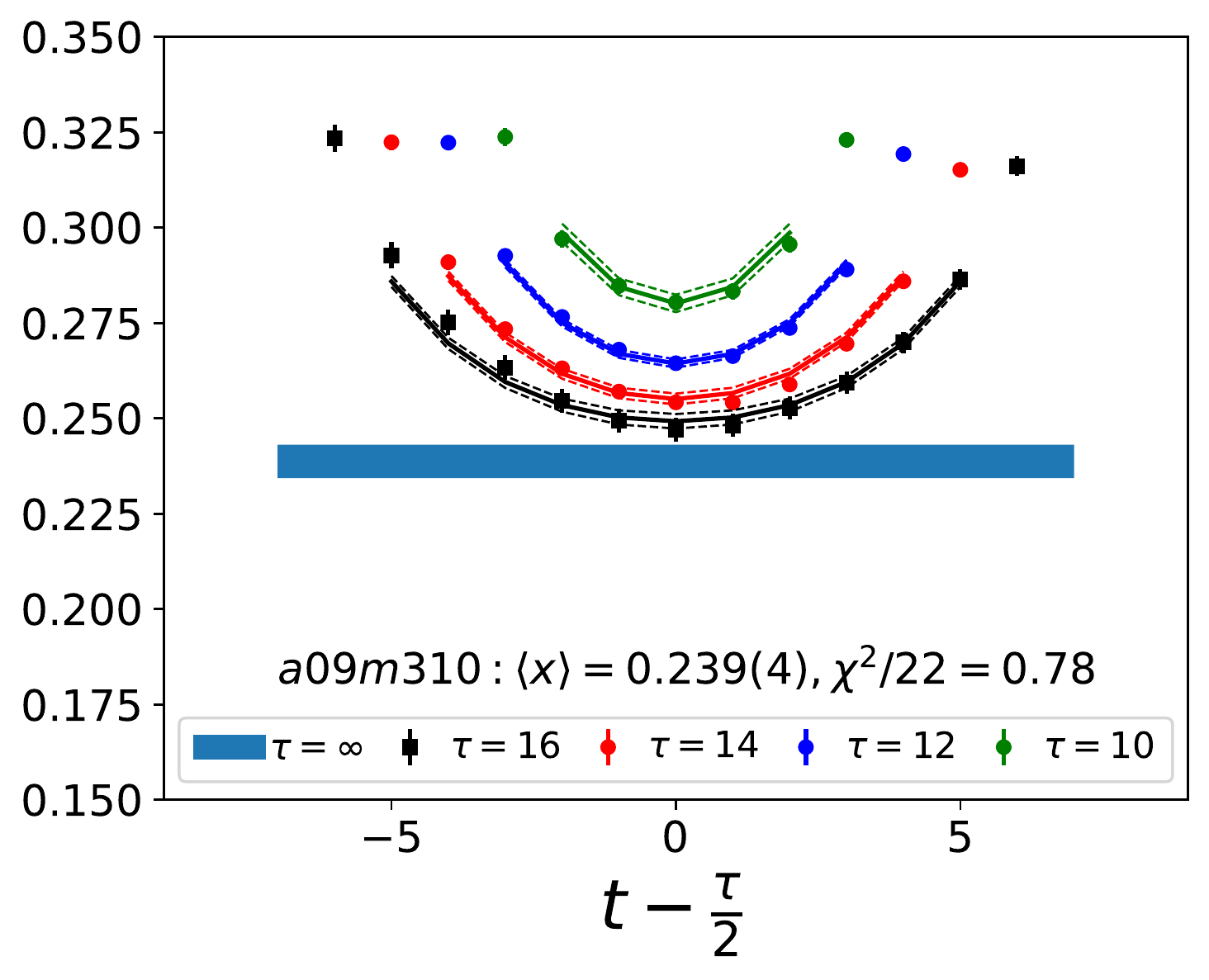}
\end{subfigure}
\begin{subfigure}
\centering
\includegraphics[angle=0,width=0.3\textwidth]{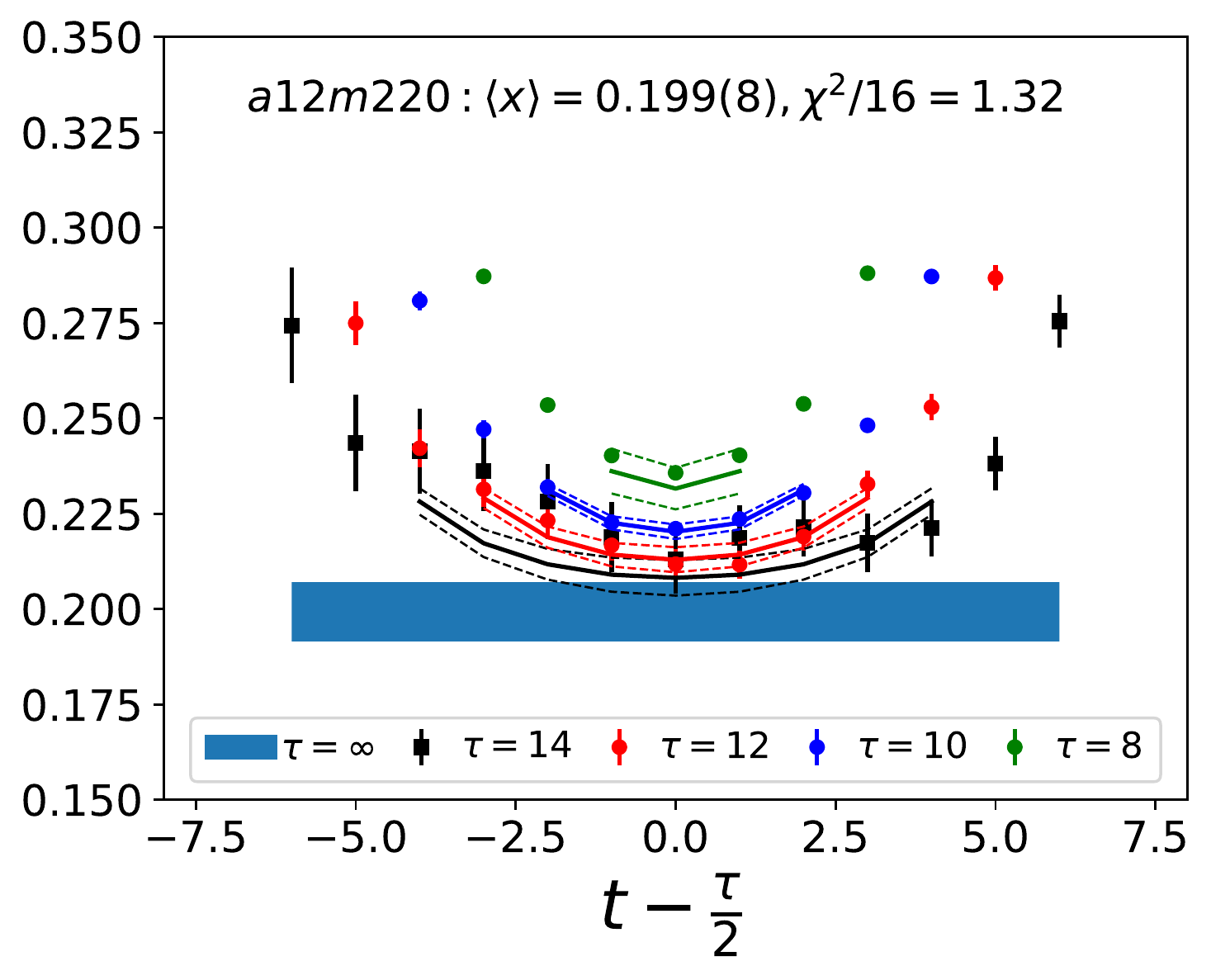}
\includegraphics[angle=0,width=0.3\textwidth]{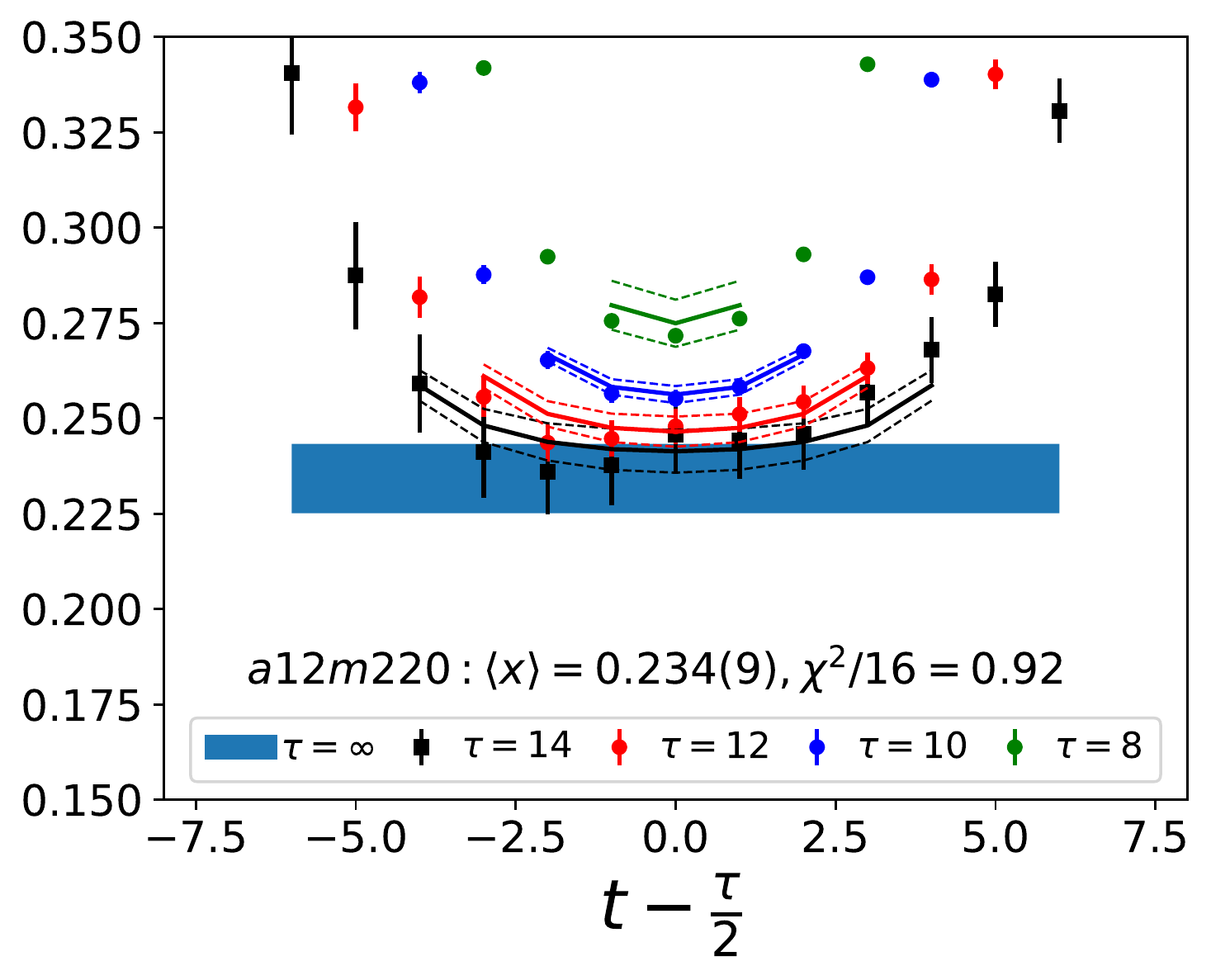}
\includegraphics[angle=0,width=0.3\textwidth]{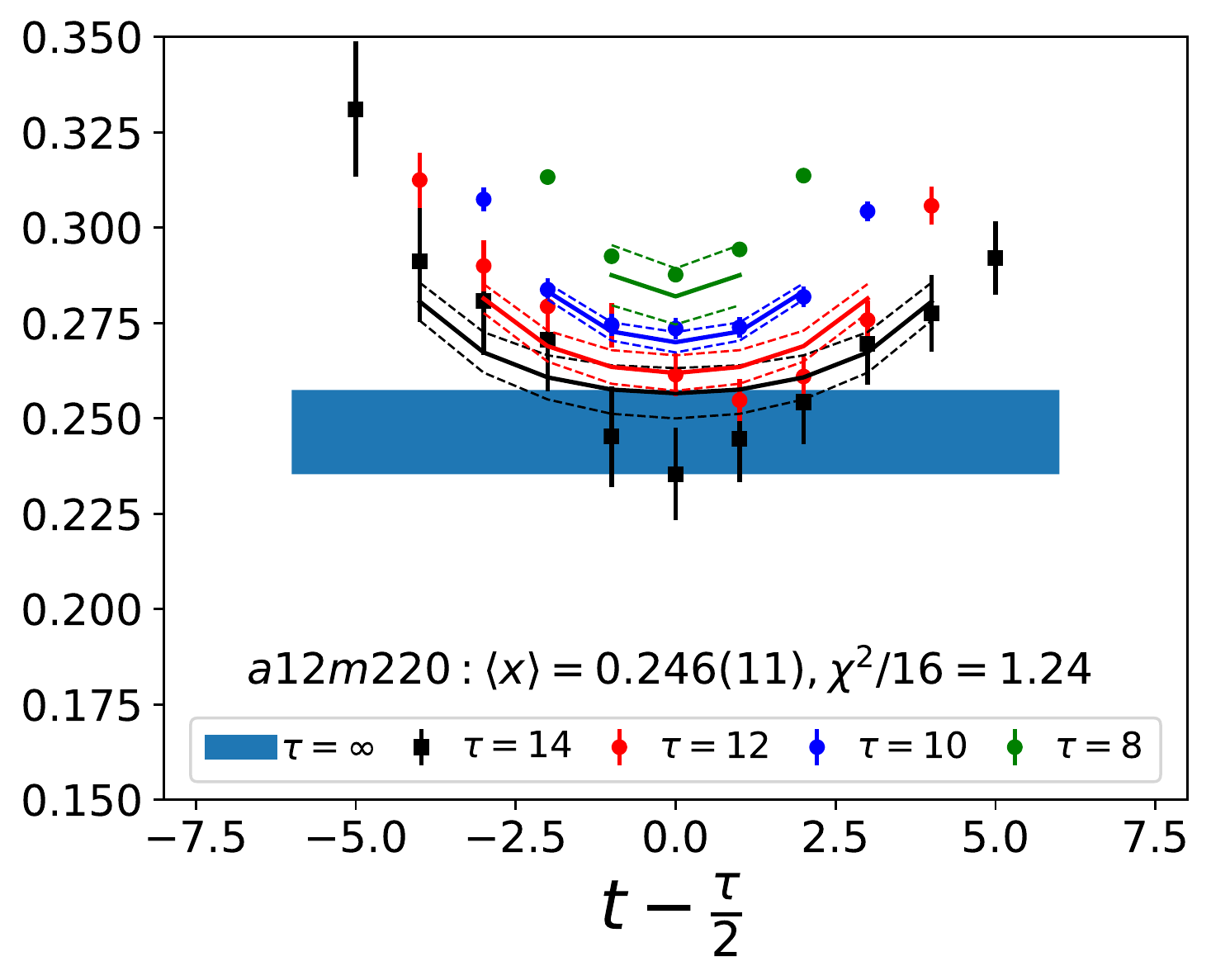}
\end{subfigure}
\begin{subfigure}
\centering
\includegraphics[angle=0,width=0.3\textwidth]{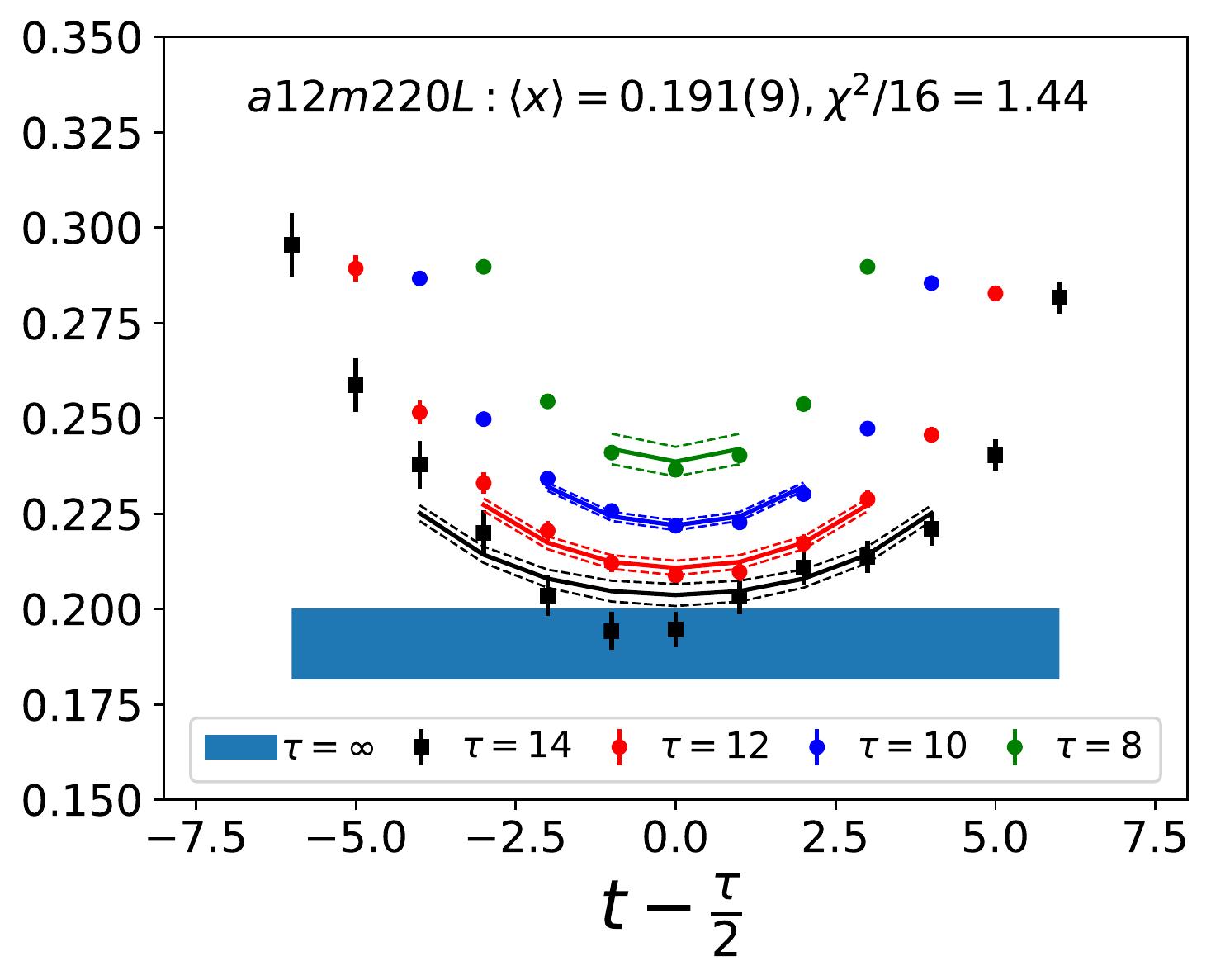}
\includegraphics[angle=0,width=0.3\textwidth]{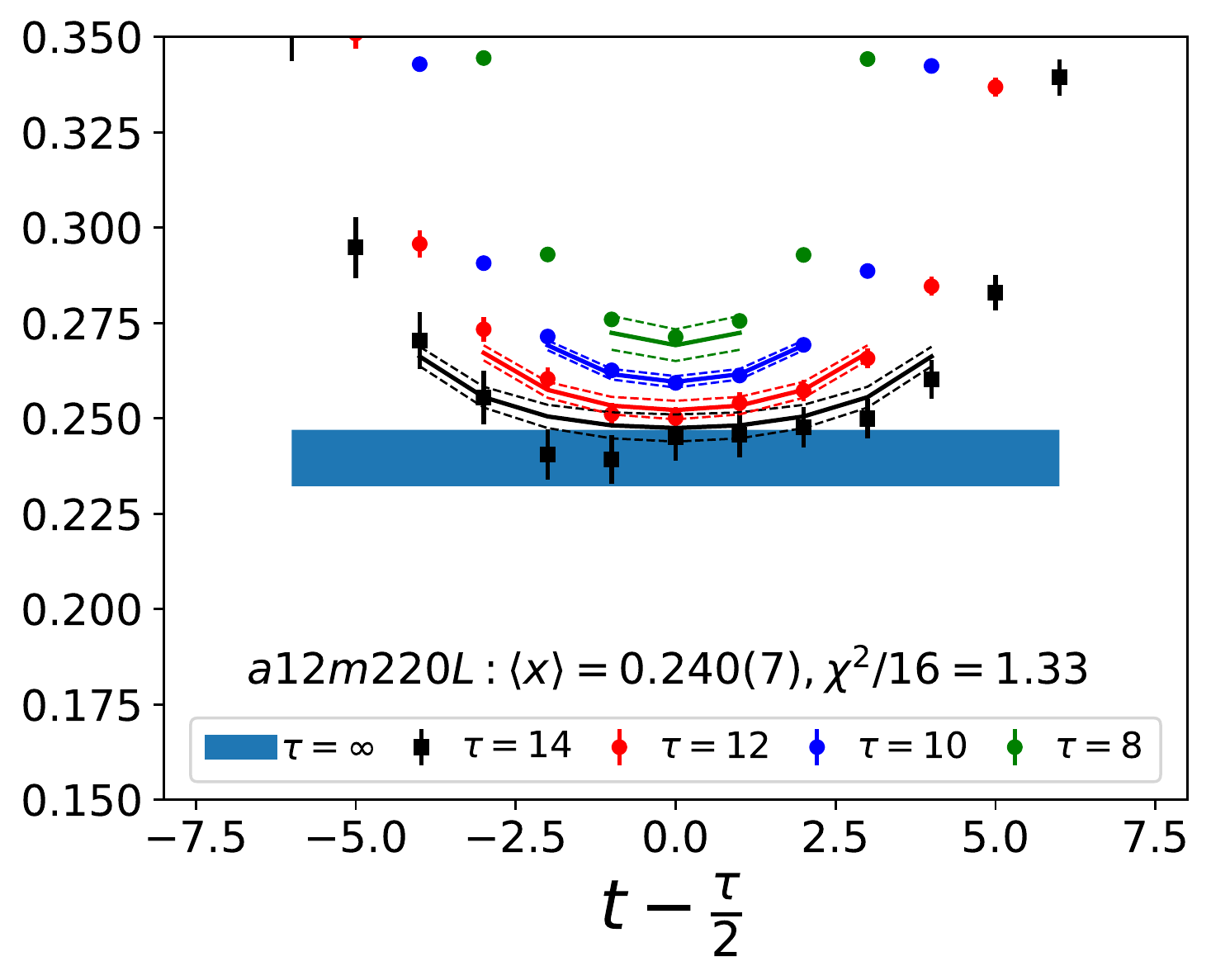}
\includegraphics[angle=0,width=0.3\textwidth]{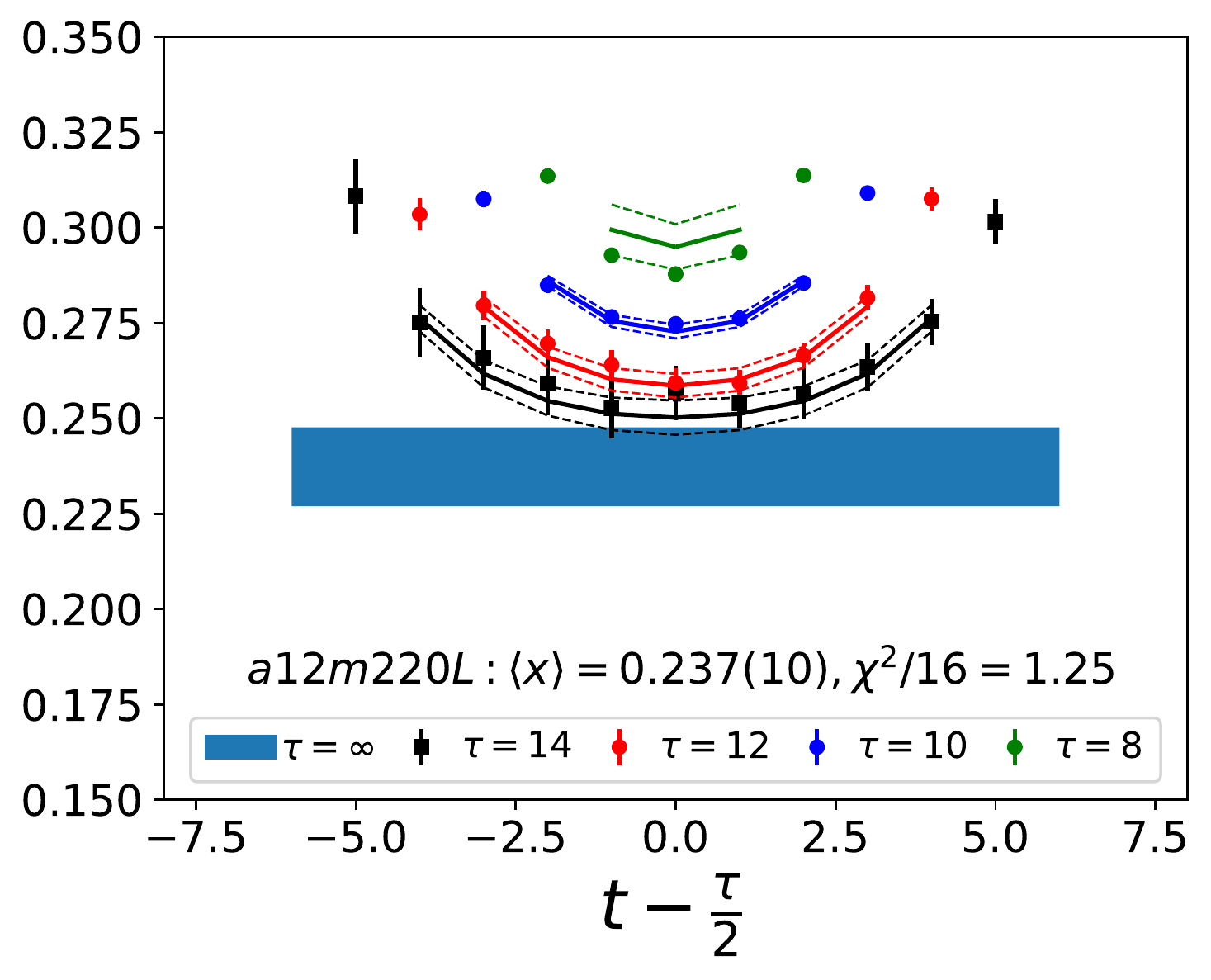}
\end{subfigure}
\begin{subfigure}
\centering
\includegraphics[angle=0,width=0.3\textwidth]{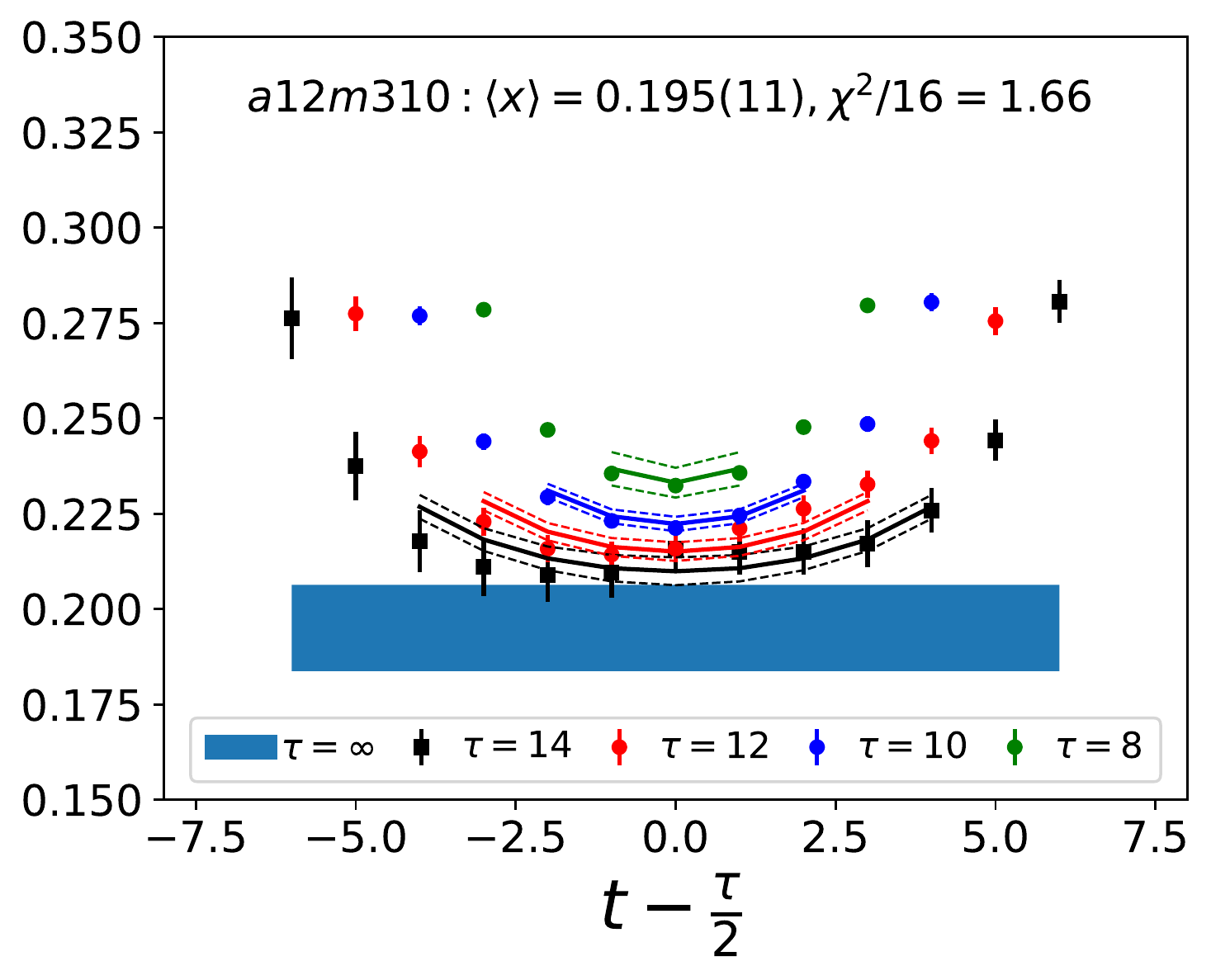}
\includegraphics[angle=0,width=0.3\textwidth]{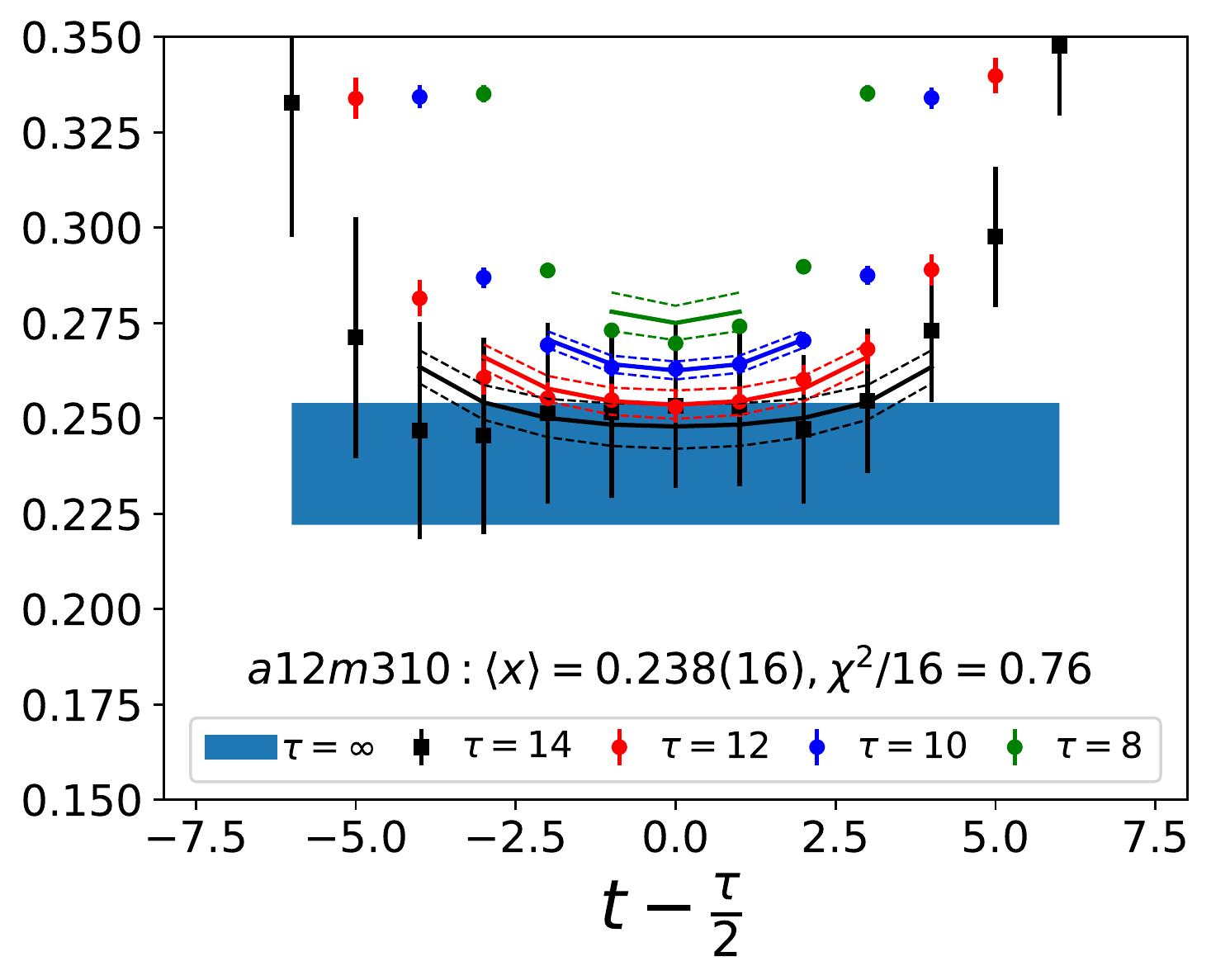}
\includegraphics[angle=0,width=0.3\textwidth]{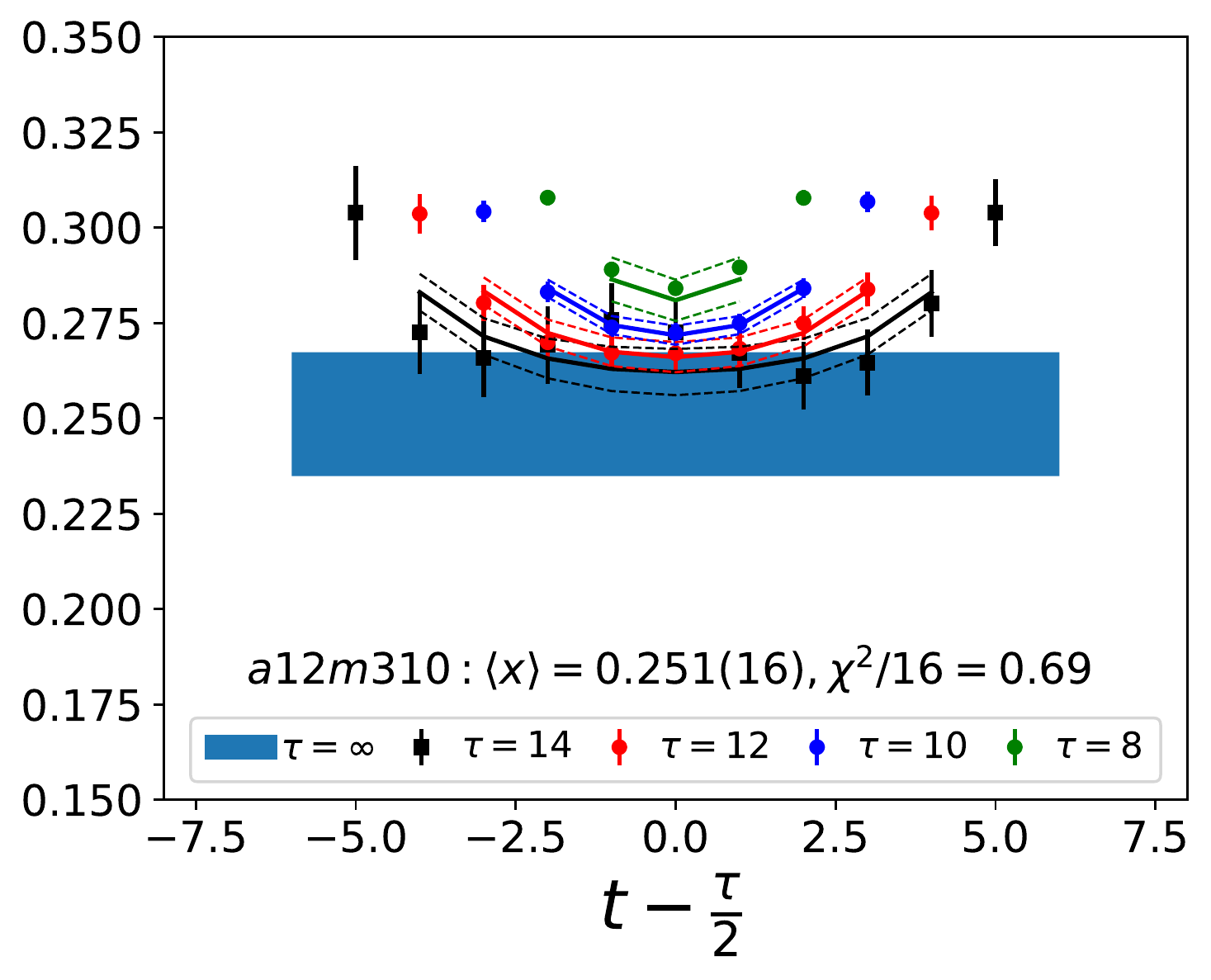}
\end{subfigure}
\begin{subfigure}
\centering
\includegraphics[angle=0,width=0.3\textwidth]{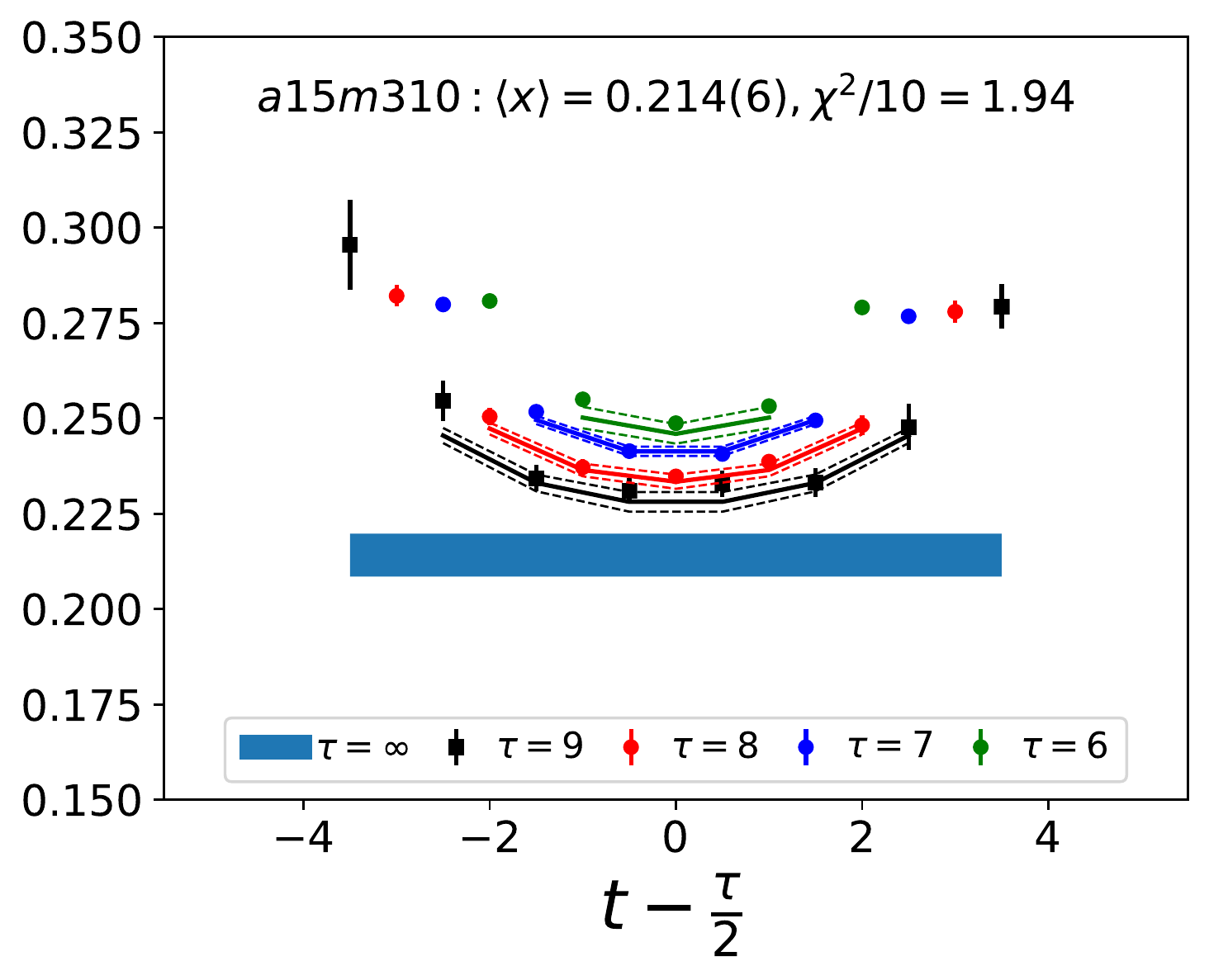}
\includegraphics[angle=0,width=0.3\textwidth]{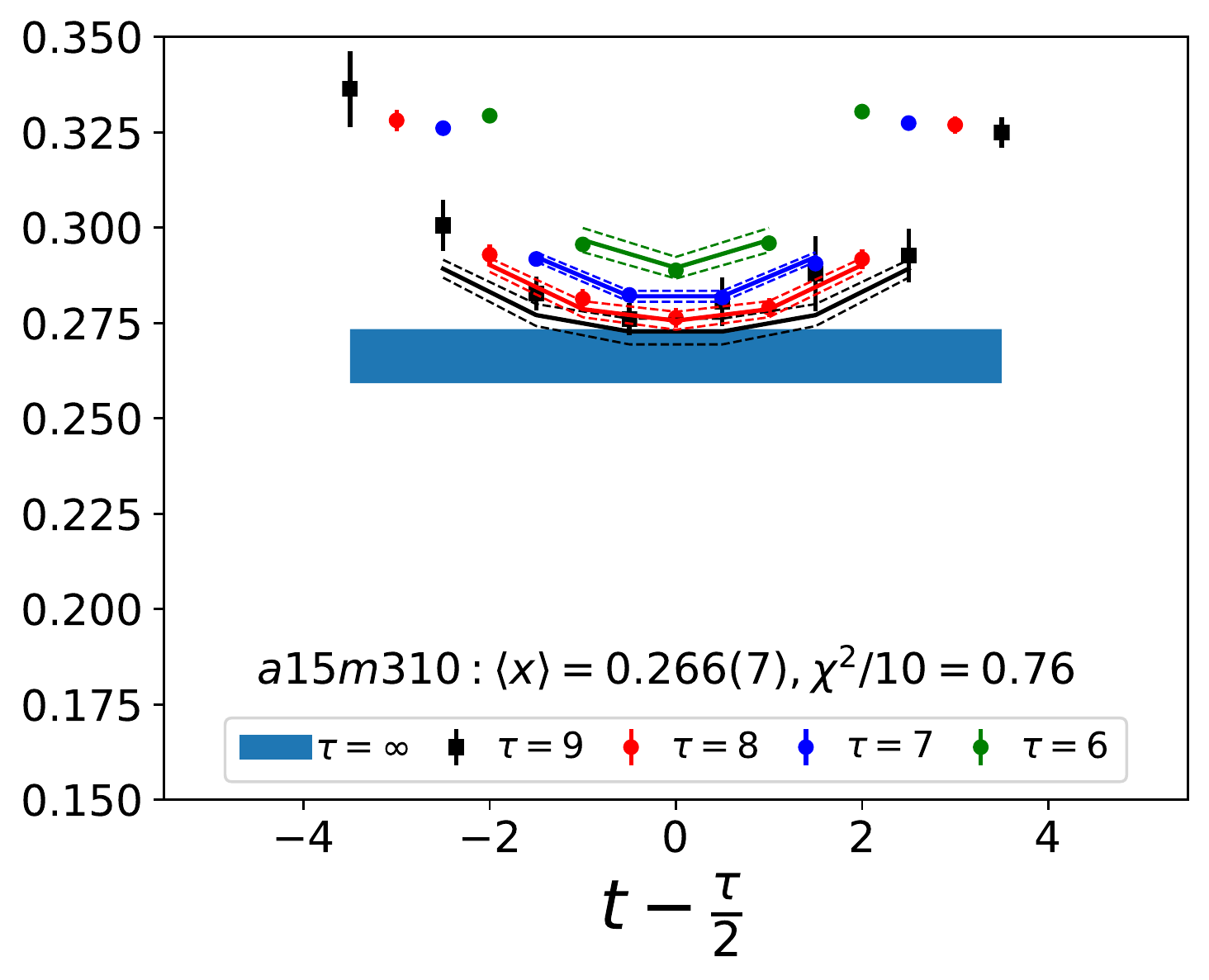}
\includegraphics[angle=0,width=0.3\textwidth]{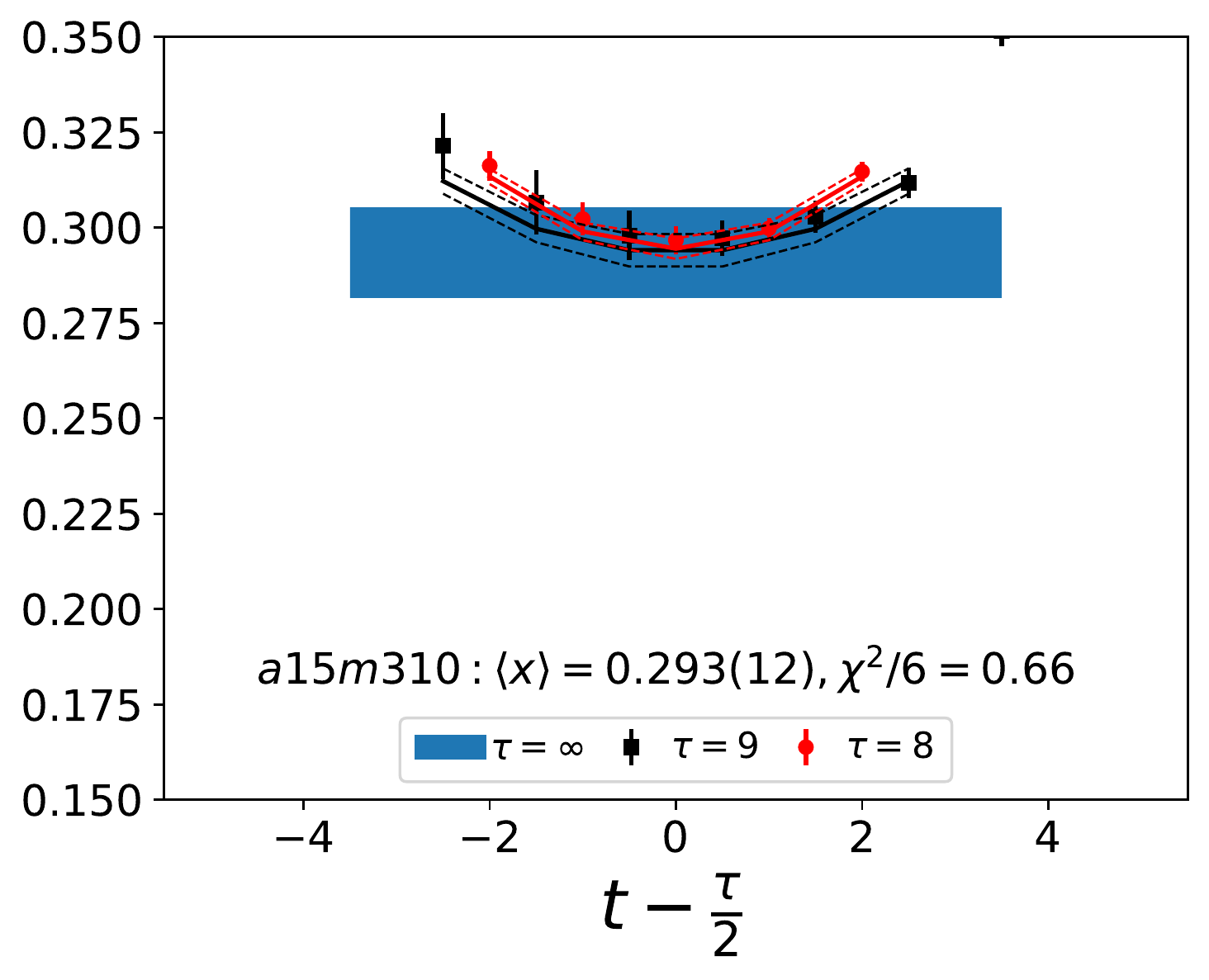}
\end{subfigure}
\caption{Data and fits for $a09m310$ (top row), $a12m220$ (second
  row), $a12m220L$ (third row), $a12m310$ (fourth row) and $a15m310$
  (bottom row) ensembles. The rest is the same as in
  Fig.~\protect\ref{fig:Ratio1}.}
\label{fig:Ratio2}
\end{figure*}

The results of the CCFV fits show that the finite volume correction
term, $c_4$, is not constrained.  We therefore, also present
results from a CC fit, i.e.,  with $c_4=0$ in Eq.~\eqref{eq:CCFV}. Results for
$c_1$ from the two fit ansatz overlap and there is a small positive
slope in both $a$ and $M_\pi^2$ for all three quantities.  The data
for both $\{4,3^\ast\}$ and $\{4,2^{\rm free}\}$, given in
Table~\ref{tab:best-fits}, are very similar, but with a systematic
shift of about 0.01--0.02 in all three cases.  This difference arises
because $\Delta M_1$ for $\{4,2^{\rm free}\}$ is larger (except in
$a09m310$) and because the convergence with respect to $\tau$ is from
above as shown in Figs.~\ref{fig:Ratio1} and~\ref{fig:Ratio2}, i.e., a larger $\Delta M_1$ implies a
smaller extrapolation and a larger $\tau \to \infty$ value.

For our final results we quote the CC fit values as the coefficient
$c_4$ of the finite-volume corrections in the CCFV fits is
undetermined. The CC results with the two strategies, $\{4,3^\ast \}$
and $\{4,2^{\rm free}\}$, are summarized in
Table~\ref{tab:finalresults}.  For our best estimates, we take the
$\{4,3^\ast \}$ results and add a second, systematic, error that is
the difference between these two strategies and represents the
uncertainty in controlling the excited-state contamination.

\begin{figure*}[htbp]
\begin{subfigure}
\centering
\includegraphics[angle=0,width=0.325\textwidth]{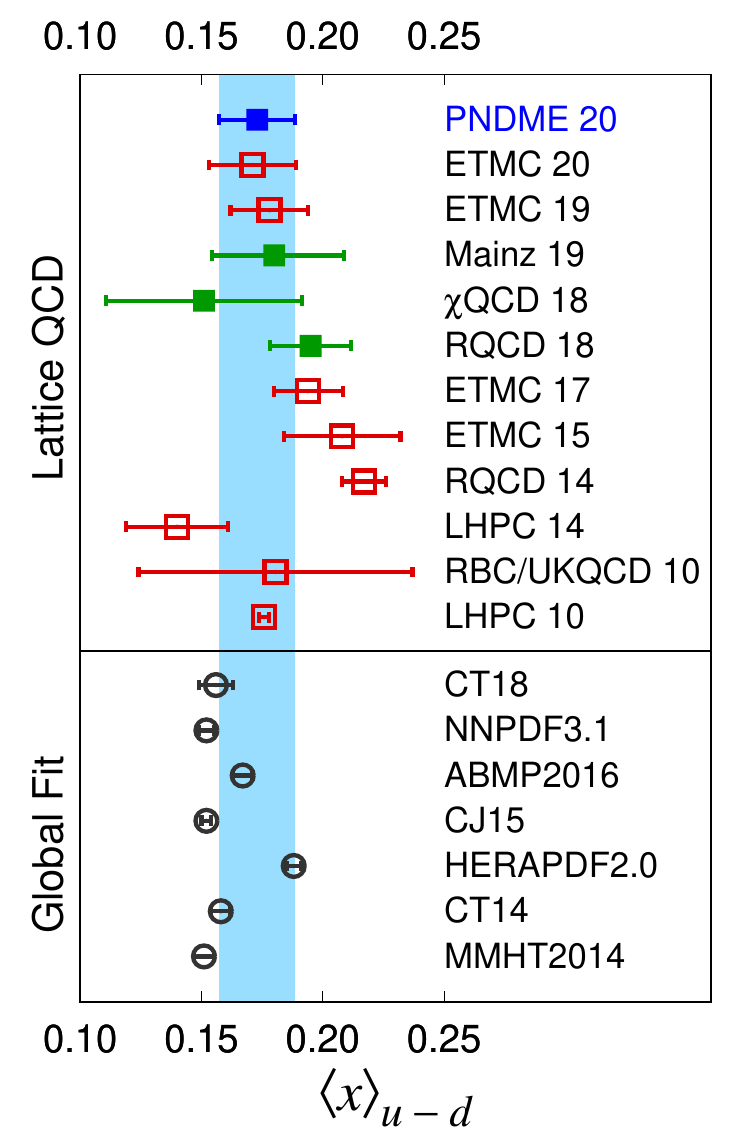}
\includegraphics[angle=0,width=0.325\textwidth]{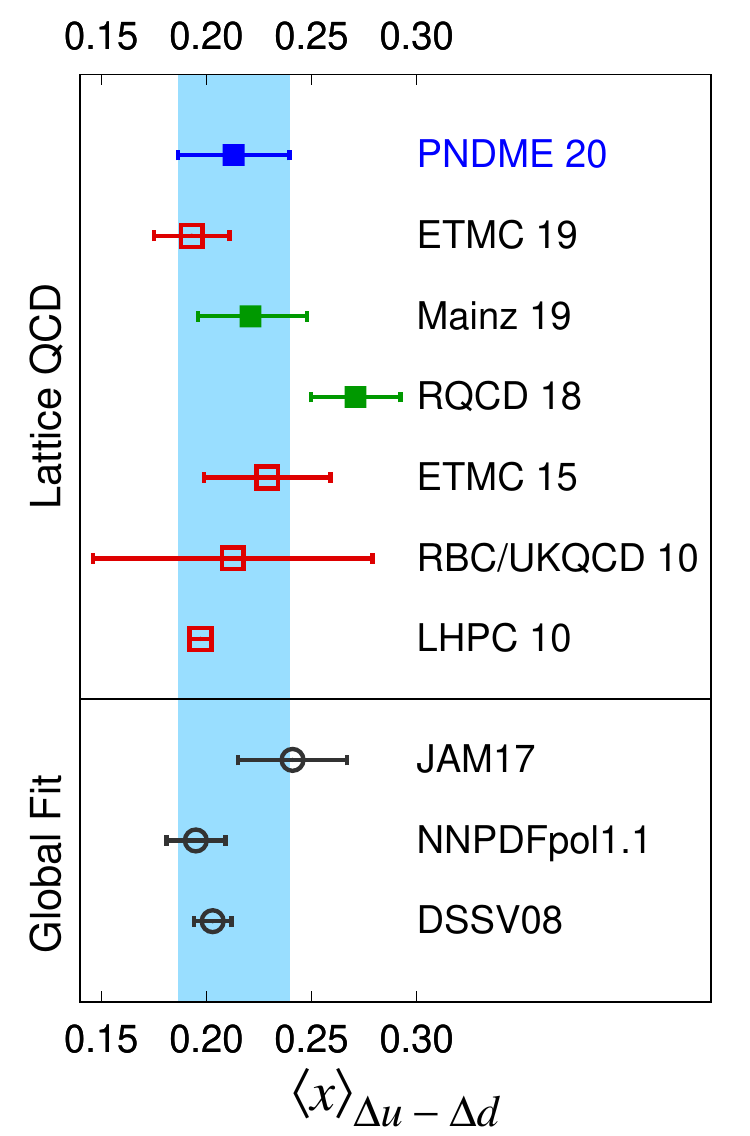}
\includegraphics[angle=0,width=0.325\textwidth]{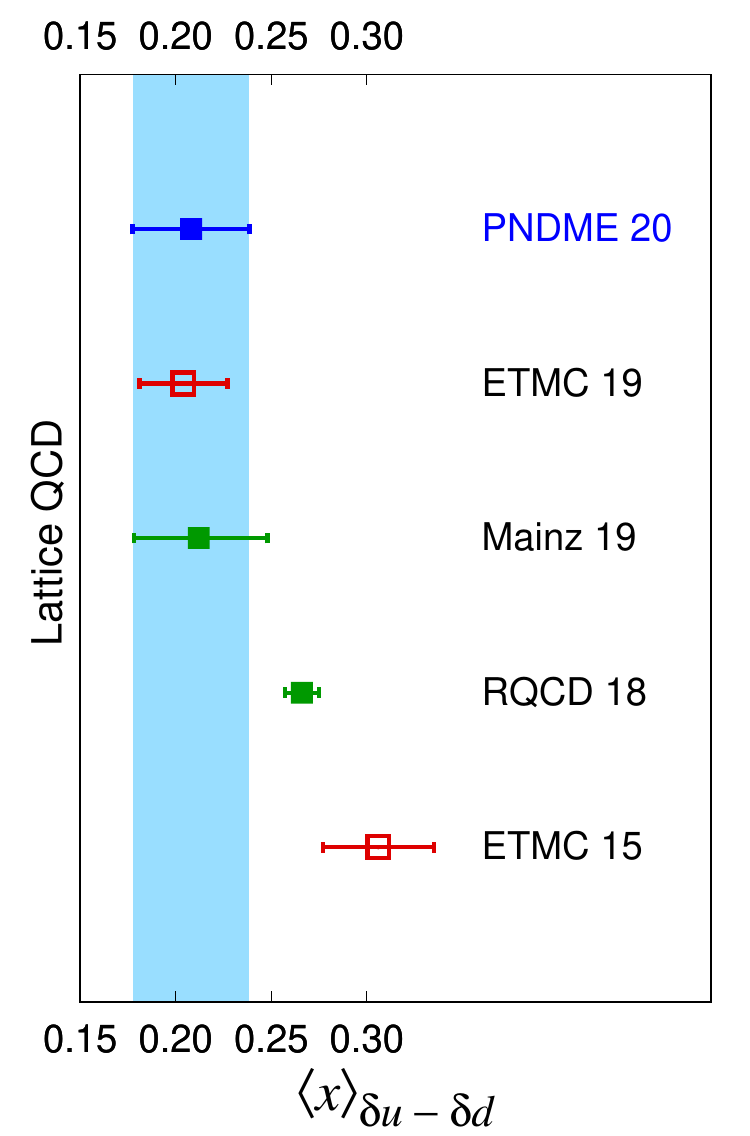}
\end{subfigure}
\vspace{-0.08in}
\caption{A comparison of results from lattice QCD calculations with dynamical fermions and
  global fits (below the black line) summarized in
  Table~\protect\ref{tab:Compare}. The left panel compares results for
  the momentum fraction, the middle for the helicity moment, and the
  right for the transversity moment. The PNDME 20 result is also shown
  as the blue band to facilitate comparison.}
\label{fig:summary}
\end{figure*}

A comparison of these results with other lattice QCD calculations on
ensembles with dynamical fermions is presented in the top half of
Table~\ref{tab:Compare} and shown in Fig.~\ref{fig:summary}. Our results agree with those from the Mainz
group~\cite{Harris:2019bih} that have also been obtained using data on
a comparable number of ensembles, but all with $M_\pi > 200$~MeV, which are used to 
perform a chiral and continuum extrapolation.  The one difference is
the slope $c_3$ of the chiral correction. For our clover-on-HISQ
formulation, we find a small positive value while the Mainz data show
a small negative value~\cite{Harris:2019bih}. Our results are also
consistent within 1$\sigma$ with the ETMC 20~\cite{Alexandrou:2020sml}
and ETMC 19~\cite{Alexandrou:2019ali} values that are from a single
physical mass ensemble.  The central value from $\chi$QCD~\cite{Yang:2018nqn}, using partially quenched 
analysis, is smaller but consistent within $1\,\sigma$. 
Results for the momentum fraction and the
helicity moment from RQCD 18~\cite{Bali:2018zgl} are taken from their
Set A with the difference between Set A and B values quoted as a
second systematic uncertainty. Their result for the transversity moment
is from a single $150$~MeV ensemble.  These values are larger,
especially for the helicity and transversity moment. Other earlier
lattice results show a spread, however, in each of these calculations,
the systematics listed in the last column of Table~\ref{tab:Compare}
have not been addressed or controlled and could, therefore, account
for the differences.

Estimates from phenomenological global fits, most of which have also
been reviewed in Ref.~\cite{Lin:2017snn}, are summarized in the bottom
of Table~\ref{tab:Compare} and shown in Fig.~\ref{fig:summary}. We find that results for the momentum
fraction from global fits are, in most cases, 1--2$\sigma$ smaller and
have much smaller errors.  Results for the helicity moment are
consistent and the size of the errors comparable.  Lattice estimates
of the transversity moment are a prediction.


\begin{table*}[htbp]
\setlength{\tabcolsep}{4pt}
\renewcommand{\arraystretch}{1.2}
\centering
\begin{tabular}{|c|c|c|c|c|c| }
\hline
Collaboration&Ref.&$\langle x \rangle_{u-d}$&$\langle x \rangle_{\Delta
u-\Delta d}$& $\langle x \rangle_{\delta u-\delta d}$& Remarks\\
\hline
\hline
PNDME 20 & &$0.173(14)(07)$&$0.213(15)(22)$&$0.208(19)(24)$& $N_f=2+1+1$\\
(this work)&&&&& clover-on-HISQ\\
\hline
ETMC 20 &\cite{Alexandrou:2020sml} &$0.171(18)$& & &$N_f=2+1+1$ twisted mass\\
&&&&& N-DIS, N-FV\\
\hline
ETMC 19 &\cite{Alexandrou:2019ali} &$0.178(16)$&$0.193(18)$ & $0.204(23)$&$N_f=2+1+1$ twisted mass\\
&&&&& N-DIS, N-FV\\
\hline
Mainz 19&\cite{Harris:2019bih}&$0.180(25)_{\rm stat}$&$0.221(25)_{\rm stat}$&$0.212(32)_{\rm stat}$&$N_f=2+1$ clover\\
        &                     &$(+14,-6)_{\rm sys}$  &$(+10,-0)_{\rm sys}$  &$(+16,-10)_{\rm sys}$ &  \\
\hline
$\chi$QCD 18&\cite{Yang:2018nqn} &$0.151(28)(29)$&        &          &$N_f=2+1$ \\
&&&&& overlap on domain wall\\
\hline
RQCD 18&\cite{Bali:2018zgl} &$0.195(07)(15)$&$0.271(14)(16)$&$0.266(08)(04)$ &$N_f=2$ clover\\
&&&&& \\
\hline
ETMC 17 &\cite{Alexandrou:2017oeh} &$0.194(9)(11)$& & &$N_f=2$ twisted mass\\
&&&&& N-DIS, N-FV\\ 
\hline
ETMC 15
&\cite{Abdel-Rehim:2015owa}&$0.208(24)$&$0.229(30)$&$0.306(29)$&$N_f=2$ twisted mass\\
&&&&&N-DIS, N-FV\\
\hline
RQCD 14&\cite{Bali:2014gha} &$0.217(9)$& & &$N_f=2$ clover\\
&&&&& N-DIS, N-CE, N-FV\\
\hline
LHPC 14 &\cite{Green:2012ud}&$0.140(21)$& & &$N_f=2+1$ clover\\
&&&&&N-DIS ($a \sim 0.12$ fm)\\
\hline
RBC/&\cite{Aoki:2010xg} &0.124--0.237&0.146--0.279& &$N_f=2+1$ domain wall\\
UKQCD 10&&&&&N-DIS, N-CE, N-ES\\
\hline  
LHPC 10 &\cite{Bratt:2010jn} &$0.1758(20)$&$0.1972(55)$& &$N_f=2+1$\\
&&&&& domain-wall-on-asqtad\\
&&&&& N-DIS, N-CE, N-NR, N-ES\\
\hline
\hline
\hline
CT18&\cite{Hou:2019efy}&$0.156(7)$& & &\\
\hline
JAM17${}^\dagger$&\protect\cite{Ethier:2017zbq,Lin:2017snn}& &0.241(26) & &\\
\hline
NNPDF3.1&\cite{Ball:2017nwa}&$0.152(3)$& & &\\
\hline
ABMP2016&\cite{Alekhin:2017kpj}&$0.167(4)$& & &\\
\hline
CJ15&\cite{Accardi:2016qay}&$0.152(2)$& & &\\
\hline
HERAPDF2.0&\cite{Abramowicz:2015mha}&$0.188(3)$& & &\\
\hline
CT14&\cite{Dulat:2015mca}&$0.158(4)$& & &\\
\hline
MMHT2014&\cite{Harland-Lang:2014zoa}&$0.151(4)$& & &\\
\hline
NNPDFpol1.1&\cite{Nocera:2014gqa}& &$0.195(14)$& &\\
\hline
DSSV08&\cite{deFlorian:2009vb,deFlorian:2008mr}& &$0.203(9)$& &\\
\hline
\end{tabular}
\caption{Our lattice QCD results are compared with other lattice
  calculations with $N_f$ flavors of dynamical fermions in rows 2--12, and with results from
  phenomenological global fits in the remainder of the table. In both
  cases, the results are arranged in reverse chronological order.  All
  results are in the $\MSbar$ scheme at scale $2$~GeV. For a
  discussion and comparison of lattice and global fit results up to
  2017 see Ref.~\protect\cite{Lin:2017snn};  and for a more recent
  comparison of $\langle x \rangle_{u-d}$ see Ref.~\cite{Hou:2019efy}. The
  JAM17${}^\dagger$ estimate for $\langle x \rangle_{\Delta u-\Delta
    d}$ is obtained from~\cite{Lin:2017snn}, where, as part of the
  review, an analysis was carried out using the data in
  \cite{Ethier:2017zbq}. The following abbreviations are used in the
  remarks column for various sources of systematic uncertainties in
  lattice calculations---DIS: discretization effects, CE: chiral
  extrapolation, FV: finite volume effects, NR: nonperturbative
  renormalization, ES: Excited state contaminations. A prefix "N-"
  means that the systematic uncertainty was not adequately controlled
  or not estimated.  }
\label{tab:Compare}
\end{table*}

\begin{figure*}[tbhp]
\begin{subfigure}
\centering
\includegraphics[angle=0,width=0.48\textwidth]{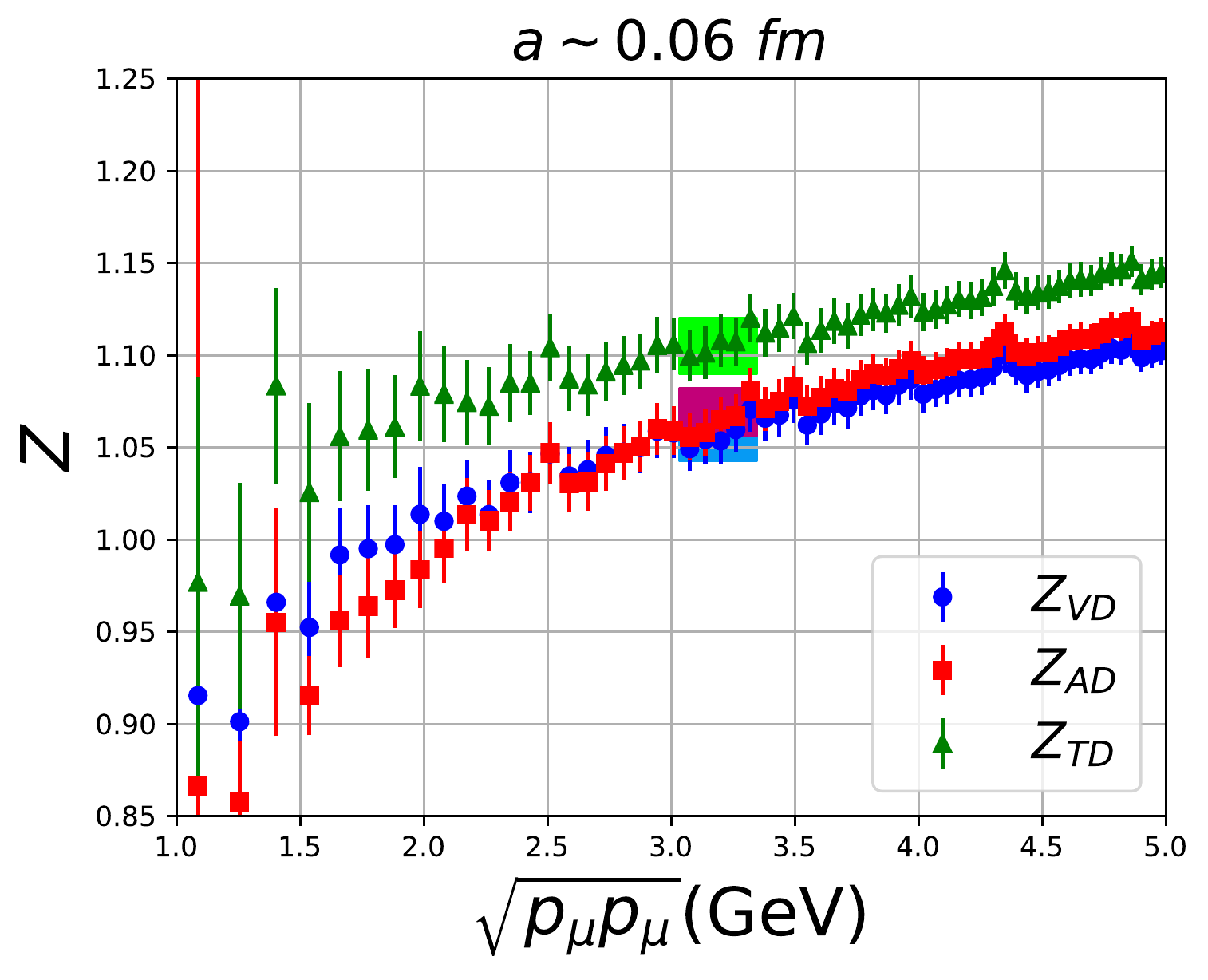}
\end{subfigure}
\begin{subfigure}
\centering
\includegraphics[angle=0,width=0.48\textwidth]{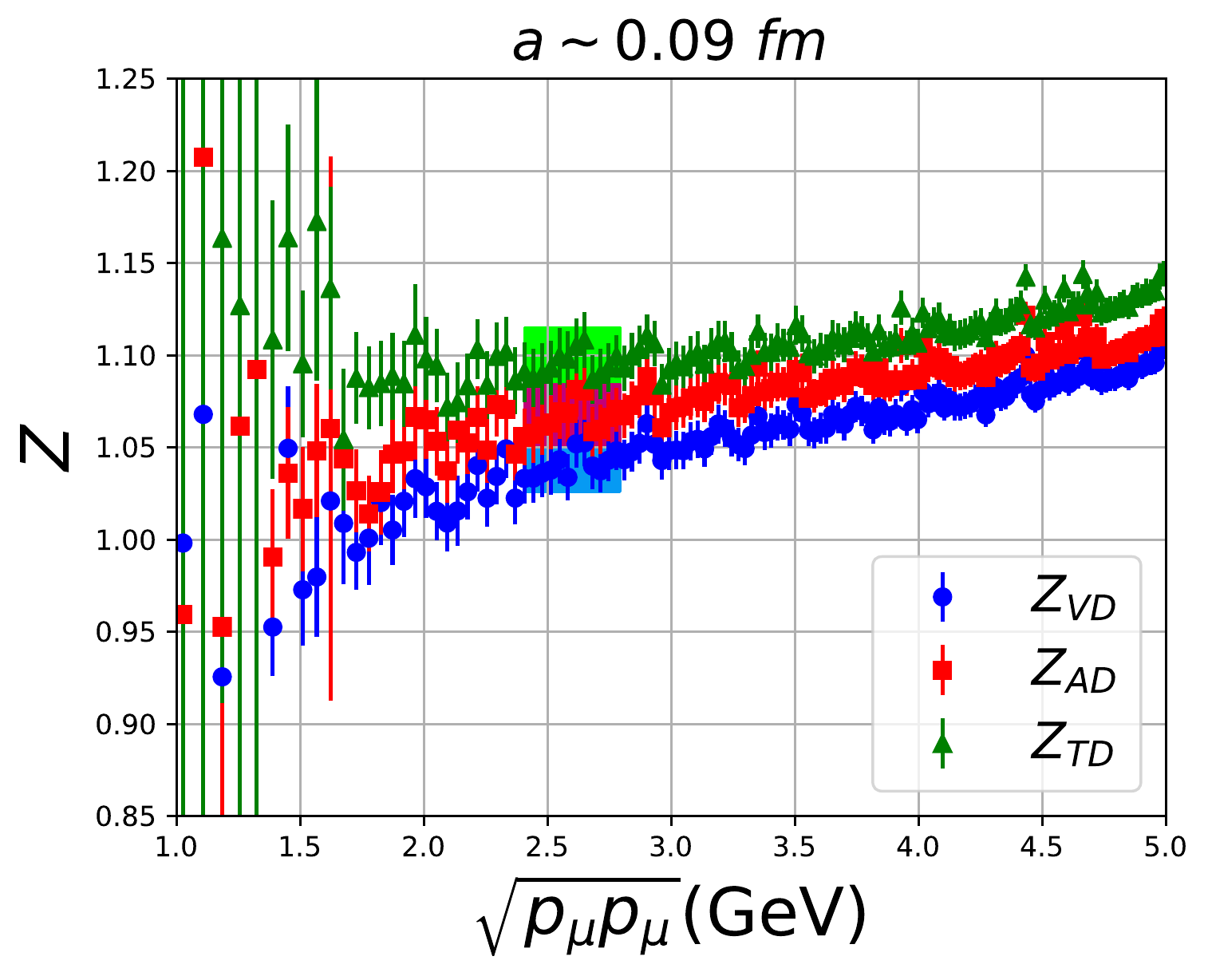}
\end{subfigure}
\begin{subfigure}
\centering
\includegraphics[angle=0,width=0.48\textwidth]{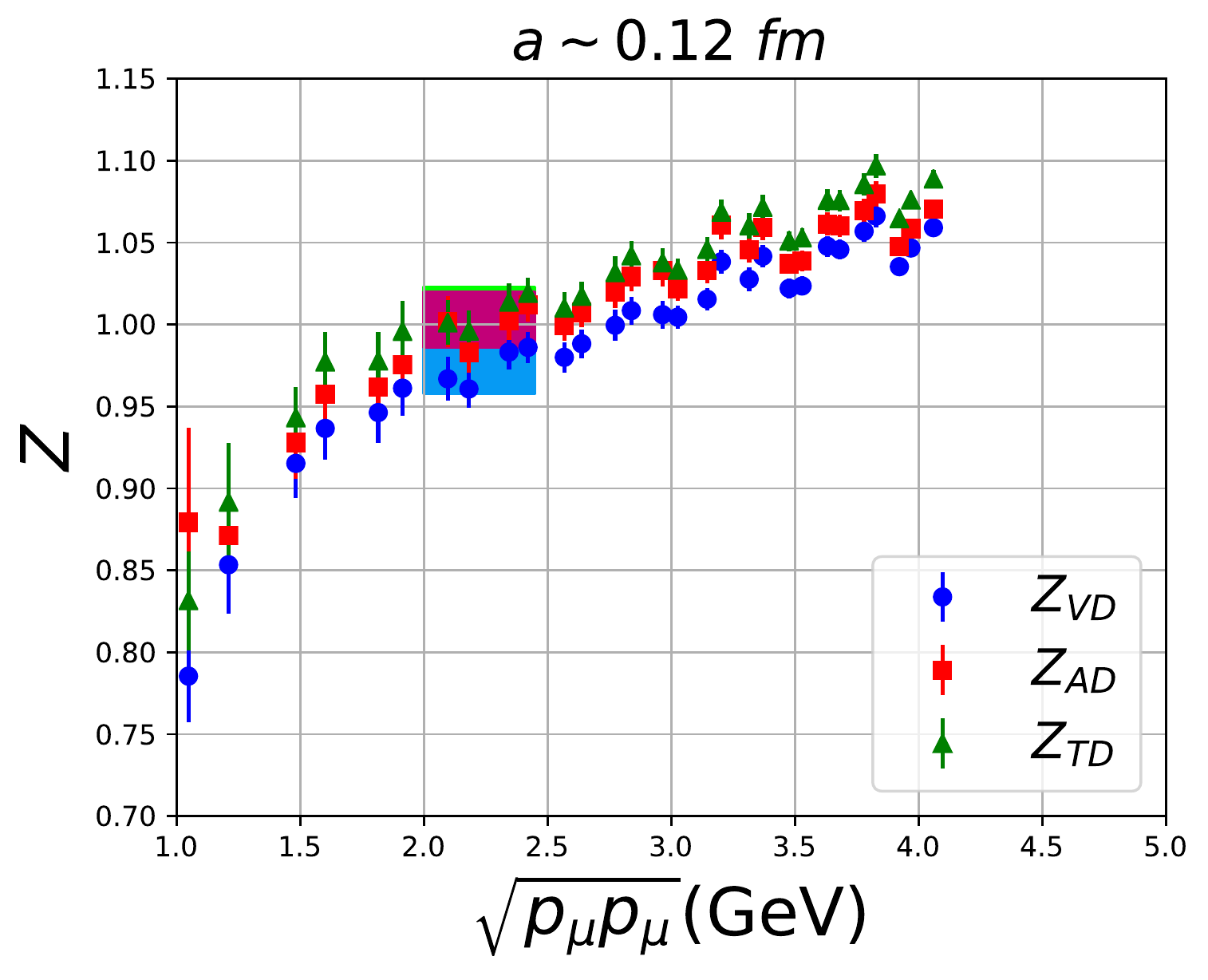}
\end{subfigure}
\begin{subfigure}
\centering
\includegraphics[angle=0,width=0.48\textwidth]{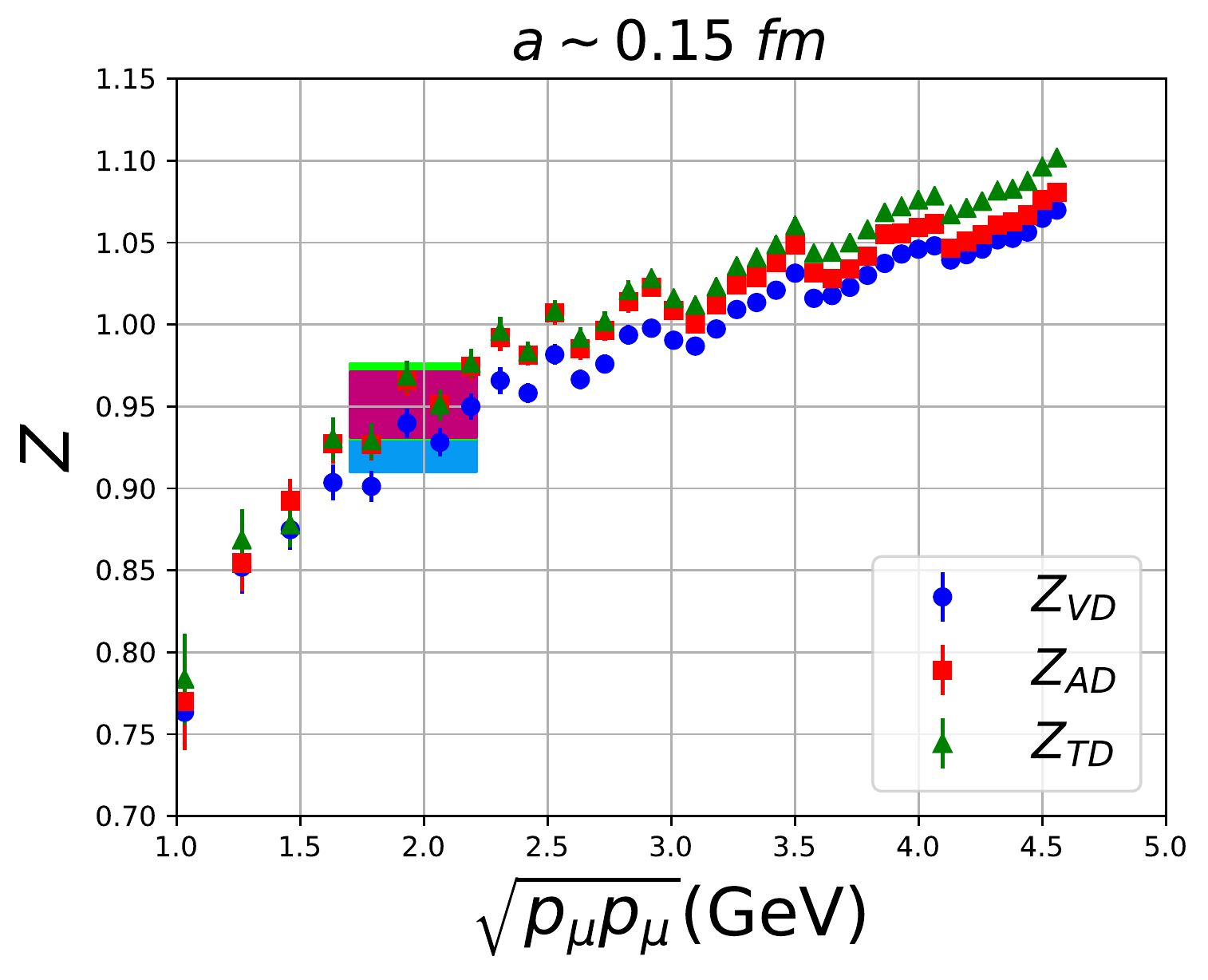}
\end{subfigure}
\caption{Nonperturbative renormalization factors for ${ \la
    x\ra_{u-d}}$, ($Z_{VD}$), ${ \la x \ra_{\Delta u - \Delta d}}$, ($Z_{AD}$),
  and ${ \la x\ra_{\delta u - \delta d}}$, ($Z_{TD}$), at the four lattice
  spacings in the ${\rm \ol{MS}}$ scheme at $\mu=2\ {\rm GeV}$.  The
  shaded bands mark the region in $\sqrt{p^2}$ that is averaged and the error in the estimate.}
\label{Z-a09m220}
\end{figure*}

\begin{figure*}[tbhp]
\begin{center}
\includegraphics[angle=0,width=0.85\textwidth]{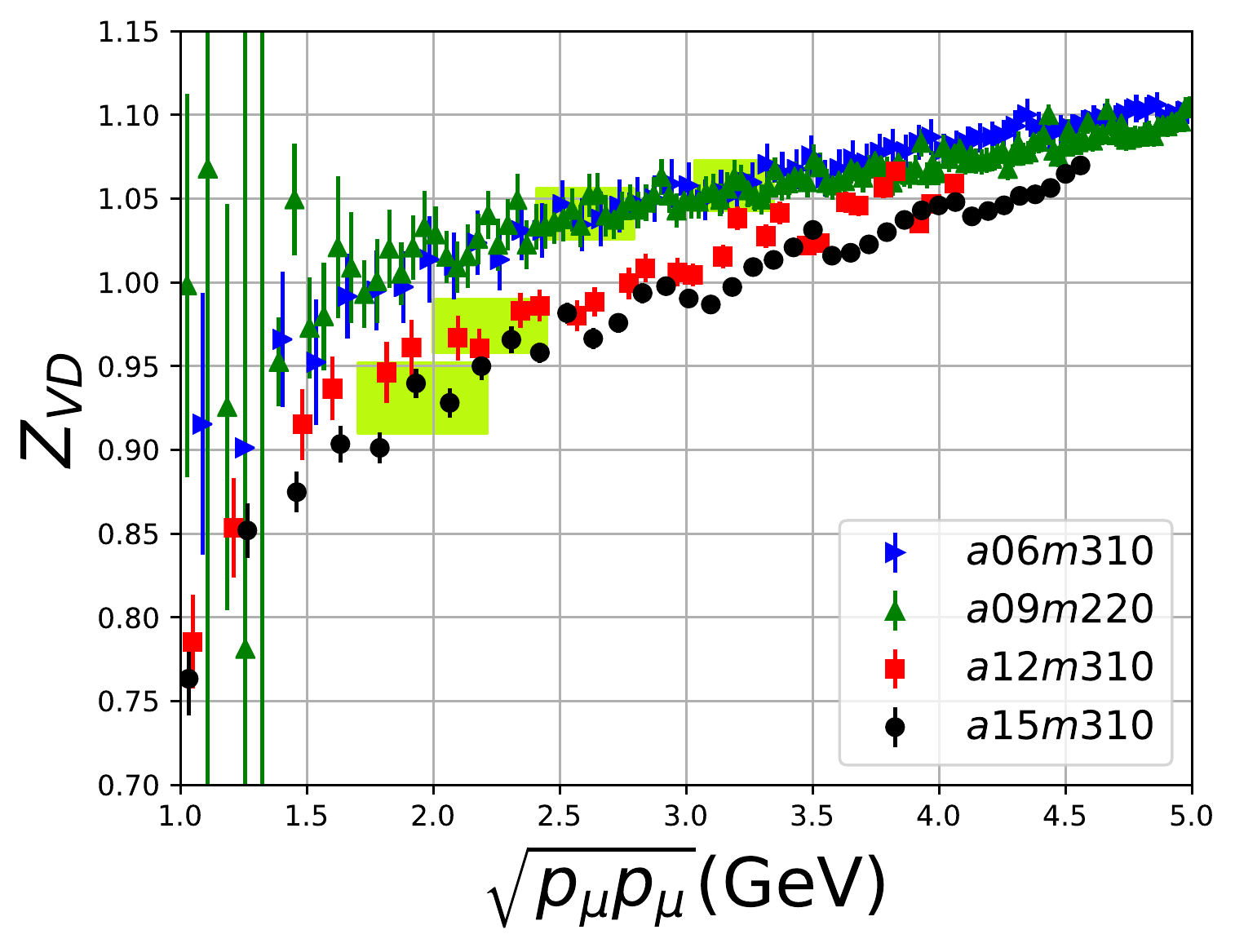}
\end{center}
\caption{Nonperturbative renormalization factor $Z_{VD}$ for ${
    \la x\ra_{u-d}}$ is calculated on four ensembles, one from each
  lattice spacing. The shaded bands give the interval in $\sqrt{p^2}$
  over which the data are averaged to get the result and the error in
  the estimate.}
\label{Z-momfrac-all-ensembles}
\end{figure*}


\section{Conclusions}
\label{sec:summary}

In this paper, we have presented results for the isovector quark
momentum fraction, $\la x \ra_{u-d}^{{\rm \ol{MS}}}$, helicity moment,
$\la x \ra^{{\rm \ol{MS}}}_{\Delta u- \Delta d}$, and transversity
moment, $\la x \ra^{{\rm \ol{MS}}}_{\delta u- \delta d}$, from a high
statistics lattice QCD calculation.  Attention has been paid to the
systematic uncertainty associated with excited-state contamination. We
have carried out the full analysis with different estimates of the
mass gaps of possible excited states, and use the difference in
results between the two strategies that give stable fits on all
ensembles as an additional systematic uncertainty to account for 
possible residual excited-state contamination.

The behavior versus $M_\pi$, the lattice spacing $a$ and finite volume
parameter $M_\pi L$ have been investigated using a simultaneous fit
that includes the leading correction in all three variables as given
in Eq.~\eqref{eq:CCFV}.  The nine data points cover the range $ 0.057
< a < 0.15$~fm, $135 < M_\pi < 320$~MeV and $3.7 < M_\pi L < 5.5$.
Over this range, all three moments, $\la x \ra_{u-d}^{{\rm \ol{MS}}}$,
$\la x \ra^{{\rm \ol{MS}}}_{\Delta u- \Delta d}$ and $\la x \ra^{{\rm
    \ol{MS}}}_{\delta u- \delta d}$, do not show a large dependence on
$a$ or $M_\pi$ or $M_\pi L$. As shown in Table~\ref{tab:CCFV},
possible dependence on the lattice size, characterized by $M_\pi L$,
is not resolved by the data, i.e., the coefficient $c_4$ is
unconstrained. We, therefore, take for our final results those
obtained from just the chiral-continuum fit. 

The small increase with $a$ and $M_\pi^2$, evident in
Figs~\ref{fig:momfrac-vs-a}--\ref{fig:transversity-vs-a}, is well fit
by the leading correction terms that are linear in these
variables. Also, for all three observables, the chirally extrapolated
value is consistent with the data from the two physical mass
ensembles. In short, the observed small dependence in all three
variables, and having two data points at $M_\pi \sim 135$~MeV to
anchor the chiral fit, allows a controlled extrapolation to the
physical point, $M_\pi = 135$~MeV and $a = 0$.

Our final results, given in Table~\ref{tab:finalresults}, are compared
with other lattice calculations and phenomenological global fit
estimates in Table~\ref{tab:Compare} and shown in Fig.~\ref{fig:summary}.  Estimates of all three
quantities are in good agreement with those from the Mainz
Collaboration~\cite{Harris:2019bih}, also obtained using a
chiral-continuum extrapolation, the ETMC
Collaboration~\cite{Alexandrou:2020sml,Alexandrou:2019ali} that are from a single
physical mass ensemble, and from the $\chi$QCD Collaboration~\cite{Yang:2018nqn}. 
On  the other hand, most global fit estimates
for the momentum fraction are about 10\% smaller and have much smaller
errors, while those for the helicity moment are consistent within
$1\sigma$. Lattice estimates for the transversity moment are a
prediction. The overall consistency of results suggests that lattice
QCD calculations of these isovector moments are now mature and future
calculations will steadily reduce the statistical and systematic
uncertainties in them.

\begin{acknowledgments}
We thank the MILC Collaboration for sharing the HISQ ensembles, and
Martha Constantinou, Giannis Koutsou, Emanuele Nocera and Juan Rojo
for discussions. The calculations used the Chroma software
suite~\cite{Edwards:2004sx}. Simulations were carried out on computer
facilities of (i) the National Energy Research Scientific Computing
Center, a DOE Office of Science User Facility supported by the Office
of Science of the U.S. Department of Energy under Contract
No. DE-AC02-05CH11231; (ii) the Oak Ridge Leadership Computing
Facility at the Oak Ridge National Laboratory, which is supported by
the Office of Science of the U.S. Department of Energy under Contract
No. DE-AC05-00OR22725; (iii) the USQCD Collaboration, which is funded
by the Office of Science of the U.S. Department of Energy, and (iv)
Institutional Computing at Los Alamos National Laboratory.
T. Bhattacharya and R. Gupta were partly supported by the
U.S. Department of Energy, Office of Science, Office of High Energy
Physics under Contract No.~DE-AC52-06NA25396.  T. Bhattacharya,
R. Gupta, S. Mondal, S. Park and B.Yoon were partly supported by the
LANL LDRD program, and S. Park by the Center for Nonlinear Studies.
The work of H.-W. Lin is partly supported by the US National Science
Foundation Grant No. PHY 1653405 ``CAREER: Constraining Parton
Distribution Functions for New-Physics Searches'', and by the Research
Corporation for Science Advancement through the Cottrell Scholar
Award.

\end{acknowledgments}

\appendix

\section{Plots of the Ratio \texorpdfstring{$C_{\mathcal{O}}^{3\text{pt}}(\tau;t)/C^{2\text{pt}}(\tau)$}{C(3pt)/C(2pt)}}
\label{sec:ratios}

In this appendix, we show in~Figs.~\ref{fig:Ratio1}
and~\ref{fig:Ratio2}, plots of the unrenormalized isovector momentum
fraction, $\la x \ra_{u-d}$, the helicity moment, $\la x \ra_{\Delta
  u-\Delta d}$, and the transversity moment, $\la x\ra_{\delta
  u-\delta d}$, for the nine ensembles. The data show the ratio
$C_\mathcal{O}^{3\text{pt}}(\tau;t)/C^{2\text{pt}}(\tau)$ multiplied
by the appropriate factor given in Eqs.~\eqref{eq:me2momentV}--\protect\eqref{eq:me2momentT} to get 
$\langle x \rangle$. The lines with the same color as the data are the
result of the fit to $C_\mathcal{O}^{3\text{pt}}(\tau;t)$ using
Eq.~\eqref{eq:3pt}. In all cases, to extract the ground state matrix
element, the fits to $C^{2\text{pt}}(\tau)$ and
$C_\mathcal{O}^{3\text{pt}}(\tau;t)$ are done within a single
jackknife loop.

\section{Renormalization}
\label{sec:renormalization}

\begin{table}[htbp]
\setlength{\tabcolsep}{4pt}
\renewcommand{\arraystretch}{1.3}
\centering
\begin{tabular}{ |c|c|c|c|c| }
\hline
$a$ &Fit-range      &$Z_{VD}$&$Z_{AD}$&$Z_{TD}$\\
$[{\rm fm}]$&$[{\rm GeV^2}]$&        &        &        \\
\hline
\hline
$0.06$&$9.2-11.2$&$1.058(30)$&$1.069(26)$&$1.105(30)$\\
\hline
$0.09$&$5.8-7.8$ &$1.041(30)$&$1.067(34)$&$1.097(36)$\\
\hline
$0.12$&$4.0-6.0$ &$0.974(32)$&$1.003(34)$&$1.007(32)$\\
\hline
$0.15$&$2.9-4.9$ &$0.931(42)$&$0.951(40)$&$0.953(46)$\\
\hline
\hline
\end{tabular}
\caption{Results for the renormalization factors, $Z_{VD,AD,TD}$, in
  the $\MSbar$ scheme at $2$~GeV.  These are calculated in the \ripmom\ 
  scheme as a function of scale $p=\sqrt{p_\mu p_\mu}$ on the lattice,
  matched to the $\MSbar$ scheme at the same scale $\mu=p$, and then
  run in the continuum $\MSbar$ scheme from $\mu$ to $2$~GeV. The
  results are the average of values over the range of $|p|$ specified
  in the second column. The final error estimate is taken to be twice
  that shown in Figs.~\protect\ref{Z-a09m220}
  and~\ref{Z-momfrac-all-ensembles}. }
\label{tab:Z-fac}
\end{table}

In this appendix, we describe the calculation of the renormalization
factors, $Z_{VD,AD,TD}$, for the three one-derivative operators. These
are determined nonperturbatively on the lattice in the
\ripmom\ scheme~\cite{Gockeler:2010yr,Constantinou:2013ada} as a
function of the lattice scale $p^2 = p^\mu p^\mu$, and then converted
to the ${\rm \ol{MS}}$ scheme using $3$-loop perturbative factors
calculated in the continuum in Ref.~\cite{Gracey:2003mr}. For data at
each $p$, we perform horizontal matching by choosing the ${\rm
  \ol{MS}}$ scale $\mu=|p|$. These numbers are then run in the
continuum ${\rm \ol{MS}}$ scheme from scale $\mu$ to $2$~GeV using
three-loop anomalous dimensions~\cite{Gracey:2003mr}.  
Note that the decomposition of the three operators into irreducible
representations given in Refs.~\cite{Gockeler:1995wg,Harris:2019bih},
shows that they can only mix with higher dimensional operators. Such
$O(a)$ effects would also be taken into account in our
CCFV fits, and removed by the continuum extrapolation.

We calculate $Z_{VD,AD,TD}$ for one value of $M_\pi$ at each
$a$.  Based on our experience with local
operators~\cite{Bhattacharya:2016zcn}, where we found insignificant
dependence of results on $M_\pi$, we assume that these results, within
the conservative error estimates we assign, give the mass-independent
renormalization factors at each $a$.  Evidence that the dependence on
$M_\pi^2$ is tiny for these 1-link operators also
comes from explicit calculations in
Refs.~\cite{Harris:2019bih,Alexandrou:2020sml}, albeit with different
latice actions. In each case, the dependence on $M_\pi^2$ is found to
be much smaller than $ 1\%$. The dominant uncertainty comes from the
dependence on $p^2$, which is discussed next.

In Fig.~\ref{Z-a09m220}, we show the behavior of the renormalization
factors $Z_{VD, AD, TD}$ in the ${\rm \ol{MS}}$ scheme at $\mu = 2$ GeV
for the four ensembles as a function of $|{p}|$---the scale of
the \ripmom\ scheme on the lattice. In
Fig.~\ref{Z-momfrac-all-ensembles} we compare $Z_{VD}$, used to
renormalize ${\la x \ra_{u-d}}$, calculated on four ensembles, one at
each lattice spacing.

For all three operators, the data do not show a window in $|{p}|$
where the results are independent of $|{p}|$. The variation in
the data is due to a combination of the breaking of full rotational
invariance on the lattice and other $p^2$ dependent artifacts. This is
the dominant uncertainty and many methods have been proposed to
control it, see for example
Refs.~\cite{Harris:2019bih,Alexandrou:2020sml,Bhattacharya:2016zcn}. In
Ref.~\cite{Bhattacharya:2016zcn}, we explored three methods that gave
consistent results, and of these we have selected the strategy labeled
``Method B'' there as it is the most straightforward. In this
approach, we take an average over the data points in an interval of
$2~ {\rm GeV^2}$ about ${p^2}= \Lambda/a$, where the scale $\Lambda =
3$ GeV is chosen to be large enough to avoid nonperturbative effects
and at which perturbation theory is expected to be reasonably well
behaved. Also, this choice satisfies both $pa \rightarrow 0$ and
$\Lambda/p \rightarrow 0$ in the continuum limit as desired.  The
window over which the data are averaged and the error (half the height
of the band) are shown by shaded bands in Figs.~\ref{Z-a09m220}
and~\ref{Z-momfrac-all-ensembles}.  Noting the large variation
with $p^2$, we take twice this error, ie, full
height of the band, for a very conservative error estimate for all three $Z'$s. 

These final estimates of $Z_{VD}$, $Z_{AD}$ and $Z_{TD}$ used to
renormalize the momentum fraction, the helicity moment and the
transversity moment, respectively, are given in Table~\ref{tab:Z-fac}.

\bibliography{ref} 

\end{document}